\documentclass[%
 reprint,
superscriptaddress,
amsmath,amssymb,
aps,
pra,
]{revtex4-1}

\usepackage{multirow}

\usepackage{graphicx}
\usepackage{dcolumn}
\usepackage{bm}

\usepackage{hyperref}
\hypersetup{
    colorlinks=true,
    linkcolor=blue,
    citecolor=blue,
    urlcolor=blue
}

\usepackage{physics}
\usepackage{float}

\usepackage{numprint}

\graphicspath{{figures/}}

\usepackage{xifthen}

\usepackage{xcolor}

\usepackage{epstopdf}

\usepackage{multirow}
\usepackage[normalem]{ulem}
\definecolor{rulecolor}{RGB}{0,71,171}

\definecolor{tableheadcolor}{gray}{0.92}

\DeclareUnicodeCharacter{3000}{\textcolor{red}{BAD!!}}

\DeclareUnicodeCharacter{2009}{\textcolor{red}{BAD!!}}

\DeclareUnicodeCharacter{03A3}{$\Sigma$}

\DeclareUnicodeCharacter{03A0}{$\Pi$}

\DeclareUnicodeCharacter{03BC}{$\mu$}

\newcommand{\Xstate}{X^1\Sigma^+}
\newcommand{\Astate}{A^1\Pi}
\newcommand{\apistate}{a^3\Pi}

\newcommand{\XAtransition}{$\Astate \leftarrow \Xstate$}

\newcommand{\transition}[4]{
\Xstate\,|v = #1, J = #2\rangle\,\leftrightarrow \Astate\,|v' = #3, J' = #4\rangle
}

\newcommand{\alcl}[2][]{\ifthenelse{\equal{#1}{}}{Al$^{#2}$Cl}{$^{#1}$Al$^{#2}$Cl}}

\newcommand{\rfig}[1]{Fig.~\ref{#1}}
\newcommand{\rsec}[1]{Section~\ref{#1}}

\newcommand{\rtab}[1]{Tab.~\ref{#1}}
\newcommand{\refcite}[1]{Ref.~\cite{#1}}
\newcommand{\refscite}[1]{Refs.~\cite{#1}}

\newcommand{\change}[1]{{#1}}

\newcommand{\remove}[1]{\ignorespaces}

\begin{document}

\title{Hyperfine structure of the $\mathbf{\Astate}$ state of AlCl and its relevance to laser cooling and trapping}

\author{John R.~Daniel}%
\thanks{These two authors contributed equally.}
\affiliation{Department of Physics and Astronomy, University of California, Riverside, California 92521, USA}

\author{Jamie C.~Shaw}%
\thanks{These two authors contributed equally.}
\affiliation{Department of Physics, University of Connecticut, 196A Auditorium Road, Unit 3046, Storrs, CT 06269-3046}

\author{Chen Wang}%
\affiliation{Department of Physics and Astronomy, University of California, Riverside, California 92521, USA}

\author{Li-Ren Liu}%
\affiliation{Department of Physics and Astronomy, University of California, Riverside, California 92521, USA}

\author{Brian K.~Kendrick}%
\affiliation{Theoretical Division (T-1, MS B221), Los Alamos National Laboratory, Los Alamos, New Mexico 87545, USA}

\author{Boerge Hemmerling}%
\email{boergeh@ucr.edu}
\affiliation{Department of Physics and Astronomy, University of California, Riverside, California 92521, USA}

\author{Daniel J.~McCarron}%
\email{daniel.mccarron@uconn.edu}
\affiliation{Department of Physics, University of Connecticut, 196A Auditorium Road, Unit 3046, Storrs, CT 06269-3046}

\date{\today}

\begin{abstract}
The majority of molecules proposed for laser cooling and trapping experiments have $\Sigma$-type ground states. Specifically, $^2\Sigma$ states have cycling transitions analogous to $D_1$-lines in alkali-metal atoms while $^1\Sigma$ states offer both strong and weak cycling transitions analogous to those in alkaline-earth atoms. Despite this proposed variety, to date, only molecules with $^2\Sigma$-type ground states have successfully been confined and cooled in magneto-optical traps. While none of the proposed $^1\Sigma$-type molecules have been successfully laser cooled and trapped, they are expected to have various advantages in terms of exhibiting a lower chemical reactivity and an internal structure that benefits the cooling schemes. Here, we present the prospects and strategies for optical cycling in AlCl -- a $^1\Sigma$ molecule -- and report on the characterization of the $\Astate$ state hyperfine structure. Based on these results, we carry out detailed simulations on the expected capture velocity of a magneto-optical trap for AlCl. Finally, using {\it ab initio} calculations, \change{we identify the photodissociation via a $3^1\Pi$ state and photoionization process via the $3^1\Sigma^+$ state as possible loss mechanisms for a magneto-optical trap of AlCl.}
\end{abstract}

\maketitle

\setcounter{page}{1}


\section{Introduction}
\label{sec:intro}

The ability to control the rich internal and external degrees of freedom of polar molecules has the prospect of enabling a large number of novel applications, including the search for new physics beyond the Standard Model and precision measurements \cite{Andreev2018,Cairncross2017,Kozyryev2017a,Hudson2011,Kozyryev2021,Kondov2019,ACMECollaboration2014,Fitch2021a,Cairncross2017,Yu2021,Hutzler2020,ORourke2019,Aggarwal2018,Uzan2003,DeMille2008,Chin2009,Kajita2009,Beloy2010,Jansen2014,Dapra2016,Kobayashi2019,Chupp2019}, controlled chemistry \cite{Krems2008,Ni2010,Ye2018,Ospelkaus2010}, and quantum simulation and computation \cite{DeMille2002,Yelin2006,Yu2019,Carr2009,Micheli2006,Bao2022,Holland2022}.
Realizing the necessary control for such applications can realistically only be achieved at low temperatures, where only a small number of quantum states are occupied, and with trapped samples that allow for long interaction times.
One way to produce ultracold molecules is to associate laser cooled atoms with carefully controlled external fields. While this method has been successful, it is limited to molecules which consist of laser coolable atoms \cite{Sage2005,Ni2008,Danzl2010,Aikawa2010,Takekoshi2014,Molony2014,Park2015,Guo2016,Liu2018,DeMarco2019}.
On the other hand, over the past two decades, a growing number of molecules have been identified with internal structures that allow for photon cycling to an extent that renders these molecules amenable to direct laser cooling and trapping \cite{Rosa2004,McCarron2018a,Tarbutt2019,Fitch2021,Langen2023,Langin2023}.
Among those species, a diverse range of diatomic molecules have been explored both theoretically \cite{
Bahns1996, 
Xiao2021, 
Gao2015,  
Gao2014, 
Xiao2022, 
Xiao2022b, 
Xiao2022c, 
Xiao2021, 
Yang2016, 
Kang2016, 
Yang2016, 
Kang2015, Kang2015b, Rodriguez2023, 
Lambo2023, Marsman2023, 
ElKork2023, 
Yang2014} 
and experimentally
\cite{
Lim2018, 
Iwata2017, 
Norrgard2017, 
Stuhl2008, 
Isaev2010, 
Xu2016,Norrgard2023, 
Chen2019,Liang2021,Bu2017,Chen2016,Bu2022,Chen2017,Albrecht2020,Kogel2021,Zhang2022,Rockenhaeuser2023, 
Truppe2019,Doppelbauer2021, 
Schnaubelt2021, 
Bahns1996, 
McNally2020, Tarallo2016}. 
Furthermore, this experience has helped to guide recent efforts extending these techniques to polyatomic species \cite{
Isaev2016,Kozyryev2016,Kozyryev2016a, 
Baum2020,Augenbraun2021, 
Mitra2020, 
Vilas2022, 
Kozyryev2017, 
Augenbraun2020,Kozyryev2017a, 
Augenbraun2023}.

Nevertheless, at present, only the diatomic molecules SrF \cite{Barry2014,McCarron2018}, CaF \cite{Anderegg2017,Truppe2017,Williams2017}, and YO \cite{Collopy2018} and the polyatomic CaOH \cite{Vilas2022} have successfully been laser cooled and confined in a magneto-optical trap (MOT), a crucial milestone towards the growing list of applications. All of these molecules possess an unpaired electronic spin and a $^2\Sigma$-type ground state with optical cycling transitions analogous to $D_1$ lines in alkali-metal atoms. By contrast, molecules with $^1\Sigma$ ground states are expected to offer certain advantages for laser cooling, as their closed shells render them less reactive and their internal structures show similarities to atoms in a type-II MOT. For $^1\Sigma$ molecules, both strong and weak optical cycling transitions may be available, analogous to the $^{1}S_{0}\rightarrow\,^{1}P_{1}$ and $^{1}S_{0}\rightarrow\,^{3}P_{J}$ transitions regularly used in alkaline-earth atoms.
To the best of our knowledge, the only three $^1\Sigma$-type species currently being experimentally studied for laser cooling are TlF \cite{Norrgard2017,Clayburn2020,Grasdijk2021}, AlF \cite{Truppe2019,Doppelbauer2021} and AlCl \cite{Rosa2004,Yang2016,Wan2016,Daniel2021}.

\change{In this paper, we spectroscopically study the AlCl hyperfine structure within the lowest three $\Astate$-state rotational levels,}
discuss the implications of its properties on laser cooling and trapping and present theoretical estimates of the expected capture velocities of a MOT for AlCl.
The metal halide AlCl has been proposed as an excellent candidate for laser cooling due to its high photon scattering rate of $\approx 2\pi \times 25$\,MHz \cite{Rogowski1987} and its almost unity Franck-Condon factors (FCFs) \cite{Yang2016,Wan2016,Langhoff1988,Ren2021,Daniel2021}.
A key challenge to laser cooling and trapping AlCl is producing sufficient laser light for the optical cycling transition at 261.5\,nm which connects the electronic ground $\Xstate$ state with the excited $\Astate$ state.
However, recent developments in UV laser technology, including work done by the authors, have shown that robust systems capable of more than 1\,W of laser power at this wavelength are now within reach \cite{Mes2003b,Ostroumov2008,Burkley2019,Burkley2021,McCarron2021}.

AlCl was first laboratory-confirmed in 1913 and has since undergone many spectroscopic and chemical studies
\cite{
Jevons1913,
Jevons1924,
Bhaduri1934,
Holst1935,
Mahanti1934,
Miescher1936,
Sharma1951,
Barrow1954,
Reddy1957,
Barrow1960,
Lide1965,
Lide1967,
Wyse1972,
Hoeft1973,
Lovas1974,
Schnoeckel1976,
Tsunoda1978,
Ram1982,
Rogowski1987,
Mahieu1989,
Mahieu1989a,
Dearden1993,
Ogilvie1994,
Hedderich1993,
Hensel1993,
Dearden1993,
Ogilvie1994,
Saksena1998,
Parvinen1998,
Brites2008b,
Pamboundom2016,
Daniel2021,
Preston2022}.
Moreover, these efforts have been complemented by many theoretical studies that explored the properties of AlCl in detail \cite{
Langhoff1988,
Langhoff1988a,
Yang2016,
Wan2016,
Andreazza2018,
Yousefi2018,
Aerts2019,
Xu2020,
Ren2021,
Zhang2021b,
Qin2021,
Daniel2021,
Bala2023,
Kaur2023}.
Its presence in the interstellar medium, carbon-rich stars, and as part of the models of exoplanets' atmospheres renders AlCl a molecule of astrophysical interest
\cite{
Cernicharo1987,
Ford2004,
Agundez2012,
Kaminski2016,
Decin2017,
Andreazza2018,
Yousefi2018,
Bernath2020,
Xu2020,
Yurchenko2023,Tennyson2020,Wang2020}.
\change{Despite this extensive work, some fundamental properties, such as the dipole moment of the ground and excited electronic states of AlCl}, remain unknown and have only been estimated theoretically \cite{Yousefi2018} or in some models substitute values from similar molecules are used \cite{Ford2004,Agundez2012}.
AlCl also has many applications outside the laboratory, for instance, it is utilized in the production of photovoltaic grade silicon \cite{Yasuda2009,Yasuda2011,Yasuda2011a}, is spectroscopically found in rocket plumes \cite{McGregor1992,McGregor1993,Oliver1992} and can be used as a probe for the detection of chlorides in drinking water \cite{Parvinen1999}.
Another interesting and unique aspect of AlCl is that the disproportionation reaction channel, which forms the stable compound AlCl$_3$, can be blocked below 180\,K and \change{solid state densities} of AlCl have been isolated using this characteristic \cite{Tacke1989}.
This property can potentially provide the ideal starting point for producing a large number of molecules in the gas phase by applying nanosecond-pulsed laser ablation to a fabricated thin precursor film of AlCl, adding to the list of advantages of AlCl as a candidate for laser cooling and trapping and the described applications.

\section{\protect\change{The AlCl $\Xstate$ and $\Astate$ structures}}
\label{sec:AlCl_H}

AlCl has two main isotopes with $^{35,37}$Cl with natural abundances of $\approx 76$ and $\approx 24$\% respectively.
The large differences in the electronegativities between Al (1.61) and Cl (3.16) form a polar bond with a theoretically predicted electric dipole moment of $1.6$\,D for the $\Xstate$ state \cite{Yousefi2018}.
\change{
The $\Xstate$ ground state is coupled to the excited, short-lived $\Astate$ state, which has a lifetime of $\approx 6$\,ns \cite{Rogowski1987}, via an electric dipole transition in the UV at 261.5\,nm. The intermediate triplet $\apistate$ state is coupled to the ground state via an electric dipole transition corresponding to a photon of wavelength $\approx 407$\,nm.
}
The meta-stable intermediate state is split into four $\Omega$-sub states whose linewidths have been estimated to be on the order of 3-90\,Hz \cite{Wan2016}, but no precise experimental data exist to date.

The $\Astate \leftarrow \Xstate$ excitation, which promotes a valence electron from the $9\sigma$- to the $4\pi$-orbital, closely resembles the $S-P$ transition in atomic aluminum \cite{Ram1982}.
The similar vibrational constants and bond lengths of the $\Xstate$ and the $\Astate$ states result in a calculated Franck-Condon factor of 99.88\% for the $v''=0$ band \cite{Daniel2021}.
The vibrational levels are split in rotational states with a rotational constant of $\approx 7.3$\,GHz.
The presence of the nuclear spins of both the chlorine atom ($I_\textrm{Cl} = 3/2$) and the aluminum atom ($I_\textrm{Al}$ = 5/2) adds  complex cascaded hyperfine splittings to each state, though the hyperfine splitting of the $\Xstate$ state is smaller than the natural linewidth and remains unresolved.

\begin{figure}[ht]
    \centering
    \includegraphics[width=\linewidth]{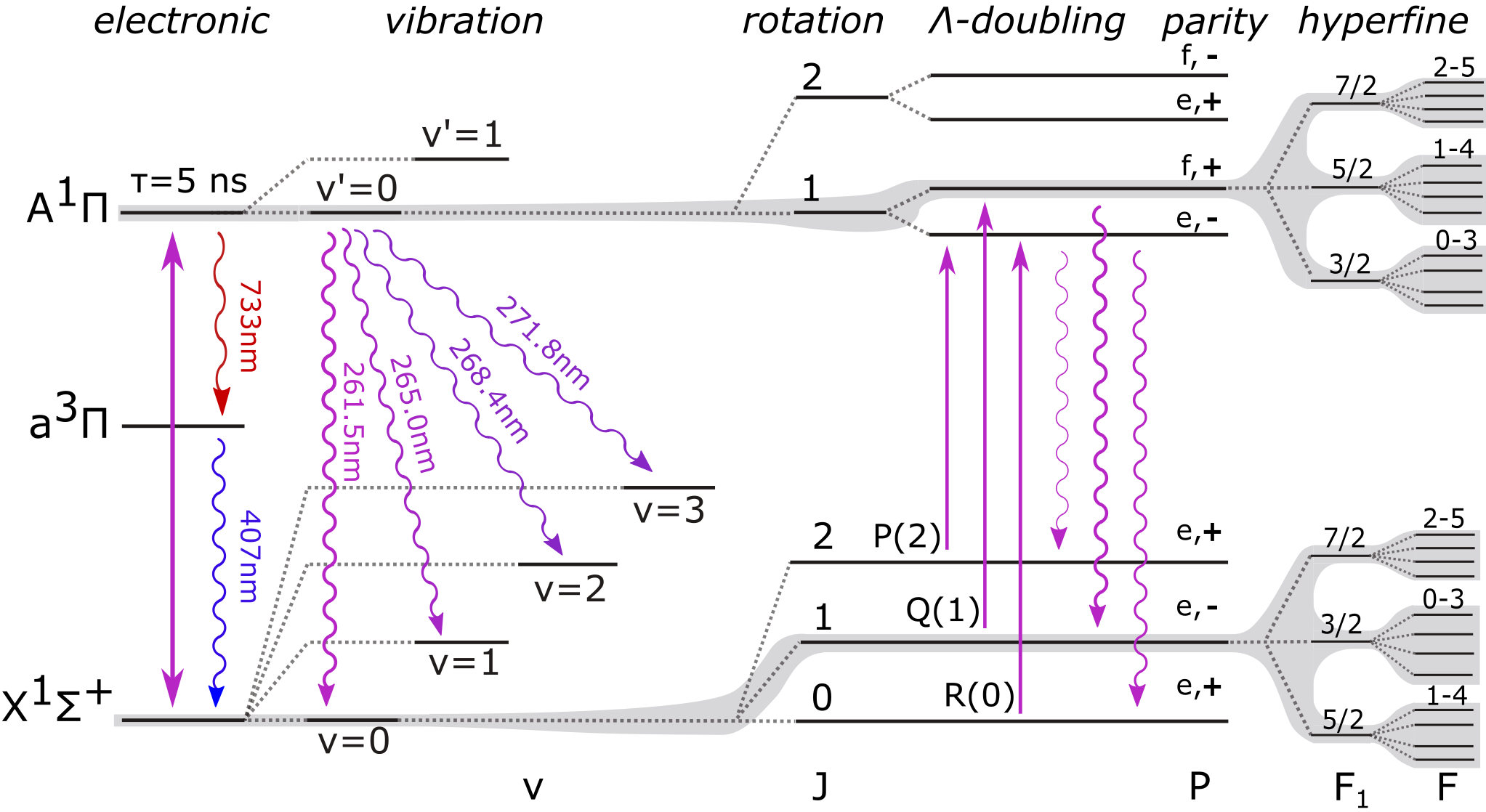}
    \caption{Electronic and rovibrational energy-level structure of AlCl. The $Q$ transitions, which are used for laser cooling, are rotationally closed, unlike the $P$ and $R$ transitions.
    \change{The rovibronic manifolds of the $\Xstate (v = 0, J = 1)$ and $\Astate (v' = 0, J' = 1)$ states include 72 and 144 hyperfine states, respectively, all of which are involved in laser cooling AlCl.}
    Adapted from \cite{Thesis_Shaw_2022}.}
    \label{fig:alcl_scheme}
\end{figure}

\rfig{fig:alcl_scheme} shows the corresponding detailed level scheme relevant to optical cycling in AlCl.
The quantum numbers of the states are defined in \rsec{sec:Xstate}.
All $Q$-type transitions ($\Delta J = 0$) can be used for laser cooling since they are rotationally closed due to dipole and parity selection rules.

\subsection{Spectroscopy on AlCl}

For the analysis in this paper, two sets of data were used from two separate experiments, one in the Hemmerling group at the University of California, Riverside (UCR), and the other in the McCarron group at the University of Connecticut (UConn).

In the UCR group, AlCl is produced in a cryogenic helium buffer-gas beam source (CBGB) \cite{Hutzler2012,Barry2011} at 3.4\,K via short-pulsed (5\,ns) laser ablation of a Al:KCl mixture target \cite{Lewis2021} with a Nd:YAG laser (Mini-Lite II, Continuum) of $\approx 10\,$mJ per shot.
More details on the source are described in \refcite{Daniel2021}.
The fluorescence data presented here were acquired $\approx 40$\,cm downstream from the source. The molecules were excited with laser light aligned orthogonal to the molecular beam direction and the induced fluorescence was collected with a photo-multiplier tube (H10722-04, Hamamatsu). The excitation laser light of a few mW at 261.5\,nm was produced by frequency-doubling the output of a 522\,nm (VALO Vecsel, Vexlum) \change{laser} with a custom-built second-harmonic generation cavity.
\change{The laser frequency was scanned and stabilized by using the frequency readout of a wavelength meter (WS-7, High-Finesse), which in turn was calibrated using a Doppler-free saturated absorption spectrum of Rubidium.
This approach yields an upper limit for the frequency accuracy of $\approx 15$\,MHz, whereas the shot-to-shot frequency stability is better than $\pm 5$\,MHz.}

At UConn, the experimental approach is similar, with pulses of cold AlCl produced from a cryogenic buffer-gas beam source at 2.7\,K via laser ablation ($\approx20$~mJ at 532~nm). The source has been described previously in \refcite{Shaw2020}. Molecules are optically addressed 94~cm downstream of the source below an EMCCD camera using $\approx100~\mu$W of laser light at 261.5~nm. This light is picked-off from a homebuilt laser system that generates $>1~$W in the fourth-harmonic from an infra-red fiber amplifier seeded by an external cavity diode laser (ECDL) at 1046 nm \cite{McCarron2021}. The EDCL is frequency stabilized and scanned using a transfer cavity locked to a frequency stabilized HeNe laser \change{and the relative frequency stability is typically $\pm2$~MHz. The absolute frequency is measured to within $\sim\,200$~MHz using a wavemeter, sufficient to identify the correct transfer cavity free spectral range}.

\subsection{Hamiltonian of AlCl}

\label{sec:Xstate}
\label{sec:Astate}

In this section, the hyperfine energy-level structures of the $X$- and $A$-electronic states are described in detail and the molecular constants, which were acquired by using data from the spectroscopy setups, are discussed. Similar to the approach taken for AlF \cite{Truppe2019}, we choose to describe both the $\Xstate$ and $\Astate$ states using a common Hund's case (a) coupling scheme.
To describe the electronic state of AlCl, \change{the total angular momentum} $\mathbf{J} = \mathbf{\Omega} + \mathbf{R}$, where $\mathbf{R}$ is the rotational angular momentum and $\mathbf{\Omega} = \mathbf{\Lambda} + \mathbf{\Sigma}$ is the sum of the projection of the electron orbital angular momentum $\mathbf{L}$ and the electron-spin angular momentum $\mathbf{S}$ on the intermolecular axis. For the $\Xstate$ state, the projections are $\Lambda = 0$ and $\Sigma = 0$. For the $\Astate$ state, $\Lambda = \pm 1$ and $\Sigma = 0$.
The hyperfine structure is accounted for by coupling $\mathbf{J}$ to the nuclear spin of the aluminum atom, $\mathbf{F_1} = \mathbf{J} + \mathbf{I}_\textrm{Al}$, which is in turn coupled to the nuclear spin of the chlorine atom, $\mathbf{F} = \mathbf{F_1} + \mathbf{I}_\textrm{Cl}$.

\paragraph{X-State}

The Hamiltonian for the $\Xstate$ state has the form \cite{Brown2003}
\begin{equation}
    H_\textrm{X} = H^\textrm{X}_0
    + H_\textrm{EQ}
    \quad,
\end{equation}
where $H^\textrm{X}_0$ includes the electronic, vibrational, and rotational energy terms, and $H_\textrm{EQ}$ is the electric quadrupole term.
$H^\textrm{X}_0$ is expressed in terms of the Dunham expansion for $E(\nu,J)$ with equilibrium constants that have previously been measured \cite{Hedderich1993}.

The electric quadrupole interaction has previously been found to be the dominant hyperfine interaction in the $\Xstate$ state \cite{Hoeft1973}, given as
\begin{eqnarray}
    H_\textrm{EQ}
    &=& \sum_\alpha \frac{\sqrt{6}(eQq_0)_\alpha}{4I_\alpha(2I_\alpha-1)} T^2_{0}(\mathbf{I}_\alpha,\mathbf{I}_\alpha)
    \quad,
\end{eqnarray}
where $\alpha$ indicates the nucleus of aluminum and chlorine.
\begin{table}[ht]
    \centering
    \begin{tabular}{|c|c|}
    \hline
        Constant & Value (MHz)  \\
        \hline
        $(eQq_0)_{Al}$ & -29.8(50)\\
        $(eQq_0)_{Cl}$ & -8.6(10)\\
        \hline
    \end{tabular}
    \caption{Experimental electric quadrupole constants $eQq_0$ for the $\Xstate$ state as measured in previous work~\cite{Hoeft1973}.}
    \label{tab:eQq0_tiemann}
\end{table}

Higher order terms, such as the nuclear-spin-rotation and the nuclear-spin-nuclear-spin interaction term are neglected, given the broad linewidth of the \XAtransition\ transition that is used in this paper and the fact that these terms are expected to be two orders of magnitude smaller than the quadrupole terms, comparable to the case for the similar molecule AlF \cite{Truppe2019}.

\paragraph{A-State}

For the $\Astate$ state, the orbital angular momentum is non-zero. The orbital degeneracy of $\Lambda = \pm 1$ is lifted due to the presence of the end-over-end rotation of the molecule and results in a splitting of the rotational states into two opposite parity states, also known as $\Lambda$-doubling, see \rfig{fig:alcl_scheme}.
The Hamiltonian for the $\Astate$ state has the form \cite{Brown2003}
\begin{equation}
    H_\textrm{A} = H^\textrm{A}_0 + H_\textrm{LI} + H_\Lambda + H_\textrm{EQ} + H_\textrm{Z}
\end{equation}
where $H^\textrm{A}_0$ includes the electronic, vibrational and rotational energy terms, $H_\textrm{LI}$ is the nuclear-spin-orbital hyperfine term, $H_\Lambda$ is the lambda-doubling term, $H_\textrm{EQ}$ is the electric quadrupole term, and $H_\textrm{Z}$ is the Zeeman term. $H^\textrm{A}_0$ is expressed in the form of the Dunham expansion with equilibrium constants that have been measured in a previous study \cite{Daniel2021}.

\begin{figure}[ht]
    \centering
    \includegraphics[width=\linewidth]{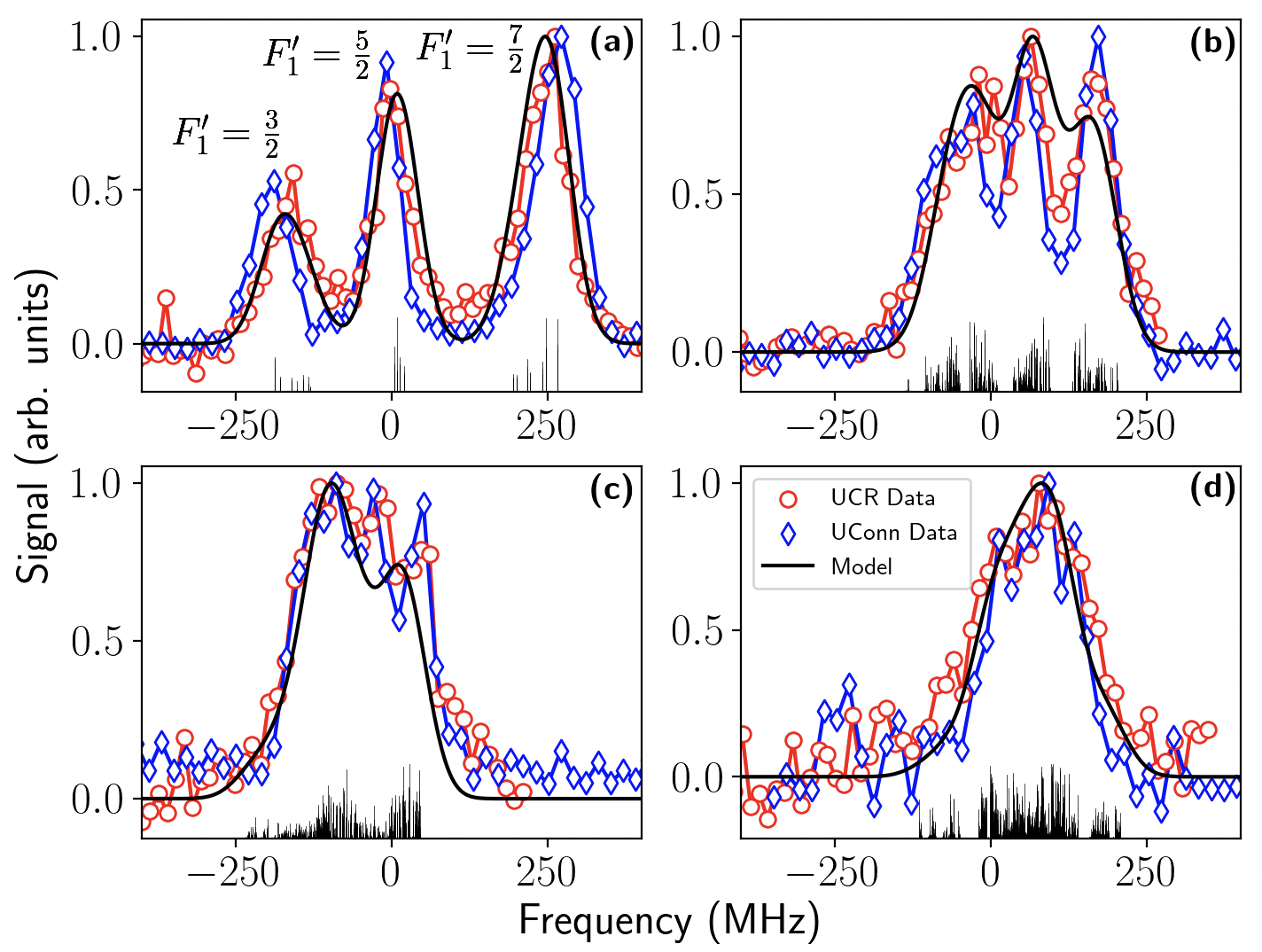}
    \caption{
     \change{Normalized fluorescence data (red circles, UCR; blue diamonds, UConn) and model (black solid line) of the R($J$) transitions,
     $\Xstate\,|v = 0, J\rangle\leftarrow \Astate\,|v' = 0, J + 1\rangle$,
     \textbf{(a)} $R(0)$, \textbf{(b)} $R(1)$, \textbf{(c)} $R(2)$, and \textbf{(d)} $R(3)$. The vertical black lines represent the different transitions predicted by our Hamiltonian model with their heights corresponding to their relative line strengths.}
    }
    \label{fig:rbranch_fit}
\end{figure}

Due to the singlet nature of the $\Astate$ state, the $\Lambda$-doubling term can be expressed as
\begin{equation}
\label{eq:A_lambda}
    H_\Lambda = -\sum_{k=\pm 1} e^{-2 \imath k \phi} q T^2_{2k}(\mathbf{J}, \mathbf{J})
\end{equation}
and the nuclear-spin-orbital hyperfine term can be expressed as
\begin{equation}
\label{eq:A_hfs}
    H_\textrm{LI} = \sum_\alpha a_\alpha T^1(\mathbf{L}) \cdot T^1(\mathbf{I}_\alpha)
    \quad.
\end{equation}
The quadrupole term for the $\Astate$ state has both a component along and a component perpendicular to the internuclear axis
\begin{eqnarray}
    \nonumber
    H_\textrm{EQ}
    &=& \sum_\alpha \frac{e Q_\alpha}{4 I_\alpha (2 I_\alpha - 1)} \left[
    \sqrt{6} q_{0,\alpha} T^2_{0}(\mathbf{I}_\alpha,\mathbf{I}_\alpha) \right.\\
    && \left. + \sum_{k=\pm 1} e^{(-2 \dot{\imath} k \phi)} q_{2,\alpha}
    T^2_{2 k}(\mathbf{I}_\alpha,\mathbf{I}_\alpha)
    \right]
    \quad.
\end{eqnarray}
The electric quadrupole constants are defined in terms of nuclear quadrupole moment, $eQ$, and electric-field-gradient at each nucleus, with $q_0$ being equal to the $V_{zz}$ component and $q_2$ being equal to $2\sqrt{6}(V_{xx} - V_{yy})$ \cite{Brown2003}.
Based on our previous {\it ab initio} calculations \cite{Daniel2021}, we performed additional {\it ab initio} calculations of the electric field gradients to get a theoretical estimate of the quadrupole constants, as shown in \rtab{tab:eQq0_ab_initio} and \rtab{tab:eQq2_ab_initio}.
Using the quadrupole moments of $147.7$\,mb for the Al nucleus \cite{Aerts2019} and $85$\,mb for the $^{35}$Cl nucleus \cite{Alonso2004}, we find reasonable agreement between the theoretical and experimental values for the $\Xstate$ state shown in \rtab{tab:eQq0_tiemann}. This result increases our confidence in our {\it ab initio} values for the $\Astate$ state.

\begin{table}[ht]
    \centering
    \begin{tabular}{|c|l|c|c|}
    \hline
        Nucleus & State & $V_{zz}$ (a.u.) & $eQq_0$ (MHz) \\\hline
        Al & $\Xstate$ & -0.809 & -28.1\\
           & $\Astate$ & -0.220 & -7.6\\\hline

        Cl & $\Xstate$ & -0.675 & -13.5\\
           & $\Astate$ & -2.554 & -51.0\\
    \hline
    \end{tabular}
    \caption{{\it Ab initio} calculations of the electric field gradients and the quadrupole constants $eQq_0$.}
    \label{tab:eQq0_ab_initio}
\end{table}

\begin{table}[ht]
    \centering
    \begin{tabular}{|c|c|c|}
    \hline
        Nucleus & $2\sqrt{6} (V_{xx} - V_{yy})$ (a.u.) & $eQq_2$ (MHz) \\\hline
        Al   & 0.605 & 102.9\\\hline

        Cl & 0.332 & 32.5\\
    \hline
    \end{tabular}
    \caption{{\it Ab initio} calculations of the electric field gradients and the quadrupole constants $eQq_2$ for the $\Astate$ state.}
    \label{tab:eQq2_ab_initio}
\end{table}

Given the broad linewidth of the \XAtransition\ transition, we use the following procedure to estimate the equilibrium constants for the $\Astate$ state.
Starting with the $R(0)$-transition, we first use a least-squares fit to extract the hyperfine constants, $a_\textrm{Al}$ and $a_\textrm{Cl}$, since the structure of this line is dominated by these parameters and much less affected by others.
Then, with the hyperfine parameters set to their optimum values, we use a least-square fit to determine an upper limit of the $\Lambda$-doubling constant $q$ by fitting the $R(0)-R(3)$ transitions simultaneously.
During the whole procedure, we keep the quadrupole constants, $eQq_0$ and $eQq_2$, for both nuclei set to the values determined by the {\it ab initio} calculations.

\begin{table}[ht]
    \centering
    \begin{tabular}{|c|c|}
    \hline
       Constant & Value (MHz) \\
    \hline
        $a_\textrm{Al}$ & $131.9\left(\substack{+3.6 \\ -3.3}\right)$\\
        $a_\textrm{Cl}$ & $42.0\left(\substack{+8.1 \\ -7.0}\right)$\\
        $q$ & \change{$(-8, -3)$}\\
    \hline
    \end{tabular}
    \caption{Molecular constants for the $\Astate$ state obtained in this paper.}
    \label{tab:fitted_A_constants}
\end{table}

The resulting equilibrium constants are presented in \rtab{tab:fitted_A_constants}. The errors of the constants correspond to a $\approx 2.5$\% deviation of the absolute value of the residuals of the fit and model from the optimal value.
\change{We note that our data only allow for constraining the value for the $\Lambda$-doubling constant between $-8$ and $-3$\,MHz, since it is mainly determined by the width of the broad features of $R(2)$ and $R(3)$.}
Our approach yields a good agreement between \change{the overlapped } independent data sets, UCR and UConn, and the Hamiltonian model using the combination of fitted and {\it ab initio} values for the molecular constants, as shown in \rfig{fig:rbranch_fit}. \change{Here the UConn data were overlapped with the UCR data via a frequency offset prior to fitting.} We note that a small discrepancy between the two data sets is shown in \rfig{fig:rbranch_fit} (a). We attribute this to the nonlinearity of the transfer cavity used in the UConn laser frequency stabilization scheme. Here, the frequency of the ECDL inherits the nonlinearity of the transfer cavity, which is exacerbated by the O-rings used in its design \cite{Thesis_Barry_2013}. This nonlinearity depends on the cavity dc value and was not a significant effect in the other $R$-line scans. For this reason, only the UCR data were used to fit the $R(0)$ transition. This highlights that, while transfer cavities are effective for frequency stabilization, care must be taken when using this approach for spectroscopy, especially where subsequent stages of second- or fourth-harmonic generation amplify these non-linearities.

\begin{figure}[ht]
    \centering
    \includegraphics[width=0.85\linewidth]{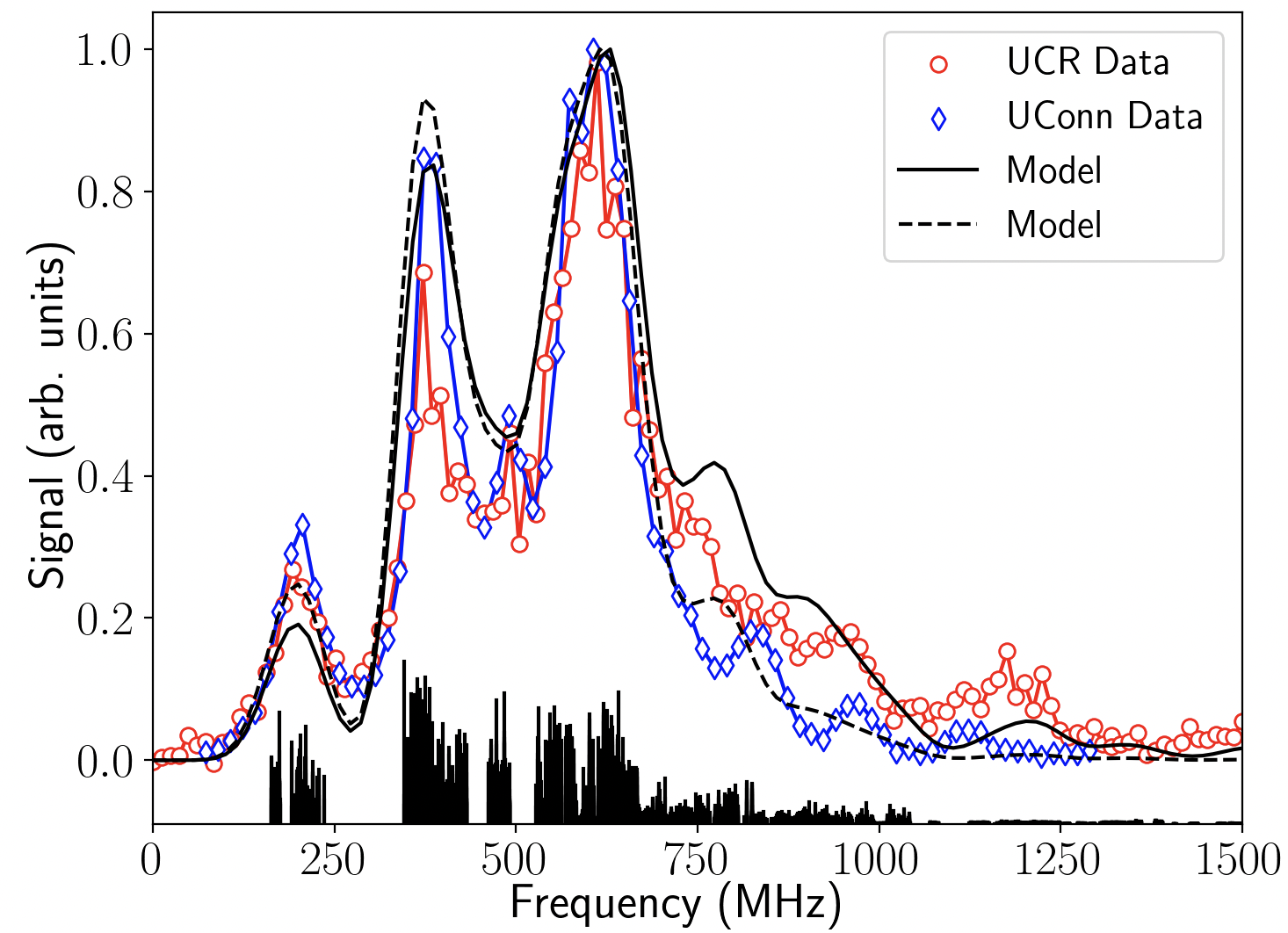}
    \caption{Normalized fluorescence data (red circles: UCR data, blue diamonds: UConn data) and model (black solid line) of the $Q(0)-Q(5)$ transitions of AlCl. \change{The models use the fitting parameters of the R-lines from \rtab{tab:fitted_A_constants} and are only optimized to reproduce the measured signal amplitude,} the rotational temperature ($T_\textrm{rot}=2.5$\,K solid line, $T_\textrm{rot}=1.6$\,K dashed line) and the absolute frequency offset.
    The vertical black lines represent the different transitions predicted by our Hamiltonian model with their heights corresponding to their relative line strengths.
    }
    \label{fig:q_lines}
\end{figure}

Finally, overlaying the model with the parameters acquired through the $R$-transitions with the fluorescence measurements of the $Q$-branch at UCR and UConn yields a reasonable agreement, as shown in \rfig{fig:q_lines}.
We note that the $Q$-branch fit has no free parameters besides the overall frequency offset, the rotational temperature [2.5\,K (1.6\,K) for the UCR (UConn) data] and the overall amplitude.
Finally, the density of the lines of the $Q$-transitions illustrates the similarity of the rotational constants of the $\Xstate$ and $\Astate$ states, which in part leads to the predicted highly diagonal Franck-Condon factors of AlCl \cite{Daniel2021}.

\subsection{Zeeman Splitting}
\label{sec:zeeman}

\begin{figure}[ht]
    \centering
    \includegraphics[width=0.85\linewidth]{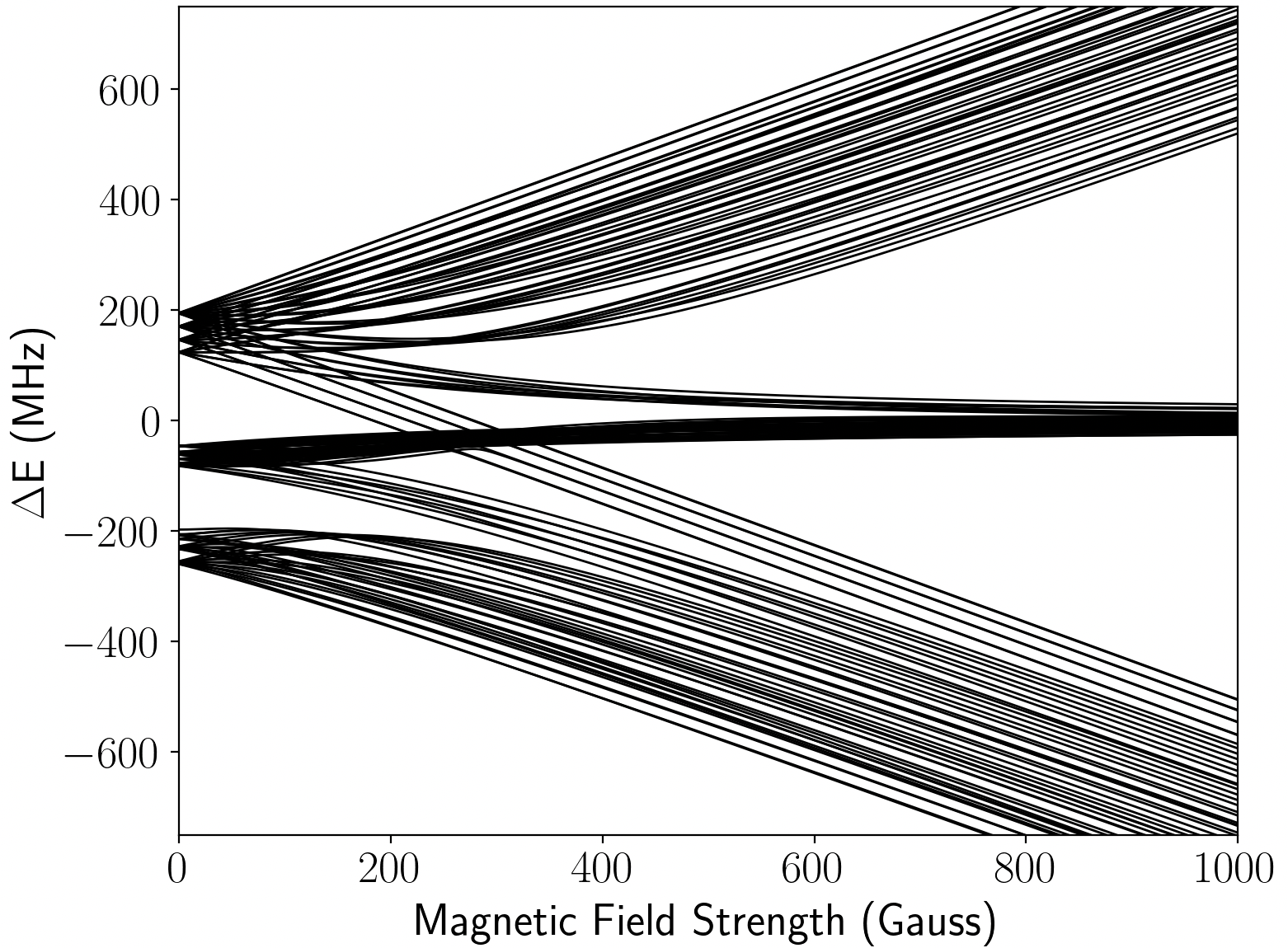}
    \caption{The Zeeman splitting of the $\Astate, (v' = 0, J'=1)$ manifold of AlCl as a function of the external magnetic field.}
    \label{fig:zeeman_splitting}
\end{figure}

To set up a magneto-optical trap for a new species, it is important to fully understand the Zeeman splitting of the cooling transition in order to design the magnetic field gradients appropriately. In the case of AlCl, the dominant Zeeman term in the $\Astate$ state is the interaction between the electron-orbital-angular-momentum, $\mathbf{L}$, and the applied magnetic field, $\mathbf{B}$. This interaction has the form
\begin{equation}
    H_\textrm{Z} = g_L\mu_BT^1(\mathbf{L})\cdot T^1(\mathbf{B})
\end{equation}
where $g_L=1$ and $\mu_B$ is the Bohr magneton.
The dominant Zeeman interaction in the $\Xstate$ state is the nuclear-spin-Zeeman interaction, $\mathbf{I}_\alpha \cdot \mathbf{B}$, which has a magnetic dipole moment that is smaller than the $\Astate$ state by a factor of $m_e/m_p$, where $m_e$ is the electron mass and $m_p$ is the proton mass. Thus, the Zeeman splitting of the $\Xstate$ state is negligible and Zeeman shifts on the cycling transition are fully determined by the splitting of the $\Astate$. We show the calculated Zeeman splitting of the $\Astate, (v'=0, J' = 1)$ state as a function of an external magnetic field in \rfig{fig:zeeman_splitting}. This is our principle target excited state for optical cycling in AlCl using the $Q(1)$ cycling transition. However, the close proximity of other $Q$-transitions means that a single laser frequency will likely address and lead to optical cycling for molecules in multiple low-lying rotational states within the electronic ground state, see \rfig{fig:q_lines}.

In the low field regime, magnetic sublevels are shifted linearly according to Land\'{e} $g$-factors that vary in both magnitude and sign for different hyperfine states, see table \rtab{tab:gF_factors} for the $g$-factors of the target even parity $\Astate, (v'=0, J'=1)$ excited state. While in principle this structure can lead to magnetically tunable transitions for use in a Zeeman slower, the large hyperfine spread of the $J'$ states within the $\Astate$ state combined with the lack of a type-I transition, can make addressing individual velocity classes challenging.

Realizing confining transitions in a MOT of AlCl to the $F_{1}=5/2$ and $7/2$ manifolds will require an orthogonal circular laser polarization compared to confining transitions to the $F_{1}=3/2$ manifold since these states have $g$-factors with opposite sign. We emphasize that a MOT of AlCl appears similar in nature to atomic type-II MOTs. Here the magnetically tunable states that enable confinement are in the electronic excited state and decay rapidly via spontaneous emission to the unresolved and unperturbed $\Xstate$ ground state. By contrast, in $^2\Sigma$-type molecules, the dominant Zeeman shift is in the ground state, which can lead to stationary magnetic dark states that require either static dual-frequencies \cite{Tarbutt2015} or rapid synchronous switching of the field gradient and laser polarizations \cite{Hummon2013,Norrgard2016} to generate substantial confining forces. While the general level structure and Zeeman shifts within molecules with $^{1}\Sigma$ ground states may simplify magneto-optical trapping methods, coherent dark states may still need to addressed, see \rsec{sec:dark_states} and \refcite{Devlin2016}.

In \rtab{tab:gF_factors}, we list the Land\'e factors for low magnetic fields ($<10$\,Gauss) of the different hyperfine states in the even parity $\Astate, (v'=0, J'=1)$ manifold. We note that a MOT of AlCl will require a large magnetic field gradient on the order of $100$\,G/cm axially to realize confining forces due to the small excited state $g$-factors and the large transition linewidth. This is similar to MOTs using strong transitions in alkaline earth and alkaline earth-like atoms \cite{Xu2003}.

\begin{table}[ht]
    \centering
\begin{tabular}{|c|c|c|c|c|c|c|c|c|}
\hline
$F_1$ & $F$ & $g_F$ &
$F_1$ & $F$ & $g_F$ &
$F_1$ & $F$ & $g_F$
\\\hline
3/2 &&& 5/2 &&& 7/2 &&\\\hline
& 0 & -    &&&&&&\\
& 1 & -0.15 && 1 & 0.15&&&\\
& 2 & -0.13 && 2 & 0.07 & & 2 & 0.23\\
& 3 & -0.12 && 3 & 0.05 & & 3 & 0.15\\
&   &      && 4 & 0.03 & & 4 & 0.12\\
&   &      &&   &      & & 5 & 0.10\\
\hline
\end{tabular}
\caption{\label{tab:gF_factors} Calculated Land\'e factors of the even parity $\Astate$ $(v' = 0, J' = 1)$ states of AlCl.}
\end{table}

Finally, in the high-field regime, the different spins decouple and the overall state structure simplifies into three manifolds, each of which has common Land\'e factors. This Paschen-Back regime has been proposed for a Zeeman slowing scheme for CaF at $\approx 300$\,Gauss \cite{Kaebert2021}.
Though, AlCl would require even higher fields and field-gradients to completely isolate the manifolds, we note that MOTs with gradients of $\approx 1$\,kGauss/cm have been realized \cite{Haubrich1996,Willems1997,Yoon2007}.

\section{Optical cycling}

\label{sec:optical_cycling}

The strong optical cycling $\Astate-\Xstate$ $(v=0,v'=0)$ transition near $261.5~$nm combines a large linewidth ($2\pi\times25~$MHz) with a large photon recoil velocity (2.5~cm/s) to offer access to strong radiative forces. These forces could potentially slow a molecular beam from a cryogenic source to below the capture velocity of a MOT in just a few centimeters, rather than the $\approx1~$m required for today's experiments with $^{2}\Sigma$ molecules \cite{Barry2014,Anderegg2017,Truppe2017,Collopy2018,Vilas2022}. Such an improvement would increase the solid-angle and trappable flux from a cryogenic source by several orders-of-magnitude and tackle inefficient MOT loading, which remains a key bottleneck in the field of molecular laser cooling and trapping.
Alternatively, new slowing techniques, such as bichromatic slowing \cite{Partlow2004,Chieda2012,Yang2016d,Corder2015,Kozyryev2018,Galica2018}, traveling-wave Stark deceleration \cite{Greenberg2021,Aggarwal2021,Shyur2018,Shyur2018a,Osterwalder2010,Meek2011,Berg2014} or Zeeman-Sisyphus deceleration \cite{Augenbraun2021}, may offer solutions to this challenge. Strong, short-wavelength optical transitions, such as those in AlCl, AlF and MgF, are attractive for laser cooling and trapping but demand high laser intensity since the saturation intensity scales as $I_{\rm{sat}}\propto\Gamma/\lambda^{3}$. The AlCl optical cycling transition at 261.5~nm is particularly fortuitous since ytterbium fiber lasers and amplifiers offer high power (10-100~W) at the fundamental of the forth-harmonic (1046~nm) and 261.5~nm is close enough to the frequency quadrupled Nd:YAG that optics are well-developed and commercially available.
In the following, we outline the prospects of optical cycling in the different ro-vibrational and hyperfine manifolds of AlCl and compare the effects to other molecules, AlF and TlF.

\subsection{Vibrational branching}

AlCl is expected to have highly diagonal Franck-Condon factors (FCFs) which limit decay into excited vibrational levels within the $\Xstate$ manifold during optical cycling. Previous work by multiple groups predict a FCF in the $v''=0$ band of $q_{00}>0.99$ \cite{Yang2016,Wan2016,Zhang2021b,Daniel2021} with approximately three lasers (one cycling laser and two repumps) being sufficient to scatter the $\sim10^4$ photons required to slow, laser cool and trap \cite{Baum2020,Vilas2022}. Experimental work to directly confirm this diagonal vibrational branching is underway \cite{Shaw2023}, similar to previous work with TlF \cite{Hunter2012}, BH \cite{Hendricks2014} and AlF \cite{Truppe2019}, other candidate molecules for laser cooling experiments with $^{1}\Sigma$ ground states. We note in passing that the single unpaired valence electron within each $^{2}\Sigma$ molecule laser cooled to-date offers an intuitive picture behind the origin of diagonal FCFs, i.e.~that the optically-addressable electron plays a negligible role in the binding of the molecule.
By contrast, to the best of our knowledge, no similar intuition can be used to identify closed-shell $^{1}\Sigma$ molecules with diagonal FCFs.

\subsection{Rotational branching}

Rotational branching within AlCl can be tamed using selection rules that dictate the allowed changes in angular momentum and wavefunction symmetry (parity) during electronic transitions \cite{Stuhl2008}. Namely, the electric dipole transitions with nonzero transition dipole matrix elements (TDME) are limited to $\Delta J=0,\pm1$ and, because the dipole operator is a rank 1 tensor (odd), the wavefunctions of the connected states must have opposite parity. For $^{1}\Sigma$ ground state molecules, such as AlCl, these selection rules enable rotational closure for all $Q$-transitions ($\Delta J = 0$) (see \rfig{fig:alcl_scheme}).

A loss channel via rotational branching can be introduced by a small electric field to mix the closely spaced opposite parity $\Lambda$-doublets in the $\Astate$ state and hence break the parity selection rule. In this case for AlCl, spontaneous emission would then populate dark rotational states via the $P$- and $R$-branches ($\Delta J = -1$ and $+1$, respectively). This loss mechanism was reported in a radio-frequency MOT of SrF molecules \cite{Norrgard2016} and, notably, has been investigated for AlF \cite{Hofsass2021}, with this loss channel becoming negligible for stray fields below $1$\,V/cm. For AlCl, the level of electric field suppression required remains unclear as our spectra can only place upper and lower limits on the $\Lambda$-doubling parameter $q$ that dictates the spacing between $\Lambda$-doublets, see \rtab{tab:fitted_A_constants}.

An additional loss mechanism, enabling transitions with $|\Delta J|>1$, can result from mixing between states with the same total angular momentum $F$ from different rotational levels within the excited electronic state. In both AlCl and AlF, this mixing in the $\Astate$ state is predominantly due to the magnetic hyperfine interaction of the Al nuclear spin and is expected to result in a small loss channel of order $10^{-6}$ \cite{Truppe2019,Thesis_Shaw_2022}. By contrast, this mixing and rotational branching in TlF can be substantial and poses a challenge to optical cycling in this molecule \cite{Norrgard2017}.

\subsection{Hyperfine Structure}

 The Al and Cl nuclear spins result in AlCl having a complex hyperfine structure, with 12 hyperfine states for $J=1$, 18 for $J=2$, 22 for $J=3$ and 24 for $J \geq 4$ (see \rfig{fig:alcl_scheme}). In the $\Xstate$ state, the lack of spin-orbit coupling results in the hyperfine structure being small and unresolved to the strong $\Astate-\Xstate$ transition ($\Gamma\approx 2\pi\times 25 $~MHz). For example, in $J=1$ all 12 hyperfine states span just $\approx 11~$\,MHz \cite{Hoeft1973}. While this allows all ground state hyperfine levels for a given $J$ to be conveniently addressed by a single laser frequency, it can also lead to the formation of slowly evolving dark states which prevent rapid optical cycling (see \rsec{sec:dark_states}).

 In the $\Astate$ state, the hyperfine structure is at best only partially resolved for low-lying rotational states (see \rfig{fig:rbranch_fit}). Our analysis shows that, similar to AlF \cite{Truppe2019}, this structure is primarily due to the nuclear spin-electron orbit interaction with the $eQq_{0}$ and $eQq_{2}$ constants dictating that the electric quadrupole interaction only plays a small role. Interestingly, the $\Astate$ $(v'=0, J'=1)$ hyperfine structure spans $\sim500~$MHz, equivalent to a Doppler spread of $\sim130$~m/s at 261.5~nm. This, combined with power broadening, may enable a single laser frequency to address the decreasing Doppler shift of molecules during laser slowing without the need for frequency chirping or phase modulating the slowing light.

For convenience, we use a Hund's case (a) basis to describe the AlCl ground and excited states using quantum numbers $F$ and $F_{1}$ (see above).
However, we emphasize that $F_{1}$ is not a good quantum number and states with common $F$ from different $F_{1}$ are mixed. In AlCl, this mixing arises in the $\Xstate$ state due to the quadrupole interaction with the Cl nuclear spin and in the $\Astate$ state due to the Cl nuclear spin-electron orbit interaction. While this mixing does not lead to loss from the optical cycle, it does skew hyperfine branching ratios away from the unmixed case \cite{Thesis_Shaw_2022}.

\subsection{Dark States}

\label{sec:dark_states}

Rotational structure in molecules dictates that optical cycling requires type-II transitions which naturally introduce dark states within the ground electronic state \cite{Berkeland2002,McCarron2018a}. In general, these can be either stationary angular momentum eigenstates, in which molecules accumulate and leave the optical cycle, or a coherent superposition of these eigenstates, which naturally precess between bright and dark states, limiting the maximum photon scattering rate \cite{Gray1978,Shakhmuratov2004}.

For $^{2}\Sigma$ molecules, it is common to remix dark states using a Zeeman shift to lift the degeneracy of ground state sublevels by $\sim\Gamma$. This approach is impractical for $^{1}\Sigma$ molecules due to their small ground state magnetic moments which, coupled with their unresolved ground state hyperfine splittings, can result in robust, slowly-evolving coherent dark states which significantly limit photon scattering rates. Fortunately, rapid polarization switching offers an alternative method to address stationary dark states and also destabilize coherent dark states, provided that the number of ground states is less than three times the number of available excited states, i.e.~a single laser polarization addresses more than $\frac{1}{3}$ of ground states.
While this can be the case for $Q$-transitions in AlCl and AlF \cite{Truppe2019}, it is not the case for TlF \cite{Norrgard2016,Grasdijk2021,Clayburn2020} where the excited state hyperfine structure is well resolved. In this case, additional switched microwave fields linking rotational states in the electronic ground state are required.

The dark state composition in AlCl was calculated following the method in \refscite{Berkeland2002,Fitch2021}, with the  number of dark states for the $Q(1)$ transition depending on the number of partially resolved excited states addressed. Assuming power broadening is adequate to address the entire $A^{1}\Sigma (v'=0,J'=1)$ state, we find that, in the absence of a magnetic field, $\pi$-transitions driven by linearly polarized light lead to $24$ coherent dark states and no stationary dark states, indicating no leakage from the optical cycle but a constraint on the maximum photon scattering rate for a fixed laser polarization.

A similar calculation for the dark state composition in AlF is described in \refcite{Hofsass2021}. Here dark states are formed by linear superpositions of states with different values of $F_1$ and the limit on the scattering rate is consistent with the smallest splitting among them. In AlCl, the dark state composition is more complicated, with dark states described instead by superpositions between states with different $F_1$, $F$, and $m_F$ quantum numbers, leading to dark states between both magnetic and hyperfine levels. Ongoing experimental work will test these results by measuring photon scattering rates in AlCl with and without polarization modulation \cite{Shaw2023}.

\section{Estimate of the capture velocity of a magneto-optical trap for AlCl}
\label{sec:mot}

To load AlCl into a magneto-optical trap, molecules need to be slowed to (or below) the MOT capture velocity. This step is necessary for molecular MOTs since, in contrast to their atomic counterparts, the beam sources that are bright enough for a realistic experiments typically are not effusive in nature. Instead, the molecular sources, e.g.~a cryogenic buffer gas beam or a super-sonic beam, have a boosted forward velocity distribution with higher average values and widths that are too narrow to provide a sufficient flux of molecules below the MOT's capture velocity.
Hence, currently a slowing stage is essential before molecules can be trapped in a MOT.
In current MOT experiments, both white-light \cite{Barry2012,Hemmerling2016} and chirped slowing \cite{Yeo2015,Truppe2017} are successfully used to prepare beams for trapping.
Since the momentum imparted by each photon recoil is small, it is necessary to cycle many photons ($ \gtrsim10^{4}$), see \rsec{sec:optical_cycling}.
We note that alternative slowing methods that avoid the need for repeated photon scattering, are being explored in the community. Examples include the bichromatic force \cite{Partlow2004,Chieda2012,Yang2016d,Corder2015,Kozyryev2018,Galica2018}, traveling-wave Stark deceleration \cite{Greenberg2021,Aggarwal2021} and Zeeman-Sisyphus deceleration \cite{Augenbraun2021}.

\change{To estimate the MOT capture velocity, we numerically simulate} the dynamics of AlCl molecules entering a MOT.
Here, we use a standard 3D-MOT configuration comprised of a quadrupole magnetic field gradient of $75$\,G/cm axially and three pairs of retro-reflected laser beams, each with a Gaussian beam profile.
We then implement the Hamiltonian from \rsec{sec:AlCl_H} and the MOT configuration in the open source Python package PyLCP \cite{Eckel2022} and solve for the time evolution of the trajectories of AlCl.

The presence of both hyperfine spins in AlCl leads to a large number of quantum states in each rotational manifold that must be included to fully describe the system.
In general, capturing the effects of coherences in simulations requires evaluating the optical Bloch equations \cite{Gordon1980, Ungar1989, Devlin2016}. However, the optical cycling transition $\transition{0}{1}{0}{1}$ involves 144 magnetic sublevels rendering a full simulation computationally challenging.
As a result, we perform the following estimates of the capture velocity using rate equations and therefore expect this approach to break down as laser intensity grows and coherent dark states begin to limit excitation.
By trading intensity for MOT beam diameter, we will operate near the saturation intensity ($\approx 232$\,mW/cm$^2$ for AlCl), where other molecules have been shown to be well described by rate equations  \cite{Tarbutt2015,Hofsass2021}.

We simulate the molecular trajectories for different initial velocities and laser powers and extract the maximum molecular velocity that is captured for a range of MOT beam diameters. The cooling lasers that address the $\transition{0}{1}{0}{1}$ transition are detuned by $-\frac{\Gamma}{2}$ from the $F_1' = 7/2$ level. We further assume that no vibrational branching occurs during the simulation, i.e.~the repump lasers have been applied accordingly. We also reduce the calculated equilibrium force by a factor of two at each timestep of the simulation to account for the $\Lambda$-system created between the cycling and the first repump lasers \cite{Shuman2009}.

\begin{figure}[ht]
    \centering
    \includegraphics[width=\linewidth]{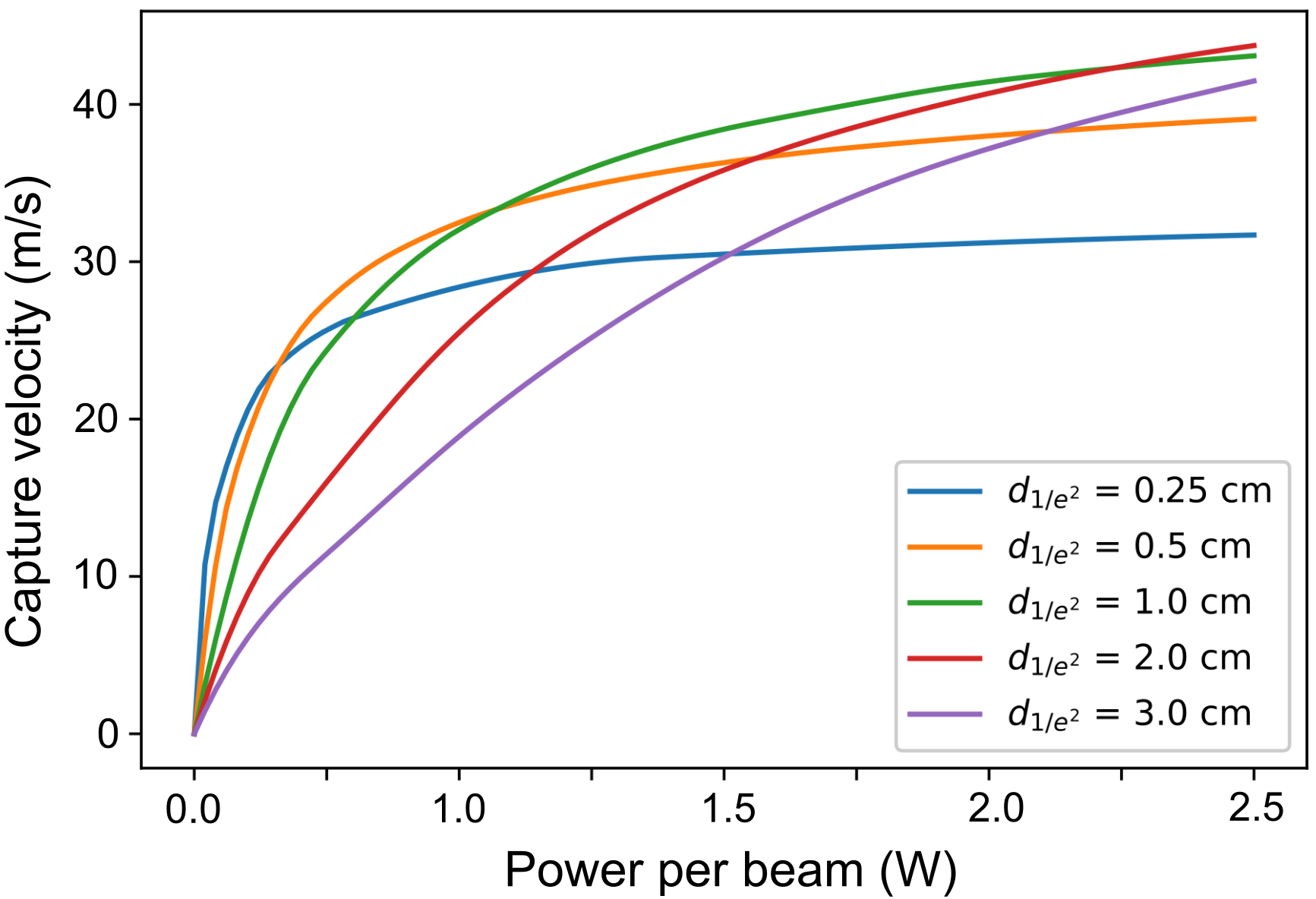}
    \caption{Simulated capture velocities plotted as a function of laser power per beam for different beam diameters. The beam diameter is defined as the $1/e^2$ diameter of the Gaussian beams. Figure reproduced from \refcite{Thesis_Shaw_2022}.}
    \label{fig:capture_velocity}
\end{figure}

The results of these simulations are shown in \rfig{fig:capture_velocity}. We find that a MOT capture velocity $v_{cap}\geq30~$m/s requires a mean intensity in the range $\sim0.1-1~$W/cm$^{2}$ per beam, depending on the MOT beam diameter ($d$), with smaller beams requiring higher intensity (and higher scattering rates, $R_{sc}$) to account for shorter interaction times. These results follow the general results of a simplified two-level model, as discussed in \refcite{Thesis_Shaw_2022}. In brief, for a fixed beam diameter $d$, $v_{cap}\propto \sqrt{d \cdot R_{sc}}$  . At low intensity $I$, for fixed laser power, $R_{sc}\propto I \propto d^{-2}$ and so $v_{cap}\propto d^{-1/2}$. At higher intensity towards saturation, this dependence is weakened since now $R_{sc}\propto d^{-m}$ where $0\le m<2$, with $m=0$ representing when the transition is fully saturated, and so $v_{cap}\propto d^{(1-m)/2}$.

While a large MOT capture velocity is desirable, it is also important to consider the spatial overlap between the MOT volume and the slowed molecular beam, which has a solid angle $\Omega_s \propto d^{2}$. In general, the number of trapped molecules $N_\textrm{MOT}\propto \Omega_s \cdot v_{cap}^{\kappa}$ where $\kappa$ is determined by the slowed beam's velocity profile. Typically, $\kappa > 1$ since there can be many more molecules available for capture at higher velocities due to reduced transverse beam divergence \cite{Barry2012} and this gives rise to two regimes. When below saturation, due to limited laser power, as could be the case in the deep UV for AlF, smaller MOT beams are desirable since here $N_\textrm{MOT} \propto d^{2-\kappa/2}$ with $(2-\kappa/2)<0$. By contrast, when transitions can be saturated using high laser power, $N_\textrm{MOT} \propto d^{2+\kappa(1-m)/2}$ with $2+\kappa(1-m)/2>0$, and large MOT beams are optimal. This high power, large diameter regime is typically where today's molecular MOTs using $^{2}\Sigma$ molecules operate when loading. Our goal is to also operate towards this regime for AlCl using $\sim1$~cm $1/e^{2}$ diameter beams each with $\sim0.5$~W. However, our ongoing work probing the scattering rate vs intensity \cite{Shaw2023} will ultimately guide our MOT beam parameters to use the available laser power \cite{Shaw2020} most efficiently to maximize $N_\textrm{MOT}$.

While high intensity and large scattering rates are beneficial for MOT loading, it is common to reduce the intensity immediately after loading. This reduces the scattering rate which both cools the trapped atoms by limiting Doppler heating and increases the trap lifetime for state preparation or transfer to a conservative trap. This step will also likely be important for an AlCl MOT since the Doppler temperature is $600~\mu K$. A blue-detuned MOT of AlCl, as recently demonstrated for YO \cite{Burau2023} and SrF \cite{Langin2023}, could potentially cool below this limit towards the recoil temperature of $5~\mu K$, though substantial vibrational closure would be needed for efficient transfer from the red-detuned MOT since these blue-detuned MOTs operate at high intensity and require $>20~$ms to load \cite{Langin2023}.

For short wavelength transitions at high intensity, one also needs to consider the possibility of significant loss via photo-dissociation and photo-ionization of the molecule. The latter effect has been known to dominate loss processes in atomic MOTs \cite{Brickman2007}. Here, we carry out {\it ab initio} calculations on the cross-sections of AlCl for these processes to characterize their effect.

\section{Photo-dissociation and ionization of AlCl}
\label{sec:dissociate}

\begin{figure}[ht]
    \centering
    \includegraphics[width=\linewidth]{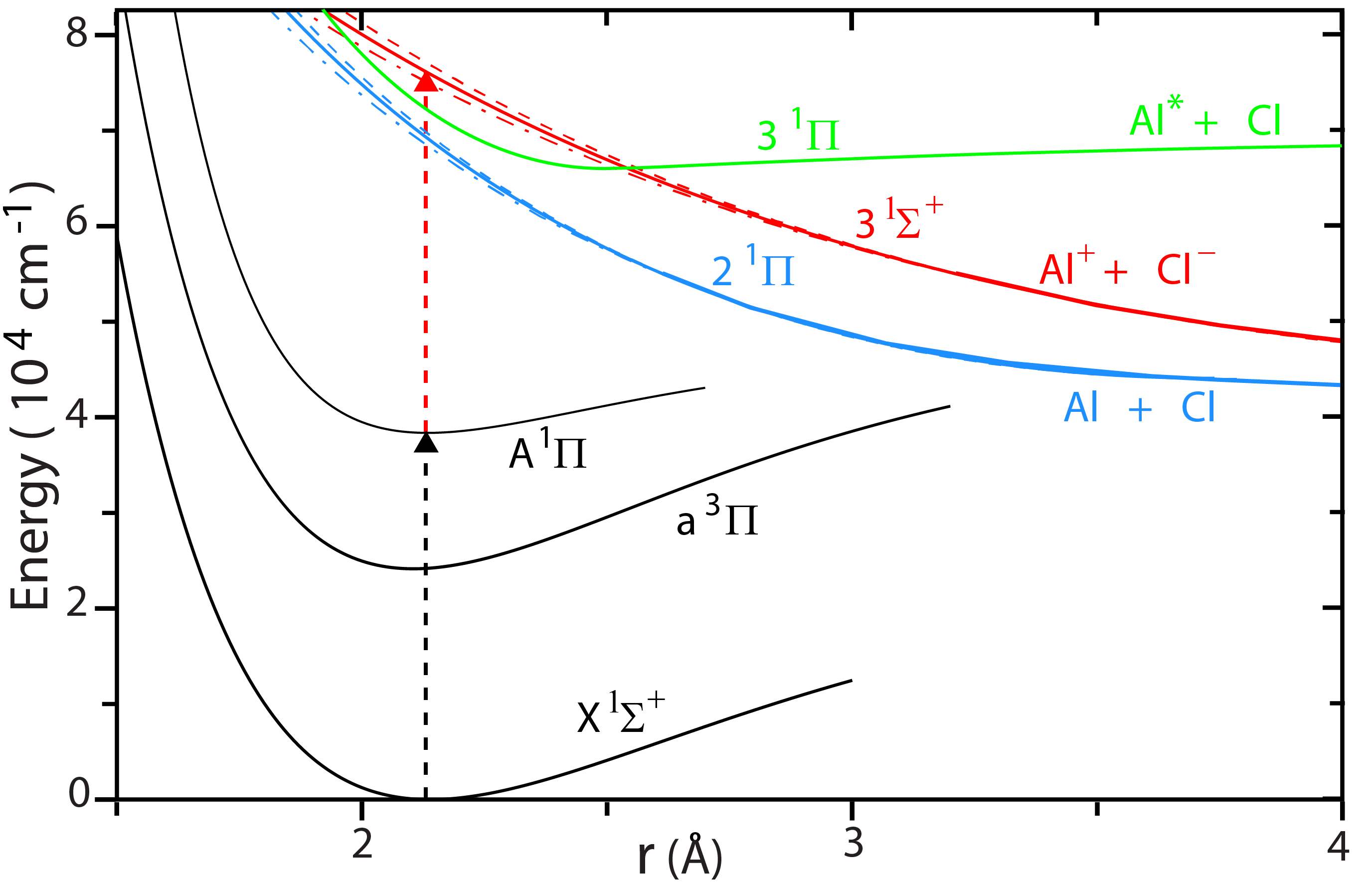}
    \caption{\change{Six} selected potential energy curves (PECs) are plotted for AlCl as a function of the internuclear distance $r$.  The three black and one blue PECs correlate to neutral Al + Cl dissociation for large $r$ whereas the red PEC leads to ionization Al$^+$ + Cl$^-$ \change{and the green PEC leads to Al$^*$ + Cl.} The two-photon process is indicated by the vertical dashed black and red arrows.\change{ The dashed and long-short dashed PECs correspond to increasing and decreasing the slopes of the PECs by 1.5\%, respectively.}}
    \label{fig:PECs}
\end{figure}

In this section, we analyze the two-photon process that can lead to photo-induced dissociation and ionization. \rfig{fig:PECs} plots selected potential energy curves (PECs) for AlCl. \change{The lowest three black curves are Morse potentials that have been fit to {\it ab initio} calculations \cite{Daniel2021}.} These PECs correlate asymptotically for large internuclear separation ($r$) to the dissociation energy of neutral Al + Cl. \change{Three} relevant excited electronic state PECs are plotted in blue, \change{red, and green for the $2\,^1\Pi$, $3\,^1\Sigma^+$, and $3\,^1\Pi$ states, respectively.} Asymptotically for large $r$, the $2\,^1\Pi$ PEC approaches the dissociation energy of neutral Al + Cl whereas the $3\,^1\Sigma^+$ PEC approaches the dissociation energy of ionic Al$^+$ + Cl$^-$, \change{and the $3\,^1\Pi$ PEC approaches the dissociation energy of Al$^*$ + Cl \cite{Ren2021,Qin2021}.} \change{The repulsive part of these} PECs are based on fitting repulsive exponential functions ($V=V_o\,{\rm exp}[\alpha(r - r_o)^2]$) to the {\it ab initio} data reported in \refcite{Ren2021,Qin2021}. The $V_o$, $\alpha$, and $r_o$ are adjustable fitting parameters and were optimized to minimize the root-mean-square error between the analytic curves and the {\it ab initio} data.

A simplex fitting algorithm was used (AMOEBA \cite{Press1986}) and the optimal parameters were determined to be $V_o=4.40249\,\times 10^4\,{\rm cm}^{-1}$, $\alpha=0.174950\,\textrm{\AA}^{-2}$,
and $r_o=3.73886\,\textrm{\AA} $ for the $2\,^1\Pi$ state and $V_o=4.53576\,\times 10^4\,{\rm cm}^{-1}$, $\alpha=6.76293\times 10^{-2} \textrm{\AA}^{-2}$,
and $r_o= 4.89511\,\textrm{\AA}$ for the $3\,^1\Sigma^+$ state \change{and $V_o=6.58982\,\times 10^4\,{\rm cm}^{-1}$, $\alpha=0.683155\,\textrm{\AA}^{-2}$,
and $r_o=2.49520\,\textrm{\AA}$ for the $3\,^1\Pi$ state} (the fitted parameters quoted above include several extra digits for numerical reasons to ensure that the potential curves can be accurately reproduced). The vertical black and red dashed arrows represent the two-photon excitation process that can lead to dissociation along the repulsive $2\,^1\Pi$ PEC (blue) \change{or the $3\,^1\Pi$ PEC (green),} or ionization along the repulsive $3\,^1\Sigma^+$ PEC (red). From these PECs, we can compute intensity profiles for the dissociation and ionization cross sections by calculating the Franck-Condon overlaps between the ground ro-vibrational eigenfunction of the $A\,^1\Pi$ state with the continuum eigenstates of the $2\,^1\Pi$, \change{$3\,^1\Pi$} and $3\,^1\Sigma^+$ states, respectively.

The reflection technique is used where the continuum eigenfunctions are represented by delta functions located at the classical turning points along the repulsive PECs \cite{herzDspectra1950}. The intensity profiles are then simply proportional to $\nu\,\psi_0^2$ where $\nu$ is the excitation energy and $\psi_0$ is the ground ro-vibrational wave function evaluated at the $r$ corresponding the classical turning point for the energy $\nu$. We can derive the classical turning points $r_c$ as a function of $\nu$ from the exponential functions given above by setting $V=\nu$ to obtain $r_c = r_o - \sqrt{{\rm ln}(\nu/V_o)/\alpha}$.

\begin{figure}[ht]
    \centering
    \includegraphics[width=\linewidth]{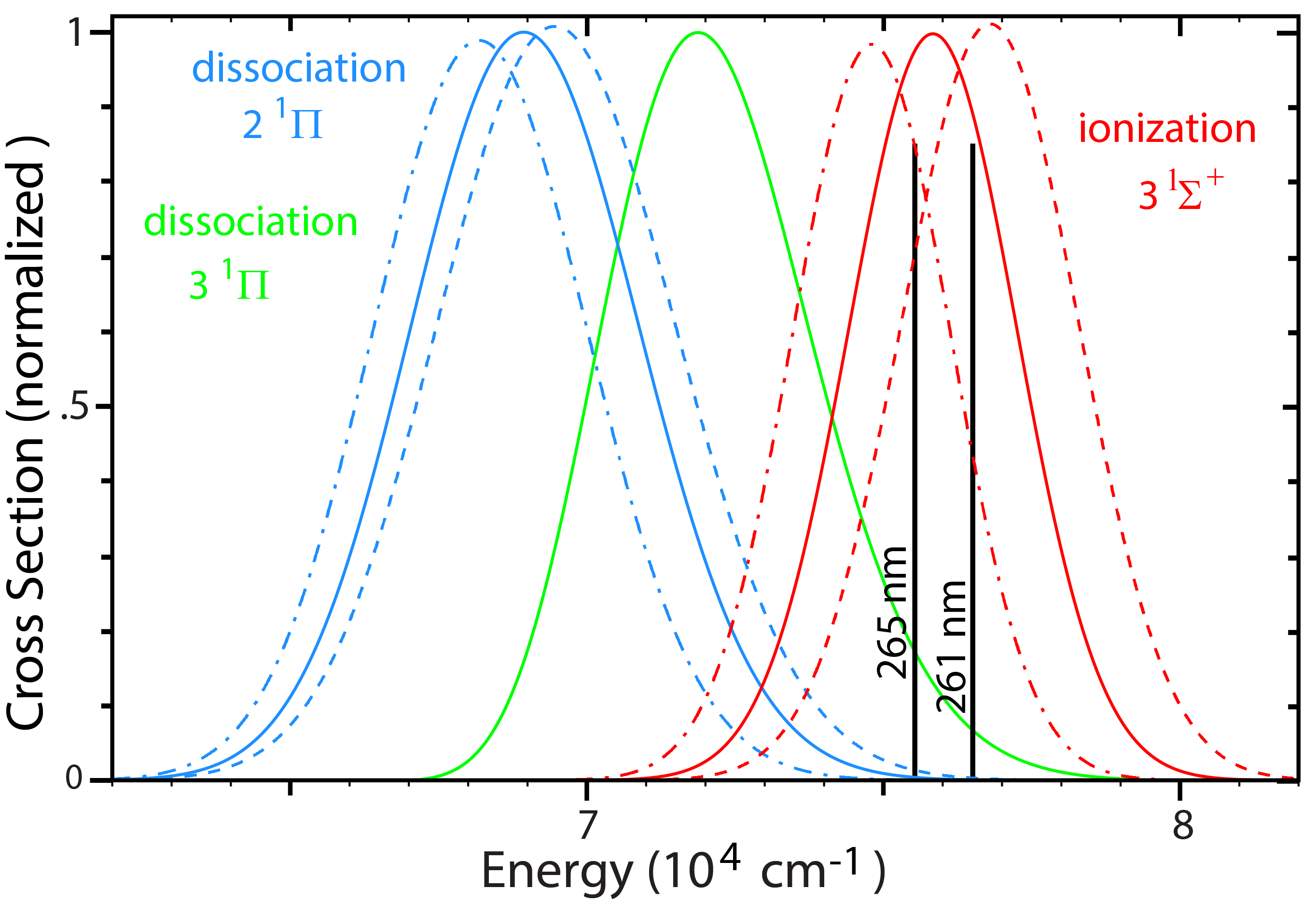}
    \caption{The normalized photo-dissociation \change{(blue =  $2\,^1\Pi$ and green = $3\,^1\Pi$)} and photo-ionization (red) cross sections are plotted as a function of the excitation energy. The energies of the two relevant laser wavelengths (261\,nm: main cooling transition, and 265\,nm: first vibrational repumper) are indicated by the vertical black lines. The dashed and short-long dashed curves quantify the sensitivity of the cross sections to the slopes of \
the potential energy curves of the repulsive excited electronic states (see \rfig{fig:PECs}).}
    \label{fig:Cross}
\end{figure}

An energy grid in $\nu$ was constructed using 100 points between 6.0 and 8.5 cm$^{-1}$ ($\times 10^4$). The ground ro-vibrational wavefunction for the $A\,^1\Pi$ state (computed in our previous work \cite{Daniel2021}) was then evaluated at each of the corresponding $r_c$ values and the resulting intensity profiles (normalized) are plotted in   \rfig{fig:Cross} for both dissociation (solid blue \change{and green)} and ionization (solid red). For reference, the energies of the two relevant laser wavelengths (265 and 261 nm) are plotted with black vertical lines. \change{The sensitivity of the profiles plotted in \rfig{fig:Cross} on the exponential fits was quantified by increasing and decreasing the slopes of the PECs by approximately $1.5\%$ (the dashed and long-short dashed red and blue curves in \rfig{fig:PECs}). The corresponding cross sections are plotted with dashed and long-short dashed red and blue curves in \rfig{fig:Cross}. The sensitivity of the
$3\,^1\Pi$ cross section (green) is similar but is not plotted for clarity.
From these normalized profiles, it is clear that these laser wavelengths could lead to photo-ionization via the excited $3^1\Sigma^+$ state but depending upon the relative magnitudes of the photo-dissociation cross sections, significant photo-dissociation could also occur.
Determining the absolute cross sections requires experimental measurements that are beyond the
scope of the present work and at these low temperatures literature values are scarce.
However, recent experiments on a similar molecule AlF report an absolute photo-ionization cross section
of $\sigma_i= 2\times 10^{-18} {\rm cm}^2$ that includes all cycling transitions \cite{Hofsass2021}.
To the authors' knowledge, the only absolute photo-dissociation cross sections reported for AlCl are the
theoretical ones in \refcite{Qin2021}: $\sigma_{2^1\Pi}= 2\times 10^{-17}{\rm cm}^2$ and
$\sigma_{3^1\Pi}= 1.5\times 10^{-16}{\rm cm}^2$.
If we assume that the absolute photo-ionization cross section for AlCl is similar to AlF, then we can estimate the relative magnitudes of the AlCl photo-dissociation cross sections: $\sigma_{2^1\Pi}$/$\sigma_i = 10$ and $\sigma_{3^1\Pi}$/$\sigma_i = 75$.
Thus, we multiply the blue curves by 10 and the green curve by 75 in \rfig{fig:Cross} and keep the red curves unchanged.
The tails of the blue and green curves at the relevant laser wavelengths are now much larger.
For the $2^1\Pi$ state (solid blue curve) the estimated relative contributions (multiplied by 10) at 265 and 261\,nm
are only 0.4 and 0.8\% of the photo-ionization cross section, respectively. However, for the $3^1\Pi$ state (solid green curve) the estimated relative contributions (multiplied by 75) are 13.2 and 5.2 times larger than the photo-ionization cross section, respectively. Thus, due to its large absolute cross section, photo-dissociation via the $3^1\Pi$ state is potentially a dominant loss mechanism.
}

\section{Conclusion}
\label{sec:conclusion}

We have characterized the hyperfine structure of the $\Astate$ state in AlCl and reported the first measurement of the nuclear spin-electron orbit interaction strength of the excited state. We discuss strategies and possible loss mechanisms for efficient optical cycling and compare the advantages of AlCl to other molecules. Both AlCl and AlF have similar internal structures and are expected to allow for similar cooling schemes. While the larger linewidth and shorter wavelength of AlF results in access to larger optical forces, the AlF saturation intensity is a factor of 4 higher than AlCl which poses significant demands on the laser technology and optics required.
An additional advantage of AlCl (and AlF) is that transitions can simultaneously be driven to multiple excited states, allowing polarization modulation to destabilize coherent dark states, in contrast to TlF. The larger ground state hyperfine splittings in AlCl and AlF also result in these dark states naturally precessing into bright states more rapidly than in the case of TlF.

Finally, the hyperfine structure in the $\Astate$ state studied in this paper spans between $\approx 250$ and $500$~MHz for $J'=1-4$ while the $\Xstate$ hyperfine structure is unresolved for the cycling transition. While such a broad excited state spread makes addressing individual velocity classes challenging using a type-II transition in, for example, a Zeeman slower, it does potentially offer two advantages.
First, increased scattering rates may be accessed by targeting different excited hyperfine states with the cycling and first repump lasers to avoid coupling these lasers and creating a $\Lambda-$system. This approach is similar to that used with alkali atoms. Second, laser slowing may be simplified since a single laser frequency can simultaneously address a broad range of velocities, similar to white light or frequency chirped slowing, but without the need to spectrally or temporally dilute the laser intensity applied to each velocity. This may allow a single laser frequency to slow molecules directly from a single-stage CBGB and reduce the technical complexity.

\change{We use a numerical simulation} of the full AlCl Hamiltonian to estimate the capture velocity of a magneto-optical trap for AlCl. Our results yield capture velocities of up to $30-40$\,m/s when using $\approx 1$\,W of laser power per MOT beam, suggesting that a significant part of a CBGB source with a slowing cell \cite{Lu2011} could be directly loaded into a MOT without slowing. This result highlights another advantage of AlCl, but should also be understood in the context of MOT capture velocities of other molecules, where magnitudes for CaF of $5-20$\,m/s \cite{Chae2017,Williams2017,Tarbutt2015,Langin2023}, for SrF of $9-13$\,m/s \cite{Langin2023}, and for MgF of $26$\,m/s \cite{Xu2019b} have been either calculated or measured. Optimization strategies for increasing these values to higher magnitudes have been explored as well \cite{Xu2021}.

Finally, a possible limit preventing a high-intensity MOT for AlCl is that two-photon excitation may lead to substantial trap loss. Using {\it ab initio} calculations of the excitation cross-sections, we identified that \change{photo-dissociation via a $3^1\Pi$ state and}
ionization via a dissociative $3^1\Sigma^+$ state could be a non-negligible loss process when using the main cycling and the first vibrational repump transition for the MOT. Experimental data on this process are required to verify if photoionization or -dissociation will indeed limit the lifetime of an AlCl MOT.

\bigskip
\begin{acknowledgements}
J.R.D., C.W., L.-R.L., and B.H. acknowledge funding from the NSF grant number 1839153 and from the AFRL grant number FA9550-21-1-0263. B.K.K.~acknowledges that part of this work was done under the auspices of the U.S. Department of Energy under Project No.~20240256ER of the Laboratory Directed Research and Development Program at Los Alamos National Laboratory. Los Alamos National Laboratory is operated by Triad National Security, LLC, for the National Nuclear Security Administration of the U.S. Department of Energy (Contract No. 89233218CNA000001). J.C.S. and D.J.M. gratefully acknowledge funding through an NSF CAREER award (Grant No.~1848435), and the University of Connecticut College of Liberal Arts and Sciences and the Office of the Vice President for Research.
\end{acknowledgements}

\bibliography{main}

\begin{thebibliography}{221}%
\makeatletter
\providecommand \@ifxundefined [1]{%
 \@ifx{#1\undefined}
}%
\providecommand \@ifnum [1]{%
 \ifnum #1\expandafter \@firstoftwo
 \else \expandafter \@secondoftwo
 \fi
}%
\providecommand \@ifx [1]{%
 \ifx #1\expandafter \@firstoftwo
 \else \expandafter \@secondoftwo
 \fi
}%
\providecommand \natexlab [1]{#1}%
\providecommand \enquote  [1]{``#1''}%
\providecommand \bibnamefont  [1]{#1}%
\providecommand \bibfnamefont [1]{#1}%
\providecommand \citenamefont [1]{#1}%
\providecommand \href@noop [0]{\@secondoftwo}%
\providecommand \href [0]{\begingroup \@sanitize@url \@href}%
\providecommand \@href[1]{\@@startlink{#1}\@@href}%
\providecommand \@@href[1]{\endgroup#1\@@endlink}%
\providecommand \@sanitize@url [0]{\catcode `\\12\catcode `\$12\catcode
  `\&12\catcode `\#12\catcode `\^12\catcode `\_12\catcode `\%12\relax}%
\providecommand \@@startlink[1]{}%
\providecommand \@@endlink[0]{}%
\providecommand \url  [0]{\begingroup\@sanitize@url \@url }%
\providecommand \@url [1]{\endgroup\@href {#1}{\urlprefix }}%
\providecommand \urlprefix  [0]{URL }%
\providecommand \Eprint [0]{\href }%
\providecommand \doibase [0]{http://dx.doi.org/}%
\providecommand \selectlanguage [0]{\@gobble}%
\providecommand \bibinfo  [0]{\@secondoftwo}%
\providecommand \bibfield  [0]{\@secondoftwo}%
\providecommand \translation [1]{[#1]}%
\providecommand \BibitemOpen [0]{}%
\providecommand \bibitemStop [0]{}%
\providecommand \bibitemNoStop [0]{.\EOS\space}%
\providecommand \EOS [0]{\spacefactor3000\relax}%
\providecommand \BibitemShut  [1]{\csname bibitem#1\endcsname}%
\let\auto@bib@innerbib\@empty
\bibitem [{\citenamefont {Andreev}\ \emph {et~al.}(2018)\citenamefont
  {Andreev}, \citenamefont {Ang}, \citenamefont {DeMille}, \citenamefont
  {Doyle}, \citenamefont {Gabrielse}, \citenamefont {Haefner}, \citenamefont
  {Hutzler}, \citenamefont {Lasner}, \citenamefont {Meisenhelder},
  \citenamefont {O’Leary}, \citenamefont {Panda}, \citenamefont {West},
  \citenamefont {West}, \citenamefont {Wu},\ and\ \citenamefont
  {Collaboration}}]{Andreev2018}%
  \BibitemOpen
  \bibfield  {author} {\bibinfo {author} {\bibfnamefont {V.}~\bibnamefont
  {Andreev}}, \bibinfo {author} {\bibfnamefont {D.~G.}\ \bibnamefont {Ang}},
  \bibinfo {author} {\bibfnamefont {D.}~\bibnamefont {DeMille}}, \bibinfo
  {author} {\bibfnamefont {J.~M.}\ \bibnamefont {Doyle}}, \bibinfo {author}
  {\bibfnamefont {G.}~\bibnamefont {Gabrielse}}, \bibinfo {author}
  {\bibfnamefont {J.}~\bibnamefont {Haefner}}, \bibinfo {author} {\bibfnamefont
  {N.~R.}\ \bibnamefont {Hutzler}}, \bibinfo {author} {\bibfnamefont
  {Z.}~\bibnamefont {Lasner}}, \bibinfo {author} {\bibfnamefont
  {C.}~\bibnamefont {Meisenhelder}}, \bibinfo {author} {\bibfnamefont {B.~R.}\
  \bibnamefont {O’Leary}}, \bibinfo {author} {\bibfnamefont {C.~D.}\
  \bibnamefont {Panda}}, \bibinfo {author} {\bibfnamefont {A.~D.}\ \bibnamefont
  {West}}, \bibinfo {author} {\bibfnamefont {E.~P.}\ \bibnamefont {West}},
  \bibinfo {author} {\bibfnamefont {X.}~\bibnamefont {Wu}}, \ and\ \bibinfo
  {author} {\bibfnamefont {A.}~\bibnamefont {Collaboration}},\ }\href {\doibase
  10.1038/s41586-018-0599-8} {\bibfield  {journal} {\bibinfo  {journal}
  {Nature}\ }\textbf {\bibinfo {volume} {562}},\ \bibinfo {pages} {355}
  (\bibinfo {year} {2018})}\BibitemShut {NoStop}%
\bibitem [{\citenamefont {Cairncross}\ \emph {et~al.}(2017)\citenamefont
  {Cairncross}, \citenamefont {Gresh}, \citenamefont {Grau}, \citenamefont
  {Cossel}, \citenamefont {Roussy}, \citenamefont {Ni}, \citenamefont {Zhou},
  \citenamefont {Ye},\ and\ \citenamefont {Cornell}}]{Cairncross2017}%
  \BibitemOpen
  \bibfield  {author} {\bibinfo {author} {\bibfnamefont {W.~B.}\ \bibnamefont
  {Cairncross}}, \bibinfo {author} {\bibfnamefont {D.~N.}\ \bibnamefont
  {Gresh}}, \bibinfo {author} {\bibfnamefont {M.}~\bibnamefont {Grau}},
  \bibinfo {author} {\bibfnamefont {K.~C.}\ \bibnamefont {Cossel}}, \bibinfo
  {author} {\bibfnamefont {T.~S.}\ \bibnamefont {Roussy}}, \bibinfo {author}
  {\bibfnamefont {Y.}~\bibnamefont {Ni}}, \bibinfo {author} {\bibfnamefont
  {Y.}~\bibnamefont {Zhou}}, \bibinfo {author} {\bibfnamefont {J.}~\bibnamefont
  {Ye}}, \ and\ \bibinfo {author} {\bibfnamefont {E.~A.}\ \bibnamefont
  {Cornell}},\ }\href {\doibase 10.1103/PhysRevLett.119.153001} {\bibfield
  {journal} {\bibinfo  {journal} {Phys. Rev. Lett.}\ }\textbf {\bibinfo
  {volume} {119}},\ \bibinfo {pages} {153001} (\bibinfo {year}
  {2017})}\BibitemShut {NoStop}%
\bibitem [{\citenamefont {Kozyryev}\ and\ \citenamefont
  {Hutzler}(2017)}]{Kozyryev2017a}%
  \BibitemOpen
  \bibfield  {author} {\bibinfo {author} {\bibfnamefont {I.}~\bibnamefont
  {Kozyryev}}\ and\ \bibinfo {author} {\bibfnamefont {N.~R.}\ \bibnamefont
  {Hutzler}},\ }\href {\doibase 10.1103/PhysRevLett.119.133002} {\bibfield
  {journal} {\bibinfo  {journal} {Physical Review Letters}\ }\textbf {\bibinfo
  {volume} {119}},\ \bibinfo {pages} {133002} (\bibinfo {year}
  {2017})}\BibitemShut {NoStop}%
\bibitem [{\citenamefont {Hudson}\ \emph {et~al.}(2011)\citenamefont {Hudson},
  \citenamefont {Kara}, \citenamefont {Smallman}, \citenamefont {Sauer},
  \citenamefont {Tarbutt},\ and\ \citenamefont {Hinds}}]{Hudson2011}%
  \BibitemOpen
  \bibfield  {author} {\bibinfo {author} {\bibfnamefont {J.~J.}\ \bibnamefont
  {Hudson}}, \bibinfo {author} {\bibfnamefont {D.~M.}\ \bibnamefont {Kara}},
  \bibinfo {author} {\bibfnamefont {I.~J.}\ \bibnamefont {Smallman}}, \bibinfo
  {author} {\bibfnamefont {B.~E.}\ \bibnamefont {Sauer}}, \bibinfo {author}
  {\bibfnamefont {M.~R.}\ \bibnamefont {Tarbutt}}, \ and\ \bibinfo {author}
  {\bibfnamefont {E.~A.}\ \bibnamefont {Hinds}},\ }\href {\doibase
  10.1038/nature10104} {\bibfield  {journal} {\bibinfo  {journal} {Nature}\
  }\textbf {\bibinfo {volume} {473}},\ \bibinfo {pages} {493} (\bibinfo {year}
  {2011})}\BibitemShut {NoStop}%
\bibitem [{\citenamefont {Kozyryev}\ \emph {et~al.}(2021)\citenamefont
  {Kozyryev}, \citenamefont {Lasner},\ and\ \citenamefont
  {Doyle}}]{Kozyryev2021}%
  \BibitemOpen
  \bibfield  {author} {\bibinfo {author} {\bibfnamefont {I.}~\bibnamefont
  {Kozyryev}}, \bibinfo {author} {\bibfnamefont {Z.}~\bibnamefont {Lasner}}, \
  and\ \bibinfo {author} {\bibfnamefont {J.~M.}\ \bibnamefont {Doyle}},\ }\href
  {\doibase 10.1103/PhysRevA.103.043313} {\bibfield  {journal} {\bibinfo
  {journal} {Physical Review A}\ }\textbf {\bibinfo {volume} {103}},\ \bibinfo
  {pages} {043313} (\bibinfo {year} {2021})}\BibitemShut {NoStop}%
\bibitem [{\citenamefont {Kondov}\ \emph {et~al.}(2019)\citenamefont {Kondov},
  \citenamefont {Lee}, \citenamefont {Leung}, \citenamefont {Liedl},
  \citenamefont {Majewska}, \citenamefont {Moszynski},\ and\ \citenamefont
  {Zelevinsky}}]{Kondov2019}%
  \BibitemOpen
  \bibfield  {author} {\bibinfo {author} {\bibfnamefont {S.~S.}\ \bibnamefont
  {Kondov}}, \bibinfo {author} {\bibfnamefont {C.-H.}\ \bibnamefont {Lee}},
  \bibinfo {author} {\bibfnamefont {K.~H.}\ \bibnamefont {Leung}}, \bibinfo
  {author} {\bibfnamefont {C.}~\bibnamefont {Liedl}}, \bibinfo {author}
  {\bibfnamefont {I.}~\bibnamefont {Majewska}}, \bibinfo {author}
  {\bibfnamefont {R.}~\bibnamefont {Moszynski}}, \ and\ \bibinfo {author}
  {\bibfnamefont {T.}~\bibnamefont {Zelevinsky}},\ }\href {\doibase
  10.1038/s41567-019-0632-3} {\bibfield  {journal} {\bibinfo  {journal} {Nature
  Physics}\ }\textbf {\bibinfo {volume} {15}},\ \bibinfo {pages} {1118}
  (\bibinfo {year} {2019})}\BibitemShut {NoStop}%
\bibitem [{\citenamefont {Collaboration}\ \emph {et~al.}(2014)\citenamefont
  {Collaboration}, \citenamefont {Baron}, \citenamefont {Campbell},
  \citenamefont {DeMille}, \citenamefont {Doyle}, \citenamefont {Gabrielse},
  \citenamefont {Gurevich}, \citenamefont {Hess}, \citenamefont {Hutzler},
  \citenamefont {Kirilov}, \citenamefont {Kozyryev}, \citenamefont {O'Leary},
  \citenamefont {Panda}, \citenamefont {Parsons}, \citenamefont {Petrik},
  \citenamefont {Spaun}, \citenamefont {Vutha},\ and\ \citenamefont
  {West}}]{ACMECollaboration2014}%
  \BibitemOpen
  \bibfield  {author} {\bibinfo {author} {\bibfnamefont {T.~A.~A.}\
  \bibnamefont {Collaboration}}, \bibinfo {author} {\bibfnamefont
  {J.}~\bibnamefont {Baron}}, \bibinfo {author} {\bibfnamefont {W.~C.}\
  \bibnamefont {Campbell}}, \bibinfo {author} {\bibfnamefont {D.}~\bibnamefont
  {DeMille}}, \bibinfo {author} {\bibfnamefont {J.~M.}\ \bibnamefont {Doyle}},
  \bibinfo {author} {\bibfnamefont {G.}~\bibnamefont {Gabrielse}}, \bibinfo
  {author} {\bibfnamefont {Y.~V.}\ \bibnamefont {Gurevich}}, \bibinfo {author}
  {\bibfnamefont {P.~W.}\ \bibnamefont {Hess}}, \bibinfo {author}
  {\bibfnamefont {N.~R.}\ \bibnamefont {Hutzler}}, \bibinfo {author}
  {\bibfnamefont {E.}~\bibnamefont {Kirilov}}, \bibinfo {author} {\bibfnamefont
  {I.}~\bibnamefont {Kozyryev}}, \bibinfo {author} {\bibfnamefont {B.~R.}\
  \bibnamefont {O'Leary}}, \bibinfo {author} {\bibfnamefont {C.~D.}\
  \bibnamefont {Panda}}, \bibinfo {author} {\bibfnamefont {M.~F.}\ \bibnamefont
  {Parsons}}, \bibinfo {author} {\bibfnamefont {E.~S.}\ \bibnamefont {Petrik}},
  \bibinfo {author} {\bibfnamefont {B.}~\bibnamefont {Spaun}}, \bibinfo
  {author} {\bibfnamefont {A.~C.}\ \bibnamefont {Vutha}}, \ and\ \bibinfo
  {author} {\bibfnamefont {A.~D.}\ \bibnamefont {West}},\ }\href {\doibase
  10.1126/science.1248213} {\bibfield  {journal} {\bibinfo  {journal}
  {Science}\ }\textbf {\bibinfo {volume} {343}},\ \bibinfo {pages} {269}
  (\bibinfo {year} {2014})}\BibitemShut {NoStop}%
\bibitem [{\citenamefont {Fitch}\ \emph {et~al.}(2021)\citenamefont {Fitch},
  \citenamefont {Lim}, \citenamefont {Hinds}, \citenamefont {Sauer},\ and\
  \citenamefont {Tarbutt}}]{Fitch2021a}%
  \BibitemOpen
  \bibfield  {author} {\bibinfo {author} {\bibfnamefont {N.~J.}\ \bibnamefont
  {Fitch}}, \bibinfo {author} {\bibfnamefont {J.}~\bibnamefont {Lim}}, \bibinfo
  {author} {\bibfnamefont {E.~A.}\ \bibnamefont {Hinds}}, \bibinfo {author}
  {\bibfnamefont {B.~E.}\ \bibnamefont {Sauer}}, \ and\ \bibinfo {author}
  {\bibfnamefont {M.~R.}\ \bibnamefont {Tarbutt}},\ }\href {\doibase
  10.1088/2058-9565/abc931} {\bibfield  {journal} {\bibinfo  {journal} {Quantum
  Sci. Technol.}\ }\textbf {\bibinfo {volume} {6}},\ \bibinfo {pages} {014006}
  (\bibinfo {year} {2021})}\BibitemShut {NoStop}%
\bibitem [{\citenamefont {Yu}\ and\ \citenamefont {Hutzler}(2021)}]{Yu2021}%
  \BibitemOpen
  \bibfield  {author} {\bibinfo {author} {\bibfnamefont {P.}~\bibnamefont
  {Yu}}\ and\ \bibinfo {author} {\bibfnamefont {N.~R.}\ \bibnamefont
  {Hutzler}},\ }\href {\doibase 10.1103/PhysRevLett.126.023003} {\bibfield
  {journal} {\bibinfo  {journal} {Physical Review Letters}\ }\textbf {\bibinfo
  {volume} {126}},\ \bibinfo {pages} {023003} (\bibinfo {year}
  {2021})}\BibitemShut {NoStop}%
\bibitem [{\citenamefont {Hutzler}(2021)}]{Hutzler2020}%
  \BibitemOpen
  \bibfield  {author} {\bibinfo {author} {\bibfnamefont {N.~R.}\ \bibnamefont
  {Hutzler}},\ }\href {\doibase 10.1088/2058-9565/abb9c5} {\bibfield  {journal}
  {\bibinfo  {journal} {Quantum Science and Technology}\ }\textbf {\bibinfo
  {volume} {5}},\ \bibinfo {pages} {44011} (\bibinfo {year}
  {2021})}\BibitemShut {NoStop}%
\bibitem [{\citenamefont {O'Rourke}\ and\ \citenamefont
  {Hutzler}(2019)}]{ORourke2019}%
  \BibitemOpen
  \bibfield  {author} {\bibinfo {author} {\bibfnamefont {M.~J.}\ \bibnamefont
  {O'Rourke}}\ and\ \bibinfo {author} {\bibfnamefont {N.~R.}\ \bibnamefont
  {Hutzler}},\ }\href {\doibase 10.1103/PhysRevA.100.022502} {\bibfield
  {journal} {\bibinfo  {journal} {Physical Review A}\ }\textbf {\bibinfo
  {volume} {100}},\ \bibinfo {pages} {022502} (\bibinfo {year}
  {2019})}\BibitemShut {NoStop}%
\bibitem [{\citenamefont {Aggarwal}\ \emph {et~al.}(2018)\citenamefont
  {Aggarwal}, \citenamefont {Bethlem}, \citenamefont {Borschevsky},
  \citenamefont {Denis}, \citenamefont {Esajas}, \citenamefont {Haase},
  \citenamefont {Hao}, \citenamefont {Hoekstra}, \citenamefont {Jungmann},
  \citenamefont {Meijknecht}, \citenamefont {Mooij}, \citenamefont
  {Timmermans}, \citenamefont {Ubachs}, \citenamefont {Willmann},\ and\
  \citenamefont {Zapara}}]{Aggarwal2018}%
  \BibitemOpen
  \bibfield  {author} {\bibinfo {author} {\bibfnamefont {P.}~\bibnamefont
  {Aggarwal}}, \bibinfo {author} {\bibfnamefont {H.~L.}\ \bibnamefont
  {Bethlem}}, \bibinfo {author} {\bibfnamefont {A.}~\bibnamefont
  {Borschevsky}}, \bibinfo {author} {\bibfnamefont {M.}~\bibnamefont {Denis}},
  \bibinfo {author} {\bibfnamefont {K.}~\bibnamefont {Esajas}}, \bibinfo
  {author} {\bibfnamefont {P.~A.}\ \bibnamefont {Haase}}, \bibinfo {author}
  {\bibfnamefont {Y.}~\bibnamefont {Hao}}, \bibinfo {author} {\bibfnamefont
  {S.}~\bibnamefont {Hoekstra}}, \bibinfo {author} {\bibfnamefont
  {K.}~\bibnamefont {Jungmann}}, \bibinfo {author} {\bibfnamefont {T.~B.}\
  \bibnamefont {Meijknecht}}, \bibinfo {author} {\bibfnamefont {M.~C.}\
  \bibnamefont {Mooij}}, \bibinfo {author} {\bibfnamefont {R.~G.}\ \bibnamefont
  {Timmermans}}, \bibinfo {author} {\bibfnamefont {W.}~\bibnamefont {Ubachs}},
  \bibinfo {author} {\bibfnamefont {L.}~\bibnamefont {Willmann}}, \ and\
  \bibinfo {author} {\bibfnamefont {A.}~\bibnamefont {Zapara}},\ }\href
  {\doibase 10.1140/epjd/e2018-90192-9} {\bibfield  {journal} {\bibinfo
  {journal} {European Physical Journal D}\ }\textbf {\bibinfo {volume} {72}},\
  \bibinfo {pages} {197} (\bibinfo {year} {2018})}\BibitemShut {NoStop}%
\bibitem [{\citenamefont {Uzan}(2003)}]{Uzan2003}%
  \BibitemOpen
  \bibfield  {author} {\bibinfo {author} {\bibfnamefont {J.-P.}\ \bibnamefont
  {Uzan}},\ }\href {\doibase 10.1103/RevModPhys.75.403} {\bibfield  {journal}
  {\bibinfo  {journal} {Reviews of Modern Physics}\ }\textbf {\bibinfo {volume}
  {75}},\ \bibinfo {pages} {403} (\bibinfo {year} {2003})}\BibitemShut
  {NoStop}%
\bibitem [{\citenamefont {DeMille}\ \emph {et~al.}(2008)\citenamefont
  {DeMille}, \citenamefont {Sainis}, \citenamefont {Sage}, \citenamefont
  {Bergeman}, \citenamefont {Kotochigova},\ and\ \citenamefont
  {Tiesinga}}]{DeMille2008}%
  \BibitemOpen
  \bibfield  {author} {\bibinfo {author} {\bibfnamefont {D.}~\bibnamefont
  {DeMille}}, \bibinfo {author} {\bibfnamefont {S.}~\bibnamefont {Sainis}},
  \bibinfo {author} {\bibfnamefont {J.}~\bibnamefont {Sage}}, \bibinfo {author}
  {\bibfnamefont {T.}~\bibnamefont {Bergeman}}, \bibinfo {author}
  {\bibfnamefont {S.}~\bibnamefont {Kotochigova}}, \ and\ \bibinfo {author}
  {\bibfnamefont {E.}~\bibnamefont {Tiesinga}},\ }\href {\doibase
  10.1103/PhysRevLett.100.043202} {\bibfield  {journal} {\bibinfo  {journal}
  {Physical Review Letters}\ }\textbf {\bibinfo {volume} {100}},\ \bibinfo
  {pages} {043202} (\bibinfo {year} {2008})}\BibitemShut {NoStop}%
\bibitem [{\citenamefont {Chin}\ \emph {et~al.}(2009)\citenamefont {Chin},
  \citenamefont {Flambaum},\ and\ \citenamefont {Kozlov}}]{Chin2009}%
  \BibitemOpen
  \bibfield  {author} {\bibinfo {author} {\bibfnamefont {C.}~\bibnamefont
  {Chin}}, \bibinfo {author} {\bibfnamefont {V.~V.}\ \bibnamefont {Flambaum}},
  \ and\ \bibinfo {author} {\bibfnamefont {M.~G.}\ \bibnamefont {Kozlov}},\
  }\href {\doibase 10.1088/1367-2630/11/5/055048} {\bibfield  {journal}
  {\bibinfo  {journal} {New Journal of Physics}\ }\textbf {\bibinfo {volume}
  {11}},\ \bibinfo {pages} {55048} (\bibinfo {year} {2009})}\BibitemShut
  {NoStop}%
\bibitem [{\citenamefont {Kajita}(2009)}]{Kajita2009}%
  \BibitemOpen
  \bibfield  {author} {\bibinfo {author} {\bibfnamefont {M.}~\bibnamefont
  {Kajita}},\ }\href {\doibase 10.1088/1367-2630/11/5/055010} {\bibfield
  {journal} {\bibinfo  {journal} {New Journal of Physics}\ }\textbf {\bibinfo
  {volume} {11}},\ \bibinfo {pages} {055010} (\bibinfo {year}
  {2009})}\BibitemShut {NoStop}%
\bibitem [{\citenamefont {Beloy}\ \emph {et~al.}(2010)\citenamefont {Beloy},
  \citenamefont {Borschevsky}, \citenamefont {Schwerdtfeger},\ and\
  \citenamefont {Flambaum}}]{Beloy2010}%
  \BibitemOpen
  \bibfield  {author} {\bibinfo {author} {\bibfnamefont {K.}~\bibnamefont
  {Beloy}}, \bibinfo {author} {\bibfnamefont {A.}~\bibnamefont {Borschevsky}},
  \bibinfo {author} {\bibfnamefont {P.}~\bibnamefont {Schwerdtfeger}}, \ and\
  \bibinfo {author} {\bibfnamefont {V.~V.}\ \bibnamefont {Flambaum}},\ }\href
  {\doibase 10.1103/PhysRevA.82.022106} {\bibfield  {journal} {\bibinfo
  {journal} {Physical Review A}\ }\textbf {\bibinfo {volume} {82}},\ \bibinfo
  {pages} {022106} (\bibinfo {year} {2010})}\BibitemShut {NoStop}%
\bibitem [{\citenamefont {Jansen}\ \emph {et~al.}(2014)\citenamefont {Jansen},
  \citenamefont {Bethlem},\ and\ \citenamefont {Ubachs}}]{Jansen2014}%
  \BibitemOpen
  \bibfield  {author} {\bibinfo {author} {\bibfnamefont {P.}~\bibnamefont
  {Jansen}}, \bibinfo {author} {\bibfnamefont {H.~L.}\ \bibnamefont {Bethlem}},
  \ and\ \bibinfo {author} {\bibfnamefont {W.}~\bibnamefont {Ubachs}},\ }\href
  {\doibase 10.1063/1.4853735} {\bibfield  {journal} {\bibinfo  {journal} {The
  Journal of Chemical Physics}\ }\textbf {\bibinfo {volume} {140}},\ \bibinfo
  {pages} {010901} (\bibinfo {year} {2014})}\BibitemShut {NoStop}%
\bibitem [{\citenamefont {Daprà}\ \emph {et~al.}(2016)\citenamefont {Daprà},
  \citenamefont {Niu}, \citenamefont {Salumbides}, \citenamefont {Murphy},\
  and\ \citenamefont {Ubachs}}]{Dapra2016}%
  \BibitemOpen
  \bibfield  {author} {\bibinfo {author} {\bibfnamefont {M.}~\bibnamefont
  {Daprà}}, \bibinfo {author} {\bibfnamefont {M.~L.}\ \bibnamefont {Niu}},
  \bibinfo {author} {\bibfnamefont {E.~J.}\ \bibnamefont {Salumbides}},
  \bibinfo {author} {\bibfnamefont {M.~T.}\ \bibnamefont {Murphy}}, \ and\
  \bibinfo {author} {\bibfnamefont {W.}~\bibnamefont {Ubachs}},\ }\href
  {\doibase 10.3847/0004-637X/826/2/192} {\bibfield  {journal} {\bibinfo
  {journal} {The Astrophysical Journal}\ }\textbf {\bibinfo {volume} {826}},\
  \bibinfo {pages} {192} (\bibinfo {year} {2016})}\BibitemShut {NoStop}%
\bibitem [{\citenamefont {Kobayashi}\ \emph {et~al.}(2019)\citenamefont
  {Kobayashi}, \citenamefont {Ogino},\ and\ \citenamefont
  {Inouye}}]{Kobayashi2019}%
  \BibitemOpen
  \bibfield  {author} {\bibinfo {author} {\bibfnamefont {J.}~\bibnamefont
  {Kobayashi}}, \bibinfo {author} {\bibfnamefont {A.}~\bibnamefont {Ogino}}, \
  and\ \bibinfo {author} {\bibfnamefont {S.}~\bibnamefont {Inouye}},\ }\href
  {\doibase 10.1038/s41467-019-11761-1} {\bibfield  {journal} {\bibinfo
  {journal} {Nature Communications}\ }\textbf {\bibinfo {volume} {10}},\
  \bibinfo {pages} {1} (\bibinfo {year} {2019})}\BibitemShut {NoStop}%
\bibitem [{\citenamefont {Chupp}\ \emph {et~al.}(2019)\citenamefont {Chupp},
  \citenamefont {Fierlinger}, \citenamefont {Ramsey-Musolf},\ and\
  \citenamefont {Singh}}]{Chupp2019}%
  \BibitemOpen
  \bibfield  {author} {\bibinfo {author} {\bibfnamefont {T.~E.}\ \bibnamefont
  {Chupp}}, \bibinfo {author} {\bibfnamefont {P.}~\bibnamefont {Fierlinger}},
  \bibinfo {author} {\bibfnamefont {M.~J.}\ \bibnamefont {Ramsey-Musolf}}, \
  and\ \bibinfo {author} {\bibfnamefont {J.~T.}\ \bibnamefont {Singh}},\ }\href
  {\doibase 10.1103/RevModPhys.91.015001} {\bibfield  {journal} {\bibinfo
  {journal} {Reviews of Modern Physics}\ }\textbf {\bibinfo {volume} {91}},\
  \bibinfo {pages} {015001} (\bibinfo {year} {2019})}\BibitemShut {NoStop}%
\bibitem [{\citenamefont {Krems}(2008)}]{Krems2008}%
  \BibitemOpen
  \bibfield  {author} {\bibinfo {author} {\bibfnamefont {R.~V.}\ \bibnamefont
  {Krems}},\ }\href {\doibase 10.1039/b802322k} {\bibfield  {journal} {\bibinfo
   {journal} {Physical Chemistry Chemical Physics}\ }\textbf {\bibinfo {volume}
  {10}},\ \bibinfo {pages} {4079} (\bibinfo {year} {2008})}\BibitemShut
  {NoStop}%
\bibitem [{\citenamefont {Ni}\ \emph {et~al.}(2010)\citenamefont {Ni},
  \citenamefont {Ospelkaus}, \citenamefont {Wang}, \citenamefont {Quéméner},
  \citenamefont {Neyenhuis}, \citenamefont {Miranda}, \citenamefont {Bohn},
  \citenamefont {Ye},\ and\ \citenamefont {Jin}}]{Ni2010}%
  \BibitemOpen
  \bibfield  {author} {\bibinfo {author} {\bibfnamefont {K.-K.~K.}\
  \bibnamefont {Ni}}, \bibinfo {author} {\bibfnamefont {S.}~\bibnamefont
  {Ospelkaus}}, \bibinfo {author} {\bibfnamefont {D.}~\bibnamefont {Wang}},
  \bibinfo {author} {\bibfnamefont {G.}~\bibnamefont {Quéméner}}, \bibinfo
  {author} {\bibfnamefont {B.}~\bibnamefont {Neyenhuis}}, \bibinfo {author}
  {\bibfnamefont {M.~H. G.~D.}\ \bibnamefont {Miranda}}, \bibinfo {author}
  {\bibfnamefont {J.~L.}\ \bibnamefont {Bohn}}, \bibinfo {author}
  {\bibfnamefont {J.}~\bibnamefont {Ye}}, \ and\ \bibinfo {author}
  {\bibfnamefont {D.~S.}\ \bibnamefont {Jin}},\ }\href {\doibase
  10.1038/nature08953} {\bibfield  {journal} {\bibinfo  {journal} {Nature}\
  }\textbf {\bibinfo {volume} {464}},\ \bibinfo {pages} {1324} (\bibinfo {year}
  {2010})}\BibitemShut {NoStop}%
\bibitem [{\citenamefont {Ye}\ \emph {et~al.}(2018)\citenamefont {Ye},
  \citenamefont {Guo}, \citenamefont {González-Martínez}, \citenamefont
  {Quéméner},\ and\ \citenamefont {Wang}}]{Ye2018}%
  \BibitemOpen
  \bibfield  {author} {\bibinfo {author} {\bibfnamefont {X.}~\bibnamefont
  {Ye}}, \bibinfo {author} {\bibfnamefont {M.}~\bibnamefont {Guo}}, \bibinfo
  {author} {\bibfnamefont {M.~L.}\ \bibnamefont {González-Martínez}},
  \bibinfo {author} {\bibfnamefont {G.}~\bibnamefont {Quéméner}}, \ and\
  \bibinfo {author} {\bibfnamefont {D.}~\bibnamefont {Wang}},\ }\href {\doibase
  10.1126/sciadv.aaq0083} {\bibfield  {journal} {\bibinfo  {journal} {Science
  Advances}\ }\textbf {\bibinfo {volume} {4}},\ \bibinfo {pages} {eaaq0083}
  (\bibinfo {year} {2018})}\BibitemShut {NoStop}%
\bibitem [{\citenamefont {Ospelkaus}\ \emph {et~al.}(2010)\citenamefont
  {Ospelkaus}, \citenamefont {Ni}, \citenamefont {Wang}, \citenamefont
  {de~Miranda}, \citenamefont {Neyenhuis}, \citenamefont {Quéméner},
  \citenamefont {Julienne}, \citenamefont {Bohn}, \citenamefont {Jin},\ and\
  \citenamefont {Ye}}]{Ospelkaus2010}%
  \BibitemOpen
  \bibfield  {author} {\bibinfo {author} {\bibfnamefont {S.}~\bibnamefont
  {Ospelkaus}}, \bibinfo {author} {\bibfnamefont {K.-K.~K.}\ \bibnamefont
  {Ni}}, \bibinfo {author} {\bibfnamefont {D.}~\bibnamefont {Wang}}, \bibinfo
  {author} {\bibfnamefont {M.~H.~G.}\ \bibnamefont {de~Miranda}}, \bibinfo
  {author} {\bibfnamefont {B.}~\bibnamefont {Neyenhuis}}, \bibinfo {author}
  {\bibfnamefont {G.}~\bibnamefont {Quéméner}}, \bibinfo {author}
  {\bibfnamefont {P.~S.}\ \bibnamefont {Julienne}}, \bibinfo {author}
  {\bibfnamefont {J.~L.}\ \bibnamefont {Bohn}}, \bibinfo {author}
  {\bibfnamefont {D.~S.}\ \bibnamefont {Jin}}, \ and\ \bibinfo {author}
  {\bibfnamefont {J.}~\bibnamefont {Ye}},\ }\href {\doibase
  10.1126/science.1184121} {\bibfield  {journal} {\bibinfo  {journal}
  {Science}\ }\textbf {\bibinfo {volume} {327}},\ \bibinfo {pages} {853}
  (\bibinfo {year} {2010})}\BibitemShut {NoStop}%
\bibitem [{\citenamefont {DeMille}(2002)}]{DeMille2002}%
  \BibitemOpen
  \bibfield  {author} {\bibinfo {author} {\bibfnamefont {D.}~\bibnamefont
  {DeMille}},\ }\href {\doibase 10.1103/PhysRevLett.88.067901} {\bibfield
  {journal} {\bibinfo  {journal} {Physical Review Letters}\ }\textbf {\bibinfo
  {volume} {88}},\ \bibinfo {pages} {067901} (\bibinfo {year}
  {2002})}\BibitemShut {NoStop}%
\bibitem [{\citenamefont {Yelin}\ \emph {et~al.}(2006)\citenamefont {Yelin},
  \citenamefont {Kirby},\ and\ \citenamefont {Côté}}]{Yelin2006}%
  \BibitemOpen
  \bibfield  {author} {\bibinfo {author} {\bibfnamefont {S.~F.}\ \bibnamefont
  {Yelin}}, \bibinfo {author} {\bibfnamefont {K.}~\bibnamefont {Kirby}}, \ and\
  \bibinfo {author} {\bibfnamefont {R.}~\bibnamefont {Côté}},\ }\href
  {\doibase 10.1103/PhysRevA.74.050301} {\bibfield  {journal} {\bibinfo
  {journal} {Physical Review A}\ }\textbf {\bibinfo {volume} {74}},\ \bibinfo
  {pages} {050301(R)} (\bibinfo {year} {2006})}\BibitemShut {NoStop}%
\bibitem [{\citenamefont {Yu}\ \emph {et~al.}(2019)\citenamefont {Yu},
  \citenamefont {Cheuk}, \citenamefont {Kozyryev},\ and\ \citenamefont
  {Doyle}}]{Yu2019}%
  \BibitemOpen
  \bibfield  {author} {\bibinfo {author} {\bibfnamefont {P.}~\bibnamefont
  {Yu}}, \bibinfo {author} {\bibfnamefont {L.~W.}\ \bibnamefont {Cheuk}},
  \bibinfo {author} {\bibfnamefont {I.}~\bibnamefont {Kozyryev}}, \ and\
  \bibinfo {author} {\bibfnamefont {J.~M.}\ \bibnamefont {Doyle}},\ }\href
  {\doibase 10.1088/1367-2630/ab428d} {\bibfield  {journal} {\bibinfo
  {journal} {New Journal of Physics}\ }\textbf {\bibinfo {volume} {21}},\
  \bibinfo {pages} {093049} (\bibinfo {year} {2019})}\BibitemShut {NoStop}%
\bibitem [{\citenamefont {Carr}\ \emph {et~al.}(2009)\citenamefont {Carr},
  \citenamefont {DeMille}, \citenamefont {Krems},\ and\ \citenamefont
  {Ye}}]{Carr2009}%
  \BibitemOpen
  \bibfield  {author} {\bibinfo {author} {\bibfnamefont {L.~D.}\ \bibnamefont
  {Carr}}, \bibinfo {author} {\bibfnamefont {D.}~\bibnamefont {DeMille}},
  \bibinfo {author} {\bibfnamefont {R.~V.}\ \bibnamefont {Krems}}, \ and\
  \bibinfo {author} {\bibfnamefont {J.}~\bibnamefont {Ye}},\ }\href {\doibase
  10.1088/1367-2630/11/5/055049} {\bibfield  {journal} {\bibinfo  {journal}
  {New Journal of Physics}\ }\textbf {\bibinfo {volume} {11}},\ \bibinfo
  {pages} {055049} (\bibinfo {year} {2009})}\BibitemShut {NoStop}%
\bibitem [{\citenamefont {Micheli}\ \emph {et~al.}(2006)\citenamefont
  {Micheli}, \citenamefont {Brennen},\ and\ \citenamefont
  {Zoller}}]{Micheli2006}%
  \BibitemOpen
  \bibfield  {author} {\bibinfo {author} {\bibfnamefont {A.}~\bibnamefont
  {Micheli}}, \bibinfo {author} {\bibfnamefont {G.~K.}\ \bibnamefont
  {Brennen}}, \ and\ \bibinfo {author} {\bibfnamefont {P.}~\bibnamefont
  {Zoller}},\ }\href {\doibase 10.1038/nphys287} {\bibfield  {journal}
  {\bibinfo  {journal} {Nature Physics}\ }\textbf {\bibinfo {volume} {2}},\
  \bibinfo {pages} {341} (\bibinfo {year} {2006})}\BibitemShut {NoStop}%
\bibitem [{\citenamefont {Bao}\ \emph {et~al.}(2022)\citenamefont {Bao},
  \citenamefont {Yu}, \citenamefont {Anderegg}, \citenamefont {Chae},
  \citenamefont {Ketterle}, \citenamefont {Ni},\ and\ \citenamefont
  {Doyle}}]{Bao2022}%
  \BibitemOpen
  \bibfield  {author} {\bibinfo {author} {\bibfnamefont {Y.}~\bibnamefont
  {Bao}}, \bibinfo {author} {\bibfnamefont {S.~S.}\ \bibnamefont {Yu}},
  \bibinfo {author} {\bibfnamefont {L.}~\bibnamefont {Anderegg}}, \bibinfo
  {author} {\bibfnamefont {E.}~\bibnamefont {Chae}}, \bibinfo {author}
  {\bibfnamefont {W.}~\bibnamefont {Ketterle}}, \bibinfo {author}
  {\bibfnamefont {K.-K.}\ \bibnamefont {Ni}}, \ and\ \bibinfo {author}
  {\bibfnamefont {J.~M.}\ \bibnamefont {Doyle}},\ }\href {\doibase
  10.48550/arXiv.2211.09780} {\bibfield  {journal} {\bibinfo  {journal}
  {arXiv:2211.09780}\ } (\bibinfo {year} {2022}),\ 10.48550/arXiv.2211.09780},\
  \bibinfo {note} {arXiv:2211.09780 [physics, physics:quant-ph] type:
  article}\BibitemShut {NoStop}%
\bibitem [{\citenamefont {Holland}\ \emph {et~al.}(2022)\citenamefont
  {Holland}, \citenamefont {Lu},\ and\ \citenamefont {Cheuk}}]{Holland2022}%
  \BibitemOpen
  \bibfield  {author} {\bibinfo {author} {\bibfnamefont {C.~M.}\ \bibnamefont
  {Holland}}, \bibinfo {author} {\bibfnamefont {Y.}~\bibnamefont {Lu}}, \ and\
  \bibinfo {author} {\bibfnamefont {L.~W.}\ \bibnamefont {Cheuk}},\ }\href
  {\doibase 10.48550/arXiv.2210.06309} {\bibfield  {journal} {\bibinfo
  {journal} {arXiv.2210.06309}\ } (\bibinfo {year} {2022}),\
  10.48550/arXiv.2210.06309},\ \bibinfo {note} {arXiv:2210.06309 [cond-mat,
  physics:physics, physics:quant-ph] type: article}\BibitemShut {NoStop}%
\bibitem [{\citenamefont {Sage}\ \emph {et~al.}(2005)\citenamefont {Sage},
  \citenamefont {Sainis}, \citenamefont {Bergeman},\ and\ \citenamefont
  {DeMille}}]{Sage2005}%
  \BibitemOpen
  \bibfield  {author} {\bibinfo {author} {\bibfnamefont {J.~M.}\ \bibnamefont
  {Sage}}, \bibinfo {author} {\bibfnamefont {S.}~\bibnamefont {Sainis}},
  \bibinfo {author} {\bibfnamefont {T.}~\bibnamefont {Bergeman}}, \ and\
  \bibinfo {author} {\bibfnamefont {D.}~\bibnamefont {DeMille}},\ }\href
  {\doibase 10.1103/PhysRevLett.94.203001} {\bibfield  {journal} {\bibinfo
  {journal} {Physical Review Letters}\ }\textbf {\bibinfo {volume} {94}},\
  \bibinfo {pages} {203001} (\bibinfo {year} {2005})}\BibitemShut {NoStop}%
\bibitem [{\citenamefont {Ni}\ \emph {et~al.}(2008)\citenamefont {Ni},
  \citenamefont {Ospelkaus}, \citenamefont {de~Miranda}, \citenamefont {Pe'er},
  \citenamefont {Neyenhuis}, \citenamefont {Zirbel}, \citenamefont
  {Kotochigova}, \citenamefont {Julienne}, \citenamefont {Jin},\ and\
  \citenamefont {Ye}}]{Ni2008}%
  \BibitemOpen
  \bibfield  {author} {\bibinfo {author} {\bibfnamefont {K.-K.}\ \bibnamefont
  {Ni}}, \bibinfo {author} {\bibfnamefont {S.}~\bibnamefont {Ospelkaus}},
  \bibinfo {author} {\bibfnamefont {M.~H.~G.}\ \bibnamefont {de~Miranda}},
  \bibinfo {author} {\bibfnamefont {A.}~\bibnamefont {Pe'er}}, \bibinfo
  {author} {\bibfnamefont {B.}~\bibnamefont {Neyenhuis}}, \bibinfo {author}
  {\bibfnamefont {J.~J.}\ \bibnamefont {Zirbel}}, \bibinfo {author}
  {\bibfnamefont {S.}~\bibnamefont {Kotochigova}}, \bibinfo {author}
  {\bibfnamefont {P.~S.}\ \bibnamefont {Julienne}}, \bibinfo {author}
  {\bibfnamefont {D.~S.}\ \bibnamefont {Jin}}, \ and\ \bibinfo {author}
  {\bibfnamefont {J.}~\bibnamefont {Ye}},\ }\href {\doibase
  10.1126/science.1163861} {\bibfield  {journal} {\bibinfo  {journal}
  {Science}\ }\textbf {\bibinfo {volume} {322}},\ \bibinfo {pages} {231}
  (\bibinfo {year} {2008})}\BibitemShut {NoStop}%
\bibitem [{\citenamefont {Danzl}\ \emph {et~al.}(2010)\citenamefont {Danzl},
  \citenamefont {Mark}, \citenamefont {Haller}, \citenamefont {Gustavsson},
  \citenamefont {Hart}, \citenamefont {Aldegunde}, \citenamefont {Hutson},\
  and\ \citenamefont {Nägerl}}]{Danzl2010}%
  \BibitemOpen
  \bibfield  {author} {\bibinfo {author} {\bibfnamefont {J.~G.}\ \bibnamefont
  {Danzl}}, \bibinfo {author} {\bibfnamefont {M.~J.}\ \bibnamefont {Mark}},
  \bibinfo {author} {\bibfnamefont {E.}~\bibnamefont {Haller}}, \bibinfo
  {author} {\bibfnamefont {M.}~\bibnamefont {Gustavsson}}, \bibinfo {author}
  {\bibfnamefont {R.}~\bibnamefont {Hart}}, \bibinfo {author} {\bibfnamefont
  {J.}~\bibnamefont {Aldegunde}}, \bibinfo {author} {\bibfnamefont {J.~M.}\
  \bibnamefont {Hutson}}, \ and\ \bibinfo {author} {\bibfnamefont {H.~C.}\
  \bibnamefont {Nägerl}},\ }\href {\doibase 10.1038/nphys1533} {\bibfield
  {journal} {\bibinfo  {journal} {Nature Physics}\ }\textbf {\bibinfo {volume}
  {6}},\ \bibinfo {pages} {265} (\bibinfo {year} {2010})}\BibitemShut {NoStop}%
\bibitem [{\citenamefont {Aikawa}\ \emph {et~al.}(2010)\citenamefont {Aikawa},
  \citenamefont {Akamatsu}, \citenamefont {Hayashi}, \citenamefont {Oasa},
  \citenamefont {Kobayashi}, \citenamefont {Naidon}, \citenamefont {Kishimoto},
  \citenamefont {Ueda},\ and\ \citenamefont {Inouye}}]{Aikawa2010}%
  \BibitemOpen
  \bibfield  {author} {\bibinfo {author} {\bibfnamefont {K.}~\bibnamefont
  {Aikawa}}, \bibinfo {author} {\bibfnamefont {D.}~\bibnamefont {Akamatsu}},
  \bibinfo {author} {\bibfnamefont {M.}~\bibnamefont {Hayashi}}, \bibinfo
  {author} {\bibfnamefont {K.}~\bibnamefont {Oasa}}, \bibinfo {author}
  {\bibfnamefont {J.}~\bibnamefont {Kobayashi}}, \bibinfo {author}
  {\bibfnamefont {P.}~\bibnamefont {Naidon}}, \bibinfo {author} {\bibfnamefont
  {T.}~\bibnamefont {Kishimoto}}, \bibinfo {author} {\bibfnamefont
  {M.}~\bibnamefont {Ueda}}, \ and\ \bibinfo {author} {\bibfnamefont
  {S.}~\bibnamefont {Inouye}},\ }\href {\doibase
  10.1103/PhysRevLett.105.203001} {\bibfield  {journal} {\bibinfo  {journal}
  {Physical Review Letters}\ }\textbf {\bibinfo {volume} {105}},\ \bibinfo
  {pages} {203001} (\bibinfo {year} {2010})}\BibitemShut {NoStop}%
\bibitem [{\citenamefont {Takekoshi}\ \emph {et~al.}(2014)\citenamefont
  {Takekoshi}, \citenamefont {Reichsöllner}, \citenamefont {Schindewolf},
  \citenamefont {Hutson}, \citenamefont {Sueur}, \citenamefont {Dulieu},
  \citenamefont {Ferlaino}, \citenamefont {Grimm},\ and\ \citenamefont
  {Nägerl}}]{Takekoshi2014}%
  \BibitemOpen
  \bibfield  {author} {\bibinfo {author} {\bibfnamefont {T.}~\bibnamefont
  {Takekoshi}}, \bibinfo {author} {\bibfnamefont {L.}~\bibnamefont
  {Reichsöllner}}, \bibinfo {author} {\bibfnamefont {A.}~\bibnamefont
  {Schindewolf}}, \bibinfo {author} {\bibfnamefont {J.~M.}\ \bibnamefont
  {Hutson}}, \bibinfo {author} {\bibfnamefont {C.~R.~L.}\ \bibnamefont
  {Sueur}}, \bibinfo {author} {\bibfnamefont {O.}~\bibnamefont {Dulieu}},
  \bibinfo {author} {\bibfnamefont {F.}~\bibnamefont {Ferlaino}}, \bibinfo
  {author} {\bibfnamefont {R.}~\bibnamefont {Grimm}}, \ and\ \bibinfo {author}
  {\bibfnamefont {H.~C.}\ \bibnamefont {Nägerl}},\ }\href {\doibase
  10.1103/PhysRevLett.113.205301} {\bibfield  {journal} {\bibinfo  {journal}
  {Physical Review Letters}\ }\textbf {\bibinfo {volume} {113}},\ \bibinfo
  {pages} {205301} (\bibinfo {year} {2014})}\BibitemShut {NoStop}%
\bibitem [{\citenamefont {Molony}\ \emph {et~al.}(2014)\citenamefont {Molony},
  \citenamefont {Gregory}, \citenamefont {Ji}, \citenamefont {Lu},
  \citenamefont {Köppinger}, \citenamefont {Sueur}, \citenamefont {Blackley},
  \citenamefont {Hutson},\ and\ \citenamefont {Cornish}}]{Molony2014}%
  \BibitemOpen
  \bibfield  {author} {\bibinfo {author} {\bibfnamefont {P.~K.}\ \bibnamefont
  {Molony}}, \bibinfo {author} {\bibfnamefont {P.~D.}\ \bibnamefont {Gregory}},
  \bibinfo {author} {\bibfnamefont {Z.}~\bibnamefont {Ji}}, \bibinfo {author}
  {\bibfnamefont {B.}~\bibnamefont {Lu}}, \bibinfo {author} {\bibfnamefont
  {M.~P.}\ \bibnamefont {Köppinger}}, \bibinfo {author} {\bibfnamefont
  {C.~R.~L.}\ \bibnamefont {Sueur}}, \bibinfo {author} {\bibfnamefont {C.~L.}\
  \bibnamefont {Blackley}}, \bibinfo {author} {\bibfnamefont {J.~M.}\
  \bibnamefont {Hutson}}, \ and\ \bibinfo {author} {\bibfnamefont {S.~L.}\
  \bibnamefont {Cornish}},\ }\href {\doibase 10.1103/PhysRevLett.113.255301}
  {\bibfield  {journal} {\bibinfo  {journal} {Physical Review Letters}\
  }\textbf {\bibinfo {volume} {113}},\ \bibinfo {pages} {255301} (\bibinfo
  {year} {2014})}\BibitemShut {NoStop}%
\bibitem [{\citenamefont {Park}\ \emph {et~al.}(2015)\citenamefont {Park},
  \citenamefont {Will},\ and\ \citenamefont {Zwierlein}}]{Park2015}%
  \BibitemOpen
  \bibfield  {author} {\bibinfo {author} {\bibfnamefont {J.~W.}\ \bibnamefont
  {Park}}, \bibinfo {author} {\bibfnamefont {S.~A.}\ \bibnamefont {Will}}, \
  and\ \bibinfo {author} {\bibfnamefont {M.~W.}\ \bibnamefont {Zwierlein}},\
  }\href {\doibase 10.1103/PhysRevLett.114.205302} {\bibfield  {journal}
  {\bibinfo  {journal} {Physical Review Letters}\ }\textbf {\bibinfo {volume}
  {114}},\ \bibinfo {pages} {205302} (\bibinfo {year} {2015})}\BibitemShut
  {NoStop}%
\bibitem [{\citenamefont {Guo}\ \emph {et~al.}(2016)\citenamefont {Guo},
  \citenamefont {Zhu}, \citenamefont {Lu}, \citenamefont {Ye}, \citenamefont
  {Wang}, \citenamefont {Vexiau}, \citenamefont {Bouloufa-Maafa}, \citenamefont
  {Quéméner}, \citenamefont {Dulieu},\ and\ \citenamefont {Wang}}]{Guo2016}%
  \BibitemOpen
  \bibfield  {author} {\bibinfo {author} {\bibfnamefont {M.}~\bibnamefont
  {Guo}}, \bibinfo {author} {\bibfnamefont {B.}~\bibnamefont {Zhu}}, \bibinfo
  {author} {\bibfnamefont {B.}~\bibnamefont {Lu}}, \bibinfo {author}
  {\bibfnamefont {X.}~\bibnamefont {Ye}}, \bibinfo {author} {\bibfnamefont
  {F.}~\bibnamefont {Wang}}, \bibinfo {author} {\bibfnamefont {R.}~\bibnamefont
  {Vexiau}}, \bibinfo {author} {\bibfnamefont {N.}~\bibnamefont
  {Bouloufa-Maafa}}, \bibinfo {author} {\bibfnamefont {G.}~\bibnamefont
  {Quéméner}}, \bibinfo {author} {\bibfnamefont {O.}~\bibnamefont {Dulieu}},
  \ and\ \bibinfo {author} {\bibfnamefont {D.}~\bibnamefont {Wang}},\ }\href
  {\doibase 10.1103/PhysRevLett.116.205303} {\bibfield  {journal} {\bibinfo
  {journal} {Physical Review Letters}\ }\textbf {\bibinfo {volume} {116}},\
  \bibinfo {pages} {205303} (\bibinfo {year} {2016})}\BibitemShut {NoStop}%
\bibitem [{\citenamefont {Liu}\ \emph {et~al.}(2018)\citenamefont {Liu},
  \citenamefont {Hood}, \citenamefont {Yu}, \citenamefont {Zhang},
  \citenamefont {Hutzler}, \citenamefont {Rosenband},\ and\ \citenamefont
  {Ni}}]{Liu2018}%
  \BibitemOpen
  \bibfield  {author} {\bibinfo {author} {\bibfnamefont {L.~R.}\ \bibnamefont
  {Liu}}, \bibinfo {author} {\bibfnamefont {J.~D.}\ \bibnamefont {Hood}},
  \bibinfo {author} {\bibfnamefont {Y.}~\bibnamefont {Yu}}, \bibinfo {author}
  {\bibfnamefont {J.~T.}\ \bibnamefont {Zhang}}, \bibinfo {author}
  {\bibfnamefont {N.~R.}\ \bibnamefont {Hutzler}}, \bibinfo {author}
  {\bibfnamefont {T.}~\bibnamefont {Rosenband}}, \ and\ \bibinfo {author}
  {\bibfnamefont {K.-K.~K.}\ \bibnamefont {Ni}},\ }\href {\doibase
  10.1126/science.aar7797} {\bibfield  {journal} {\bibinfo  {journal}
  {Science}\ }\textbf {\bibinfo {volume} {360}},\ \bibinfo {pages} {900}
  (\bibinfo {year} {2018})}\BibitemShut {NoStop}%
\bibitem [{\citenamefont {Marco}\ \emph {et~al.}(2019)\citenamefont {Marco},
  \citenamefont {Valtolina}, \citenamefont {Matsuda}, \citenamefont {Tobias},
  \citenamefont {Covey},\ and\ \citenamefont {Ye}}]{DeMarco2019}%
  \BibitemOpen
  \bibfield  {author} {\bibinfo {author} {\bibfnamefont {L.~D.}\ \bibnamefont
  {Marco}}, \bibinfo {author} {\bibfnamefont {G.}~\bibnamefont {Valtolina}},
  \bibinfo {author} {\bibfnamefont {K.}~\bibnamefont {Matsuda}}, \bibinfo
  {author} {\bibfnamefont {W.~G.}\ \bibnamefont {Tobias}}, \bibinfo {author}
  {\bibfnamefont {J.~P.}\ \bibnamefont {Covey}}, \ and\ \bibinfo {author}
  {\bibfnamefont {J.}~\bibnamefont {Ye}},\ }\href {\doibase
  10.1126/science.aau7230} {\bibfield  {journal} {\bibinfo  {journal}
  {Science}\ }\textbf {\bibinfo {volume} {363}},\ \bibinfo {pages} {853}
  (\bibinfo {year} {2019})}\BibitemShut {NoStop}%
\bibitem [{\citenamefont {Rosa}(2004)}]{Rosa2004}%
  \BibitemOpen
  \bibfield  {author} {\bibinfo {author} {\bibfnamefont {M.~D.~D.}\
  \bibnamefont {Rosa}},\ }\href {\doibase 10.1140/epjd/e2004-00167-2}
  {\bibfield  {journal} {\bibinfo  {journal} {The European Physical Journal D}\
  }\textbf {\bibinfo {volume} {31}},\ \bibinfo {pages} {395} (\bibinfo {year}
  {2004})}\BibitemShut {NoStop}%
\bibitem [{\citenamefont {McCarron}(2018)}]{McCarron2018a}%
  \BibitemOpen
  \bibfield  {author} {\bibinfo {author} {\bibfnamefont {D.}~\bibnamefont
  {McCarron}},\ }\href {\doibase 10.1088/1361-6455/aadfba} {\bibfield
  {journal} {\bibinfo  {journal} {Journal of Physics B: Atomic, Molecular and
  Optical Physics}\ }\textbf {\bibinfo {volume} {51}},\ \bibinfo {pages}
  {212001} (\bibinfo {year} {2018})}\BibitemShut {NoStop}%
\bibitem [{\citenamefont {Tarbutt}(2018)}]{Tarbutt2019}%
  \BibitemOpen
  \bibfield  {author} {\bibinfo {author} {\bibfnamefont {M.~R.}\ \bibnamefont
  {Tarbutt}},\ }\href {\doibase 10.1080/00107514.2018.1576338} {\bibfield
  {journal} {\bibinfo  {journal} {Contemporary Physics}\ }\textbf {\bibinfo
  {volume} {59}},\ \bibinfo {pages} {356} (\bibinfo {year} {2018})}\BibitemShut
  {NoStop}%
\bibitem [{\citenamefont {Fitch}\ and\ \citenamefont
  {Tarbutt}(2021)}]{Fitch2021}%
  \BibitemOpen
  \bibfield  {author} {\bibinfo {author} {\bibfnamefont {N.~J.}\ \bibnamefont
  {Fitch}}\ and\ \bibinfo {author} {\bibfnamefont {M.~R.}\ \bibnamefont
  {Tarbutt}},\ }\href {\doibase 10.1016/bs.aamop.2021.04.003} {\bibfield
  {journal} {\bibinfo  {journal} {Advances In Atomic, Molecular, and Optical
  Physics}\ }\textbf {\bibinfo {volume} {70}},\ \bibinfo {pages} {157}
  (\bibinfo {year} {2021})}\BibitemShut {NoStop}%
\bibitem [{\citenamefont {Langen}\ \emph {et~al.}(2023)\citenamefont {Langen},
  \citenamefont {Valtolina}, \citenamefont {Wang},\ and\ \citenamefont
  {Ye}}]{Langen2023}%
  \BibitemOpen
  \bibfield  {author} {\bibinfo {author} {\bibfnamefont {T.}~\bibnamefont
  {Langen}}, \bibinfo {author} {\bibfnamefont {G.}~\bibnamefont {Valtolina}},
  \bibinfo {author} {\bibfnamefont {D.}~\bibnamefont {Wang}}, \ and\ \bibinfo
  {author} {\bibfnamefont {J.}~\bibnamefont {Ye}},\ }\href {\doibase
  10.48550/arXiv.2305.13445} {\bibfield  {journal} {\bibinfo  {journal}
  {ArXiv}\ } (\bibinfo {year} {2023}),\ 10.48550/arXiv.2305.13445}\BibitemShut
  {NoStop}%
\bibitem [{\citenamefont {Langin}\ and\ \citenamefont
  {DeMille}(2023)}]{Langin2023}%
  \BibitemOpen
  \bibfield  {author} {\bibinfo {author} {\bibfnamefont {T.~K.}\ \bibnamefont
  {Langin}}\ and\ \bibinfo {author} {\bibfnamefont {D.}~\bibnamefont
  {DeMille}},\ }\href {\doibase 10.1088/1367-2630/acc34d} {\bibfield  {journal}
  {\bibinfo  {journal} {New Journal of Physics}\ }\textbf {\bibinfo {volume}
  {25}},\ \bibinfo {pages} {043005} (\bibinfo {year} {2023})}\BibitemShut
  {NoStop}%
\bibitem [{\citenamefont {Bahns}\ \emph {et~al.}(1996)\citenamefont {Bahns},
  \citenamefont {Stwalley},\ and\ \citenamefont {Gould}}]{Bahns1996}%
  \BibitemOpen
  \bibfield  {author} {\bibinfo {author} {\bibfnamefont {J.~T.}\ \bibnamefont
  {Bahns}}, \bibinfo {author} {\bibfnamefont {W.~C.}\ \bibnamefont {Stwalley}},
  \ and\ \bibinfo {author} {\bibfnamefont {P.~L.}\ \bibnamefont {Gould}},\
  }\href {\doibase 10.1063/1.471731} {\bibfield  {journal} {\bibinfo  {journal}
  {Journal of Chemical Physics}\ }\textbf {\bibinfo {volume} {104}},\ \bibinfo
  {pages} {9689} (\bibinfo {year} {1996})}\BibitemShut {NoStop}%
\bibitem [{\citenamefont {Xiao}\ \emph {et~al.}(2021)\citenamefont {Xiao},
  \citenamefont {Yang}, \citenamefont {Zhu},\ and\ \citenamefont
  {Gao}}]{Xiao2021}%
  \BibitemOpen
  \bibfield  {author} {\bibinfo {author} {\bibfnamefont {H.}~\bibnamefont
  {Xiao}}, \bibinfo {author} {\bibfnamefont {Q.~S.}\ \bibnamefont {Yang}},
  \bibinfo {author} {\bibfnamefont {J.}~\bibnamefont {Zhu}}, \ and\ \bibinfo
  {author} {\bibfnamefont {T.}~\bibnamefont {Gao}},\ }\href {\doibase
  10.1016/J.SAA.2020.118998} {\bibfield  {journal} {\bibinfo  {journal}
  {Spectrochimica Acta Part A: Molecular and Biomolecular Spectroscopy}\
  }\textbf {\bibinfo {volume} {246}},\ \bibinfo {pages} {118998} (\bibinfo
  {year} {2021})}\BibitemShut {NoStop}%
\bibitem [{\citenamefont {Gao}\ and\ \citenamefont {Gao}(2015)}]{Gao2015}%
  \BibitemOpen
  \bibfield  {author} {\bibinfo {author} {\bibfnamefont {Y.~F.}\ \bibnamefont
  {Gao}}\ and\ \bibinfo {author} {\bibfnamefont {T.}~\bibnamefont {Gao}},\
  }\href {\doibase 10.1039/C5CP00025D} {\bibfield  {journal} {\bibinfo
  {journal} {Physical Chemistry Chemical Physics}\ }\textbf {\bibinfo {volume}
  {17}},\ \bibinfo {pages} {10830} (\bibinfo {year} {2015})}\BibitemShut
  {NoStop}%
\bibitem [{\citenamefont {Gao}\ and\ \citenamefont {Gao}(2014)}]{Gao2014}%
  \BibitemOpen
  \bibfield  {author} {\bibinfo {author} {\bibfnamefont {Y.}~\bibnamefont
  {Gao}}\ and\ \bibinfo {author} {\bibfnamefont {T.}~\bibnamefont {Gao}},\
  }\href {\doibase 10.1016/J.SAA.2013.07.009} {\bibfield  {journal} {\bibinfo
  {journal} {Spectrochimica Acta Part A: Molecular and Biomolecular
  Spectroscopy}\ }\textbf {\bibinfo {volume} {118}},\ \bibinfo {pages} {308}
  (\bibinfo {year} {2014})}\BibitemShut {NoStop}%
\bibitem [{\citenamefont {Xiao}\ \emph
  {et~al.}(2022{\natexlab{a}})\citenamefont {Xiao}, \citenamefont {Dong},\ and\
  \citenamefont {Gao}}]{Xiao2022}%
  \BibitemOpen
  \bibfield  {author} {\bibinfo {author} {\bibfnamefont {H.}~\bibnamefont
  {Xiao}}, \bibinfo {author} {\bibfnamefont {S.}~\bibnamefont {Dong}}, \ and\
  \bibinfo {author} {\bibfnamefont {T.}~\bibnamefont {Gao}},\ }\href {\doibase
  10.1016/J.CPLETT.2022.139407} {\bibfield  {journal} {\bibinfo  {journal}
  {Chemical Physics Letters}\ }\textbf {\bibinfo {volume} {793}},\ \bibinfo
  {pages} {139407} (\bibinfo {year} {2022}{\natexlab{a}})}\BibitemShut
  {NoStop}%
\bibitem [{\citenamefont {Xiao}\ \emph
  {et~al.}(2022{\natexlab{b}})\citenamefont {Xiao}, \citenamefont {Zhang},
  \citenamefont {Ma},\ and\ \citenamefont {Gao}}]{Xiao2022b}%
  \BibitemOpen
  \bibfield  {author} {\bibinfo {author} {\bibfnamefont {H.}~\bibnamefont
  {Xiao}}, \bibinfo {author} {\bibfnamefont {R.}~\bibnamefont {Zhang}},
  \bibinfo {author} {\bibfnamefont {H.}~\bibnamefont {Ma}}, \ and\ \bibinfo
  {author} {\bibfnamefont {T.}~\bibnamefont {Gao}},\ }\href {\doibase
  10.1016/J.SAA.2022.121279} {\bibfield  {journal} {\bibinfo  {journal}
  {Spectrochimica Acta Part A: Molecular and Biomolecular Spectroscopy}\
  }\textbf {\bibinfo {volume} {277}},\ \bibinfo {pages} {121279} (\bibinfo
  {year} {2022}{\natexlab{b}})}\BibitemShut {NoStop}%
\bibitem [{\citenamefont {Xiao}\ \emph
  {et~al.}(2022{\natexlab{c}})\citenamefont {Xiao}, \citenamefont {Wang},
  \citenamefont {Zhang}, \citenamefont {Shan},\ and\ \citenamefont
  {Gao}}]{Xiao2022c}%
  \BibitemOpen
  \bibfield  {author} {\bibinfo {author} {\bibfnamefont {H.}~\bibnamefont
  {Xiao}}, \bibinfo {author} {\bibfnamefont {J.}~\bibnamefont {Wang}}, \bibinfo
  {author} {\bibfnamefont {R.}~\bibnamefont {Zhang}}, \bibinfo {author}
  {\bibfnamefont {N.}~\bibnamefont {Shan}}, \ and\ \bibinfo {author}
  {\bibfnamefont {T.}~\bibnamefont {Gao}},\ }\href {\doibase
  10.1016/J.SAA.2022.121679} {\bibfield  {journal} {\bibinfo  {journal}
  {Spectrochimica Acta Part A: Molecular and Biomolecular Spectroscopy}\
  }\textbf {\bibinfo {volume} {282}},\ \bibinfo {pages} {121679} (\bibinfo
  {year} {2022}{\natexlab{c}})}\BibitemShut {NoStop}%
\bibitem [{\citenamefont {Yang}\ \emph
  {et~al.}(2016{\natexlab{a}})\citenamefont {Yang}, \citenamefont {Tang},\ and\
  \citenamefont {Gao}}]{Yang2016}%
  \BibitemOpen
  \bibfield  {author} {\bibinfo {author} {\bibfnamefont {R.}~\bibnamefont
  {Yang}}, \bibinfo {author} {\bibfnamefont {B.}~\bibnamefont {Tang}}, \ and\
  \bibinfo {author} {\bibfnamefont {T.}~\bibnamefont {Gao}},\ }\href {\doibase
  10.1088/1674-1056/25/4/043101} {\bibfield  {journal} {\bibinfo  {journal}
  {Chinese Physics B}\ }\textbf {\bibinfo {volume} {25}},\ \bibinfo {pages}
  {043101} (\bibinfo {year} {2016}{\natexlab{a}})}\BibitemShut {NoStop}%
\bibitem [{\citenamefont {Kang}\ \emph {et~al.}(2016)\citenamefont {Kang},
  \citenamefont {Kuang}, \citenamefont {Jiang},\ and\ \citenamefont
  {Du}}]{Kang2016}%
  \BibitemOpen
  \bibfield  {author} {\bibinfo {author} {\bibfnamefont {S.}~\bibnamefont
  {Kang}}, \bibinfo {author} {\bibfnamefont {F.}~\bibnamefont {Kuang}},
  \bibinfo {author} {\bibfnamefont {G.}~\bibnamefont {Jiang}}, \ and\ \bibinfo
  {author} {\bibfnamefont {J.}~\bibnamefont {Du}},\ }\href {\doibase
  10.1080/00268976.2015.1121294} {\bibfield  {journal} {\bibinfo  {journal}
  {Molecular Physics}\ }\textbf {\bibinfo {volume} {114}},\ \bibinfo {pages}
  {810} (\bibinfo {year} {2016})}\BibitemShut {NoStop}%
\bibitem [{\citenamefont {Kang}\ \emph
  {et~al.}(2015{\natexlab{a}})\citenamefont {Kang}, \citenamefont {Gao},
  \citenamefont {Kuang}, \citenamefont {Gao}, \citenamefont {Du},\ and\
  \citenamefont {Jiang}}]{Kang2015}%
  \BibitemOpen
  \bibfield  {author} {\bibinfo {author} {\bibfnamefont {S.}~\bibnamefont
  {Kang}}, \bibinfo {author} {\bibfnamefont {Y.}~\bibnamefont {Gao}}, \bibinfo
  {author} {\bibfnamefont {F.}~\bibnamefont {Kuang}}, \bibinfo {author}
  {\bibfnamefont {T.}~\bibnamefont {Gao}}, \bibinfo {author} {\bibfnamefont
  {J.}~\bibnamefont {Du}}, \ and\ \bibinfo {author} {\bibfnamefont
  {G.}~\bibnamefont {Jiang}},\ }\href {\doibase 10.1103/PHYSREVA.92.069902}
  {\bibfield  {journal} {\bibinfo  {journal} {Physical Review A}\ }\textbf
  {\bibinfo {volume} {92}},\ \bibinfo {pages} {069902} (\bibinfo {year}
  {2015}{\natexlab{a}})}\BibitemShut {NoStop}%
\bibitem [{\citenamefont {Kang}\ \emph
  {et~al.}(2015{\natexlab{b}})\citenamefont {Kang}, \citenamefont {Gao},
  \citenamefont {Kuang}, \citenamefont {Gao}, \citenamefont {Du},\ and\
  \citenamefont {Jiang}}]{Kang2015b}%
  \BibitemOpen
  \bibfield  {author} {\bibinfo {author} {\bibfnamefont {S.}~\bibnamefont
  {Kang}}, \bibinfo {author} {\bibfnamefont {Y.}~\bibnamefont {Gao}}, \bibinfo
  {author} {\bibfnamefont {F.}~\bibnamefont {Kuang}}, \bibinfo {author}
  {\bibfnamefont {T.}~\bibnamefont {Gao}}, \bibinfo {author} {\bibfnamefont
  {J.}~\bibnamefont {Du}}, \ and\ \bibinfo {author} {\bibfnamefont
  {G.}~\bibnamefont {Jiang}},\ }\href {\doibase
  10.1103/PHYSREVA.91.042511/FIGURES/5/MEDIUM} {\bibfield  {journal} {\bibinfo
  {journal} {Physical Review A - Atomic, Molecular, and Optical Physics}\
  }\textbf {\bibinfo {volume} {91}},\ \bibinfo {pages} {042511} (\bibinfo
  {year} {2015}{\natexlab{b}})}\BibitemShut {NoStop}%
\bibitem [{\citenamefont {Rodriguez}\ \emph {et~al.}(2023)\citenamefont
  {Rodriguez}, \citenamefont {Pilgram}, \citenamefont {Barker}, \citenamefont
  {Eckel},\ and\ \citenamefont {Norrgard}}]{Rodriguez2023}%
  \BibitemOpen
  \bibfield  {author} {\bibinfo {author} {\bibfnamefont {K.~J.}\ \bibnamefont
  {Rodriguez}}, \bibinfo {author} {\bibfnamefont {N.~H.}\ \bibnamefont
  {Pilgram}}, \bibinfo {author} {\bibfnamefont {D.~S.}\ \bibnamefont {Barker}},
  \bibinfo {author} {\bibfnamefont {S.~P.}\ \bibnamefont {Eckel}}, \ and\
  \bibinfo {author} {\bibfnamefont {E.~B.}\ \bibnamefont {Norrgard}},\ }\href
  {\doibase 10.48550/arXiv.2305.04879} {\bibfield  {journal} {\bibinfo
  {journal} {ArXiv}\ } (\bibinfo {year} {2023}),\ 10.48550/arXiv.2305.04879},\
  \bibinfo {note} {arXiv:2305.04879 [physics] type: article}\BibitemShut
  {NoStop}%
\bibitem [{\citenamefont {Lambo}\ \emph {et~al.}(2023)\citenamefont {Lambo},
  \citenamefont {Koyanagi}, \citenamefont {Ragyanszki}, \citenamefont
  {Horbatsch}, \citenamefont {Fournier},\ and\ \citenamefont
  {Hessels}}]{Lambo2023}%
  \BibitemOpen
  \bibfield  {author} {\bibinfo {author} {\bibfnamefont {R.~L.}\ \bibnamefont
  {Lambo}}, \bibinfo {author} {\bibfnamefont {G.~K.}\ \bibnamefont {Koyanagi}},
  \bibinfo {author} {\bibfnamefont {A.}~\bibnamefont {Ragyanszki}}, \bibinfo
  {author} {\bibfnamefont {M.}~\bibnamefont {Horbatsch}}, \bibinfo {author}
  {\bibfnamefont {R.}~\bibnamefont {Fournier}}, \ and\ \bibinfo {author}
  {\bibfnamefont {E.~A.}\ \bibnamefont {Hessels}},\ }\href {\doibase
  10.1080/00268976.2023.2198044} {\bibfield  {journal} {\bibinfo  {journal}
  {Molecular Physics}\ }\textbf {\bibinfo {volume} {121}},\ \bibinfo {pages}
  {e2198044} (\bibinfo {year} {2023})}\BibitemShut {NoStop}%
\bibitem [{\citenamefont {Marsman}\ \emph {et~al.}(2023)\citenamefont
  {Marsman}, \citenamefont {Horbatsch},\ and\ \citenamefont
  {Hessels}}]{Marsman2023}%
  \BibitemOpen
  \bibfield  {author} {\bibinfo {author} {\bibfnamefont {A.}~\bibnamefont
  {Marsman}}, \bibinfo {author} {\bibfnamefont {M.}~\bibnamefont {Horbatsch}},
  \ and\ \bibinfo {author} {\bibfnamefont {E.~A.}\ \bibnamefont {Hessels}},\
  }\href {\doibase 10.48550/arXiv.2305.10641} {\bibfield  {journal} {\bibinfo
  {journal} {ArXiv}\ } (\bibinfo {year} {2023}),\
  10.48550/arXiv.2305.10641}\BibitemShut {NoStop}%
\bibitem [{\citenamefont {El-Kork}\ \emph {et~al.}(2023)\citenamefont
  {El-Kork}, \citenamefont {AlMasri~Alwan}, \citenamefont {Abu El~Kher},
  \citenamefont {Assaf}, \citenamefont {Ayari}, \citenamefont {Alhseinat},\
  and\ \citenamefont {Korek}}]{ElKork2023}%
  \BibitemOpen
  \bibfield  {author} {\bibinfo {author} {\bibfnamefont {N.}~\bibnamefont
  {El-Kork}}, \bibinfo {author} {\bibfnamefont {A.}~\bibnamefont
  {AlMasri~Alwan}}, \bibinfo {author} {\bibfnamefont {N.}~\bibnamefont {Abu
  El~Kher}}, \bibinfo {author} {\bibfnamefont {J.}~\bibnamefont {Assaf}},
  \bibinfo {author} {\bibfnamefont {T.}~\bibnamefont {Ayari}}, \bibinfo
  {author} {\bibfnamefont {E.}~\bibnamefont {Alhseinat}}, \ and\ \bibinfo
  {author} {\bibfnamefont {M.}~\bibnamefont {Korek}},\ }\href {\doibase
  10.1038/s41598-023-32439-1} {\bibfield  {journal} {\bibinfo  {journal}
  {Scientific Reports}\ }\textbf {\bibinfo {volume} {13}},\ \bibinfo {pages}
  {7087} (\bibinfo {year} {2023})}\BibitemShut {NoStop}%
\bibitem [{\citenamefont {Yang}\ \emph {et~al.}(2014)\citenamefont {Yang},
  \citenamefont {Gao}, \citenamefont {Tang},\ and\ \citenamefont
  {Gao}}]{Yang2014}%
  \BibitemOpen
  \bibfield  {author} {\bibinfo {author} {\bibfnamefont {R.}~\bibnamefont
  {Yang}}, \bibinfo {author} {\bibfnamefont {Y.}~\bibnamefont {Gao}}, \bibinfo
  {author} {\bibfnamefont {B.}~\bibnamefont {Tang}}, \ and\ \bibinfo {author}
  {\bibfnamefont {T.}~\bibnamefont {Gao}},\ }\href {\doibase
  10.1039/C4CP04781H} {\bibfield  {journal} {\bibinfo  {journal} {Physical
  Chemistry Chemical Physics}\ }\textbf {\bibinfo {volume} {17}},\ \bibinfo
  {pages} {1900} (\bibinfo {year} {2014})}\BibitemShut {NoStop}%
\bibitem [{\citenamefont {Lim}\ \emph {et~al.}(2018)\citenamefont {Lim},
  \citenamefont {Almond}, \citenamefont {Trigatzis}, \citenamefont {Devlin},
  \citenamefont {Fitch}, \citenamefont {Sauer}, \citenamefont {Tarbutt},\ and\
  \citenamefont {Hinds}}]{Lim2018}%
  \BibitemOpen
  \bibfield  {author} {\bibinfo {author} {\bibfnamefont {J.}~\bibnamefont
  {Lim}}, \bibinfo {author} {\bibfnamefont {J.~R.}\ \bibnamefont {Almond}},
  \bibinfo {author} {\bibfnamefont {M.}~\bibnamefont {Trigatzis}}, \bibinfo
  {author} {\bibfnamefont {J.}~\bibnamefont {Devlin}}, \bibinfo {author}
  {\bibfnamefont {N.}~\bibnamefont {Fitch}}, \bibinfo {author} {\bibfnamefont
  {B.}~\bibnamefont {Sauer}}, \bibinfo {author} {\bibfnamefont
  {M.}~\bibnamefont {Tarbutt}}, \ and\ \bibinfo {author} {\bibfnamefont
  {E.}~\bibnamefont {Hinds}},\ }\href {\doibase 10.1103/PhysRevLett.120.123201}
  {\bibfield  {journal} {\bibinfo  {journal} {Physical Review Letters}\
  }\textbf {\bibinfo {volume} {120}},\ \bibinfo {pages} {123201} (\bibinfo
  {year} {2018})}\BibitemShut {NoStop}%
\bibitem [{\citenamefont {Iwata}\ \emph {et~al.}(2017)\citenamefont {Iwata},
  \citenamefont {McNally},\ and\ \citenamefont {Zelevinsky}}]{Iwata2017}%
  \BibitemOpen
  \bibfield  {author} {\bibinfo {author} {\bibfnamefont {G.~Z.}\ \bibnamefont
  {Iwata}}, \bibinfo {author} {\bibfnamefont {R.~L.}\ \bibnamefont {McNally}},
  \ and\ \bibinfo {author} {\bibfnamefont {T.}~\bibnamefont {Zelevinsky}},\
  }\href {\doibase 10.1103/PhysRevA.96.022509} {\bibfield  {journal} {\bibinfo
  {journal} {Physical Review A}\ }\textbf {\bibinfo {volume} {96}},\ \bibinfo
  {pages} {022509} (\bibinfo {year} {2017})}\BibitemShut {NoStop}%
\bibitem [{\citenamefont {Norrgard}\ \emph {et~al.}(2017)\citenamefont
  {Norrgard}, \citenamefont {Edwards}, \citenamefont {McCarron}, \citenamefont
  {Steinecker}, \citenamefont {DeMille}, \citenamefont {Alam}, \citenamefont
  {Peck}, \citenamefont {Wadia},\ and\ \citenamefont {Hunter}}]{Norrgard2017}%
  \BibitemOpen
  \bibfield  {author} {\bibinfo {author} {\bibfnamefont {E.~B.}\ \bibnamefont
  {Norrgard}}, \bibinfo {author} {\bibfnamefont {E.~R.}\ \bibnamefont
  {Edwards}}, \bibinfo {author} {\bibfnamefont {D.~J.}\ \bibnamefont
  {McCarron}}, \bibinfo {author} {\bibfnamefont {M.~H.}\ \bibnamefont
  {Steinecker}}, \bibinfo {author} {\bibfnamefont {D.}~\bibnamefont {DeMille}},
  \bibinfo {author} {\bibfnamefont {S.~S.}\ \bibnamefont {Alam}}, \bibinfo
  {author} {\bibfnamefont {S.~K.}\ \bibnamefont {Peck}}, \bibinfo {author}
  {\bibfnamefont {N.~S.}\ \bibnamefont {Wadia}}, \ and\ \bibinfo {author}
  {\bibfnamefont {L.~R.}\ \bibnamefont {Hunter}},\ }\href {\doibase
  10.1103/PhysRevA.95.062506} {\bibfield  {journal} {\bibinfo  {journal}
  {Physical Review A}\ }\textbf {\bibinfo {volume} {95}},\ \bibinfo {pages}
  {062506} (\bibinfo {year} {2017})}\BibitemShut {NoStop}%
\bibitem [{\citenamefont {Stuhl}\ \emph {et~al.}(2008)\citenamefont {Stuhl},
  \citenamefont {Sawyer}, \citenamefont {Wang},\ and\ \citenamefont
  {Ye}}]{Stuhl2008}%
  \BibitemOpen
  \bibfield  {author} {\bibinfo {author} {\bibfnamefont {B.~K.}\ \bibnamefont
  {Stuhl}}, \bibinfo {author} {\bibfnamefont {B.~C.}\ \bibnamefont {Sawyer}},
  \bibinfo {author} {\bibfnamefont {D.}~\bibnamefont {Wang}}, \ and\ \bibinfo
  {author} {\bibfnamefont {J.}~\bibnamefont {Ye}},\ }\href {\doibase
  10.1103/PhysRevLett.101.243002} {\bibfield  {journal} {\bibinfo  {journal}
  {Physical Review Letters}\ }\textbf {\bibinfo {volume} {101}},\ \bibinfo
  {pages} {243002} (\bibinfo {year} {2008})}\BibitemShut {NoStop}%
\bibitem [{\citenamefont {Isaev}\ \emph {et~al.}(2010)\citenamefont {Isaev},
  \citenamefont {Hoekstra},\ and\ \citenamefont {Berger}}]{Isaev2010}%
  \BibitemOpen
  \bibfield  {author} {\bibinfo {author} {\bibfnamefont {T.~A.}\ \bibnamefont
  {Isaev}}, \bibinfo {author} {\bibfnamefont {S.}~\bibnamefont {Hoekstra}}, \
  and\ \bibinfo {author} {\bibfnamefont {R.}~\bibnamefont {Berger}},\ }\href
  {\doibase 10.1103/PhysRevA.82.052521} {\bibfield  {journal} {\bibinfo
  {journal} {Physical Review A}\ }\textbf {\bibinfo {volume} {82}},\ \bibinfo
  {pages} {052521} (\bibinfo {year} {2010})}\BibitemShut {NoStop}%
\bibitem [{\citenamefont {Xu}\ \emph {et~al.}(2016)\citenamefont {Xu},
  \citenamefont {Yin}, \citenamefont {Wei}, \citenamefont {Xia},\ and\
  \citenamefont {Yin}}]{Xu2016}%
  \BibitemOpen
  \bibfield  {author} {\bibinfo {author} {\bibfnamefont {L.}~\bibnamefont
  {Xu}}, \bibinfo {author} {\bibfnamefont {Y.}~\bibnamefont {Yin}}, \bibinfo
  {author} {\bibfnamefont {B.}~\bibnamefont {Wei}}, \bibinfo {author}
  {\bibfnamefont {Y.}~\bibnamefont {Xia}}, \ and\ \bibinfo {author}
  {\bibfnamefont {J.}~\bibnamefont {Yin}},\ }\href {\doibase
  10.1103/PhysRevA.93.013408} {\bibfield  {journal} {\bibinfo  {journal}
  {Physical Review A}\ }\textbf {\bibinfo {volume} {93}},\ \bibinfo {pages}
  {013408} (\bibinfo {year} {2016})}\BibitemShut {NoStop}%
\bibitem [{\citenamefont {Norrgard}\ \emph {et~al.}(2023)\citenamefont
  {Norrgard}, \citenamefont {Chamorro}, \citenamefont {Cooksey}, \citenamefont
  {Eckel}, \citenamefont {Pilgram}, \citenamefont {Rodriguez}, \citenamefont
  {Yoon}, \citenamefont {Pasteka},\ and\ \citenamefont
  {Borschevsky}}]{Norrgard2023}%
  \BibitemOpen
  \bibfield  {author} {\bibinfo {author} {\bibfnamefont {E.}~\bibnamefont
  {Norrgard}}, \bibinfo {author} {\bibfnamefont {Y.}~\bibnamefont {Chamorro}},
  \bibinfo {author} {\bibfnamefont {C.}~\bibnamefont {Cooksey}}, \bibinfo
  {author} {\bibfnamefont {S.}~\bibnamefont {Eckel}}, \bibinfo {author}
  {\bibfnamefont {N.}~\bibnamefont {Pilgram}}, \bibinfo {author} {\bibfnamefont
  {K.}~\bibnamefont {Rodriguez}}, \bibinfo {author} {\bibfnamefont
  {H.}~\bibnamefont {Yoon}}, \bibinfo {author} {\bibfnamefont {L.}~\bibnamefont
  {Pasteka}}, \ and\ \bibinfo {author} {\bibfnamefont {A.}~\bibnamefont
  {Borschevsky}},\ }\href {\doibase
  https://doi.org/10.1103/PhysRevA.108.032809} {\bibfield  {journal} {\bibinfo
  {journal} {PRA}\ }\textbf {\bibinfo {volume} {108}},\ \bibinfo {pages}
  {032809} (\bibinfo {year} {2023})}\BibitemShut {NoStop}%
\bibitem [{\citenamefont {Chen}\ \emph {et~al.}(2019)\citenamefont {Chen},
  \citenamefont {Bu},\ and\ \citenamefont {Yan}}]{Chen2019}%
  \BibitemOpen
  \bibfield  {author} {\bibinfo {author} {\bibfnamefont {T.}~\bibnamefont
  {Chen}}, \bibinfo {author} {\bibfnamefont {W.}~\bibnamefont {Bu}}, \ and\
  \bibinfo {author} {\bibfnamefont {B.}~\bibnamefont {Yan}},\ }\href {\doibase
  10.1103/PHYSREVA.100.029901} {\bibfield  {journal} {\bibinfo  {journal}
  {Physical Review A}\ }\textbf {\bibinfo {volume} {100}},\ \bibinfo {pages}
  {029901} (\bibinfo {year} {2019})}\BibitemShut {NoStop}%
\bibitem [{\citenamefont {Liang}\ \emph {et~al.}(2021)\citenamefont {Liang},
  \citenamefont {Chen}, \citenamefont {Bu}, \citenamefont {Zhang},\ and\
  \citenamefont {Yan}}]{Liang2021}%
  \BibitemOpen
  \bibfield  {author} {\bibinfo {author} {\bibfnamefont {Q.}~\bibnamefont
  {Liang}}, \bibinfo {author} {\bibfnamefont {T.}~\bibnamefont {Chen}},
  \bibinfo {author} {\bibfnamefont {W.~H.}\ \bibnamefont {Bu}}, \bibinfo
  {author} {\bibfnamefont {Y.~H.}\ \bibnamefont {Zhang}}, \ and\ \bibinfo
  {author} {\bibfnamefont {B.}~\bibnamefont {Yan}},\ }\href {\doibase
  10.1007/S11467-020-1019-8} {\bibfield  {journal} {\bibinfo  {journal}
  {Frontiers of Physics}\ }\textbf {\bibinfo {volume} {16}},\ \bibinfo {pages}
  {32501} (\bibinfo {year} {2021})}\BibitemShut {NoStop}%
\bibitem [{\citenamefont {Bu}\ \emph {et~al.}(2017)\citenamefont {Bu},
  \citenamefont {Chen}, \citenamefont {Lv},\ and\ \citenamefont
  {Yan}}]{Bu2017}%
  \BibitemOpen
  \bibfield  {author} {\bibinfo {author} {\bibfnamefont {W.}~\bibnamefont
  {Bu}}, \bibinfo {author} {\bibfnamefont {T.}~\bibnamefont {Chen}}, \bibinfo
  {author} {\bibfnamefont {G.}~\bibnamefont {Lv}}, \ and\ \bibinfo {author}
  {\bibfnamefont {B.}~\bibnamefont {Yan}},\ }\href {\doibase
  10.1103/PhysRevA.95.032701} {\bibfield  {journal} {\bibinfo  {journal}
  {Physical Review A}\ }\textbf {\bibinfo {volume} {95}},\ \bibinfo {pages}
  {032701} (\bibinfo {year} {2017})}\BibitemShut {NoStop}%
\bibitem [{\citenamefont {Chen}\ \emph {et~al.}(2016)\citenamefont {Chen},
  \citenamefont {Bu},\ and\ \citenamefont {Yan}}]{Chen2016}%
  \BibitemOpen
  \bibfield  {author} {\bibinfo {author} {\bibfnamefont {T.}~\bibnamefont
  {Chen}}, \bibinfo {author} {\bibfnamefont {W.}~\bibnamefont {Bu}}, \ and\
  \bibinfo {author} {\bibfnamefont {B.}~\bibnamefont {Yan}},\ }\href {\doibase
  10.1103/PhysRevA.94.063415} {\bibfield  {journal} {\bibinfo  {journal}
  {Physical Review A}\ }\textbf {\bibinfo {volume} {94}},\ \bibinfo {pages}
  {063415} (\bibinfo {year} {2016})}\BibitemShut {NoStop}%
\bibitem [{\citenamefont {Bu}\ \emph {et~al.}(2022)\citenamefont {Bu},
  \citenamefont {Zhang}, \citenamefont {Liang}, \citenamefont {Chen},\ and\
  \citenamefont {Yan}}]{Bu2022}%
  \BibitemOpen
  \bibfield  {author} {\bibinfo {author} {\bibfnamefont {W.}~\bibnamefont
  {Bu}}, \bibinfo {author} {\bibfnamefont {Y.}~\bibnamefont {Zhang}}, \bibinfo
  {author} {\bibfnamefont {Q.}~\bibnamefont {Liang}}, \bibinfo {author}
  {\bibfnamefont {T.}~\bibnamefont {Chen}}, \ and\ \bibinfo {author}
  {\bibfnamefont {B.}~\bibnamefont {Yan}},\ }\href {\doibase
  10.1007/S11467-022-1194-X/METRICS} {\bibfield  {journal} {\bibinfo  {journal}
  {Frontiers of Physics}\ }\textbf {\bibinfo {volume} {17}},\ \bibinfo {pages}
  {1} (\bibinfo {year} {2022})}\BibitemShut {NoStop}%
\bibitem [{\citenamefont {Chen}\ \emph {et~al.}(2017)\citenamefont {Chen},
  \citenamefont {Bu},\ and\ \citenamefont {Yan}}]{Chen2017}%
  \BibitemOpen
  \bibfield  {author} {\bibinfo {author} {\bibfnamefont {T.}~\bibnamefont
  {Chen}}, \bibinfo {author} {\bibfnamefont {W.}~\bibnamefont {Bu}}, \ and\
  \bibinfo {author} {\bibfnamefont {B.}~\bibnamefont {Yan}},\ }\href {\doibase
  10.1103/PhysRevA.96.053401} {\bibfield  {journal} {\bibinfo  {journal}
  {Physical Review A}\ }\textbf {\bibinfo {volume} {96}},\ \bibinfo {pages}
  {053401} (\bibinfo {year} {2017})}\BibitemShut {NoStop}%
\bibitem [{\citenamefont {Albrecht}\ \emph {et~al.}(2020)\citenamefont
  {Albrecht}, \citenamefont {Scharwaechter}, \citenamefont {Sixt},
  \citenamefont {Hofer},\ and\ \citenamefont {Langen}}]{Albrecht2020}%
  \BibitemOpen
  \bibfield  {author} {\bibinfo {author} {\bibfnamefont {R.}~\bibnamefont
  {Albrecht}}, \bibinfo {author} {\bibfnamefont {M.}~\bibnamefont
  {Scharwaechter}}, \bibinfo {author} {\bibfnamefont {T.}~\bibnamefont {Sixt}},
  \bibinfo {author} {\bibfnamefont {L.}~\bibnamefont {Hofer}}, \ and\ \bibinfo
  {author} {\bibfnamefont {T.}~\bibnamefont {Langen}},\ }\href {\doibase
  10.1103/PhysRevA.101.013413} {\bibfield  {journal} {\bibinfo  {journal}
  {Physical Review A}\ }\textbf {\bibinfo {volume} {101}},\ \bibinfo {pages}
  {013413} (\bibinfo {year} {2020})}\BibitemShut {NoStop}%
\bibitem [{\citenamefont {Kogel}\ \emph {et~al.}(2021)\citenamefont {Kogel},
  \citenamefont {Rockenh\"auser}, \citenamefont {Albrecht},\ and\ \citenamefont
  {Langen}}]{Kogel2021}%
  \BibitemOpen
  \bibfield  {author} {\bibinfo {author} {\bibfnamefont {F.}~\bibnamefont
  {Kogel}}, \bibinfo {author} {\bibfnamefont {M.}~\bibnamefont
  {Rockenh\"auser}}, \bibinfo {author} {\bibfnamefont {R.}~\bibnamefont
  {Albrecht}}, \ and\ \bibinfo {author} {\bibfnamefont {T.}~\bibnamefont
  {Langen}},\ }\href {\doibase 10.1088/1367-2630/AC1DF2} {\bibfield  {journal}
  {\bibinfo  {journal} {New Journal of Physics}\ }\textbf {\bibinfo {volume}
  {23}},\ \bibinfo {pages} {095003} (\bibinfo {year} {2021})}\BibitemShut
  {NoStop}%
\bibitem [{\citenamefont {Zhang}\ \emph {et~al.}(2022)\citenamefont {Zhang},
  \citenamefont {Zeng}, \citenamefont {Liang}, \citenamefont {Bu},\ and\
  \citenamefont {Yan}}]{Zhang2022}%
  \BibitemOpen
  \bibfield  {author} {\bibinfo {author} {\bibfnamefont {Y.}~\bibnamefont
  {Zhang}}, \bibinfo {author} {\bibfnamefont {Z.}~\bibnamefont {Zeng}},
  \bibinfo {author} {\bibfnamefont {Q.}~\bibnamefont {Liang}}, \bibinfo
  {author} {\bibfnamefont {W.}~\bibnamefont {Bu}}, \ and\ \bibinfo {author}
  {\bibfnamefont {B.}~\bibnamefont {Yan}},\ }\href {\doibase
  10.1103/PHYSREVA.105.033307/FIGURES/6/MEDIUM} {\bibfield  {journal} {\bibinfo
   {journal} {Physical Review A}\ }\textbf {\bibinfo {volume} {105}},\ \bibinfo
  {pages} {033307} (\bibinfo {year} {2022})}\BibitemShut {NoStop}%
\bibitem [{\citenamefont {Rockenhäuser}\ \emph {et~al.}(2023)\citenamefont
  {Rockenhäuser}, \citenamefont {Kogel}, \citenamefont {Pultinevicius},\ and\
  \citenamefont {Langen}}]{Rockenhaeuser2023}%
  \BibitemOpen
  \bibfield  {author} {\bibinfo {author} {\bibfnamefont {M.}~\bibnamefont
  {Rockenhäuser}}, \bibinfo {author} {\bibfnamefont {F.}~\bibnamefont
  {Kogel}}, \bibinfo {author} {\bibfnamefont {E.}~\bibnamefont
  {Pultinevicius}}, \ and\ \bibinfo {author} {\bibfnamefont {T.}~\bibnamefont
  {Langen}},\ }\href {\doibase https://doi.org/10.1016/bs.aamop.2023.04.005}
  {\bibfield  {journal} {\bibinfo  {journal} {Physical Review A}\ }\textbf
  {\bibinfo {volume} {108}},\ \bibinfo {pages} {062812} (\bibinfo {year}
  {2023})}\BibitemShut {NoStop}%
\bibitem [{\citenamefont {Truppe}\ \emph {et~al.}(2019)\citenamefont {Truppe},
  \citenamefont {Marx}, \citenamefont {Kray}, \citenamefont {Doppelbauer},
  \citenamefont {Hofsäss}, \citenamefont {Schewe}, \citenamefont {Walter},
  \citenamefont {Pérez-Riós}, \citenamefont {Sartakov},\ and\ \citenamefont
  {Meijer}}]{Truppe2019}%
  \BibitemOpen
  \bibfield  {author} {\bibinfo {author} {\bibfnamefont {S.}~\bibnamefont
  {Truppe}}, \bibinfo {author} {\bibfnamefont {S.}~\bibnamefont {Marx}},
  \bibinfo {author} {\bibfnamefont {S.}~\bibnamefont {Kray}}, \bibinfo {author}
  {\bibfnamefont {M.}~\bibnamefont {Doppelbauer}}, \bibinfo {author}
  {\bibfnamefont {S.}~\bibnamefont {Hofsäss}}, \bibinfo {author}
  {\bibfnamefont {H.~C.}\ \bibnamefont {Schewe}}, \bibinfo {author}
  {\bibfnamefont {N.}~\bibnamefont {Walter}}, \bibinfo {author} {\bibfnamefont
  {J.}~\bibnamefont {Pérez-Riós}}, \bibinfo {author} {\bibfnamefont {B.~G.}\
  \bibnamefont {Sartakov}}, \ and\ \bibinfo {author} {\bibfnamefont
  {G.}~\bibnamefont {Meijer}},\ }\href {\doibase 10.1103/PhysRevA.100.052513}
  {\bibfield  {journal} {\bibinfo  {journal} {Physical Review A}\ }\textbf
  {\bibinfo {volume} {100}},\ \bibinfo {pages} {052513} (\bibinfo {year}
  {2019})}\BibitemShut {NoStop}%
\bibitem [{\citenamefont {Doppelbauer}\ \emph {et~al.}(2021)\citenamefont
  {Doppelbauer}, \citenamefont {Walter}, \citenamefont {Hofs\"ass},
  \citenamefont {Marx}, \citenamefont {Schewe}, \citenamefont {Kray},
  \citenamefont {P\'erez-R\'ios}, \citenamefont {Sartakov}, \citenamefont
  {Truppe},\ and\ \citenamefont {Meijer}}]{Doppelbauer2021}%
  \BibitemOpen
  \bibfield  {author} {\bibinfo {author} {\bibfnamefont {M.}~\bibnamefont
  {Doppelbauer}}, \bibinfo {author} {\bibfnamefont {N.}~\bibnamefont {Walter}},
  \bibinfo {author} {\bibfnamefont {S.}~\bibnamefont {Hofs\"ass}}, \bibinfo
  {author} {\bibfnamefont {S.}~\bibnamefont {Marx}}, \bibinfo {author}
  {\bibfnamefont {H.~C.}\ \bibnamefont {Schewe}}, \bibinfo {author}
  {\bibfnamefont {S.}~\bibnamefont {Kray}}, \bibinfo {author} {\bibfnamefont
  {J.}~\bibnamefont {P\'erez-R\'ios}}, \bibinfo {author} {\bibfnamefont
  {B.~G.}\ \bibnamefont {Sartakov}}, \bibinfo {author} {\bibfnamefont
  {S.}~\bibnamefont {Truppe}}, \ and\ \bibinfo {author} {\bibfnamefont
  {G.}~\bibnamefont {Meijer}},\ }\href {\doibase 10.1080/00268976.2020.1810351}
  {\bibfield  {journal} {\bibinfo  {journal} {Molecular Physics}\ }\textbf
  {\bibinfo {volume} {119}},\ \bibinfo {pages} {e1810351} (\bibinfo {year}
  {2021})}\BibitemShut {NoStop}%
\bibitem [{\citenamefont {Schnaubelt}\ \emph {et~al.}(2021)\citenamefont
  {Schnaubelt}, \citenamefont {Shaw},\ and\ \citenamefont
  {McCarron}}]{Schnaubelt2021}%
  \BibitemOpen
  \bibfield  {author} {\bibinfo {author} {\bibfnamefont {J.~C.}\ \bibnamefont
  {Schnaubelt}}, \bibinfo {author} {\bibfnamefont {J.~C.}\ \bibnamefont
  {Shaw}}, \ and\ \bibinfo {author} {\bibfnamefont {D.~J.}\ \bibnamefont
  {McCarron}},\ }\href {https://arxiv.org/abs/2109.03953v1} {\bibfield
  {journal} {\bibinfo  {journal} {arXiv:2109.03953}\ } (\bibinfo {year}
  {2021})}\BibitemShut {NoStop}%
\bibitem [{\citenamefont {McNally}\ \emph {et~al.}(2020)\citenamefont
  {McNally}, \citenamefont {Kozyryev}, \citenamefont {Vazquez-Carson},
  \citenamefont {Wenz}, \citenamefont {Wang},\ and\ \citenamefont
  {Zelevinsky}}]{McNally2020}%
  \BibitemOpen
  \bibfield  {author} {\bibinfo {author} {\bibfnamefont {R.~L.}\ \bibnamefont
  {McNally}}, \bibinfo {author} {\bibfnamefont {I.}~\bibnamefont {Kozyryev}},
  \bibinfo {author} {\bibfnamefont {S.}~\bibnamefont {Vazquez-Carson}},
  \bibinfo {author} {\bibfnamefont {K.}~\bibnamefont {Wenz}}, \bibinfo {author}
  {\bibfnamefont {T.}~\bibnamefont {Wang}}, \ and\ \bibinfo {author}
  {\bibfnamefont {T.}~\bibnamefont {Zelevinsky}},\ }\href {\doibase
  10.1088/1367-2630/ABA3E9} {\bibfield  {journal} {\bibinfo  {journal} {New
  Journal of Physics}\ }\textbf {\bibinfo {volume} {22}},\ \bibinfo {pages}
  {083047} (\bibinfo {year} {2020})}\BibitemShut {NoStop}%
\bibitem [{\citenamefont {Tarallo}\ \emph {et~al.}(2016)\citenamefont
  {Tarallo}, \citenamefont {Iwata},\ and\ \citenamefont
  {Zelevinsky}}]{Tarallo2016}%
  \BibitemOpen
  \bibfield  {author} {\bibinfo {author} {\bibfnamefont {M.~G.}\ \bibnamefont
  {Tarallo}}, \bibinfo {author} {\bibfnamefont {G.~Z.}\ \bibnamefont {Iwata}},
  \ and\ \bibinfo {author} {\bibfnamefont {T.}~\bibnamefont {Zelevinsky}},\
  }\href {\doibase 10.1103/PHYSREVA.93.032509/FIGURES/10/MEDIUM} {\bibfield
  {journal} {\bibinfo  {journal} {Physical Review A}\ }\textbf {\bibinfo
  {volume} {93}},\ \bibinfo {pages} {032509} (\bibinfo {year}
  {2016})}\BibitemShut {NoStop}%
\bibitem [{\citenamefont {Isaev}\ and\ \citenamefont
  {Berger}(2016)}]{Isaev2016}%
  \BibitemOpen
  \bibfield  {author} {\bibinfo {author} {\bibfnamefont {T.~A.}\ \bibnamefont
  {Isaev}}\ and\ \bibinfo {author} {\bibfnamefont {R.}~\bibnamefont {Berger}},\
  }\href {\doibase 10.1103/PhysRevLett.116.063006} {\bibfield  {journal}
  {\bibinfo  {journal} {Physical Review Letters}\ }\textbf {\bibinfo {volume}
  {116}},\ \bibinfo {pages} {063006} (\bibinfo {year} {2016})}\BibitemShut
  {NoStop}%
\bibitem [{\citenamefont {Kozyryev}\ \emph
  {et~al.}(2016{\natexlab{a}})\citenamefont {Kozyryev}, \citenamefont {Baum},
  \citenamefont {Matsuda}, \citenamefont {Hemmerling},\ and\ \citenamefont
  {Doyle}}]{Kozyryev2016}%
  \BibitemOpen
  \bibfield  {author} {\bibinfo {author} {\bibfnamefont {I.}~\bibnamefont
  {Kozyryev}}, \bibinfo {author} {\bibfnamefont {L.}~\bibnamefont {Baum}},
  \bibinfo {author} {\bibfnamefont {K.}~\bibnamefont {Matsuda}}, \bibinfo
  {author} {\bibfnamefont {B.}~\bibnamefont {Hemmerling}}, \ and\ \bibinfo
  {author} {\bibfnamefont {J.~M.}\ \bibnamefont {Doyle}},\ }\href {\doibase
  10.1088/0953-4075/49/13/134002} {\bibfield  {journal} {\bibinfo  {journal}
  {Journal of Physics B: Atomic, Molecular and Optical Physics}\ }\textbf
  {\bibinfo {volume} {49}},\ \bibinfo {pages} {134002} (\bibinfo {year}
  {2016}{\natexlab{a}})}\BibitemShut {NoStop}%
\bibitem [{\citenamefont {Kozyryev}\ \emph
  {et~al.}(2016{\natexlab{b}})\citenamefont {Kozyryev}, \citenamefont {Baum},
  \citenamefont {Matsuda},\ and\ \citenamefont {Doyle}}]{Kozyryev2016a}%
  \BibitemOpen
  \bibfield  {author} {\bibinfo {author} {\bibfnamefont {I.}~\bibnamefont
  {Kozyryev}}, \bibinfo {author} {\bibfnamefont {L.}~\bibnamefont {Baum}},
  \bibinfo {author} {\bibfnamefont {K.}~\bibnamefont {Matsuda}}, \ and\
  \bibinfo {author} {\bibfnamefont {J.~M.}\ \bibnamefont {Doyle}},\ }\href
  {\doibase 10.1002/cphc.201601051} {\bibfield  {journal} {\bibinfo  {journal}
  {ChemPhysChem}\ }\textbf {\bibinfo {volume} {17}},\ \bibinfo {pages} {3641}
  (\bibinfo {year} {2016}{\natexlab{b}})}\BibitemShut {NoStop}%
\bibitem [{\citenamefont {Baum}\ \emph {et~al.}(2020)\citenamefont {Baum},
  \citenamefont {Vilas}, \citenamefont {Hallas}, \citenamefont {Augenbraun},
  \citenamefont {Raval}, \citenamefont {Mitra},\ and\ \citenamefont
  {Doyle}}]{Baum2020}%
  \BibitemOpen
  \bibfield  {author} {\bibinfo {author} {\bibfnamefont {L.}~\bibnamefont
  {Baum}}, \bibinfo {author} {\bibfnamefont {N.~B.}\ \bibnamefont {Vilas}},
  \bibinfo {author} {\bibfnamefont {C.}~\bibnamefont {Hallas}}, \bibinfo
  {author} {\bibfnamefont {B.~L.}\ \bibnamefont {Augenbraun}}, \bibinfo
  {author} {\bibfnamefont {S.}~\bibnamefont {Raval}}, \bibinfo {author}
  {\bibfnamefont {D.}~\bibnamefont {Mitra}}, \ and\ \bibinfo {author}
  {\bibfnamefont {J.~M.}\ \bibnamefont {Doyle}},\ }\href {\doibase
  10.1103/PhysRevLett.124.133201} {\bibfield  {journal} {\bibinfo  {journal}
  {Physical Review Letters}\ }\textbf {\bibinfo {volume} {124}},\ \bibinfo
  {pages} {133201} (\bibinfo {year} {2020})}\BibitemShut {NoStop}%
\bibitem [{\citenamefont {Augenbraun}\ \emph {et~al.}(2021)\citenamefont
  {Augenbraun}, \citenamefont {Frenett}, \citenamefont {Sawaoka}, \citenamefont
  {Hallas}, \citenamefont {Vilas}, \citenamefont {Nasir}, \citenamefont
  {Lasner},\ and\ \citenamefont {Doyle}}]{Augenbraun2021}%
  \BibitemOpen
  \bibfield  {author} {\bibinfo {author} {\bibfnamefont {B.~L.}\ \bibnamefont
  {Augenbraun}}, \bibinfo {author} {\bibfnamefont {A.}~\bibnamefont {Frenett}},
  \bibinfo {author} {\bibfnamefont {H.}~\bibnamefont {Sawaoka}}, \bibinfo
  {author} {\bibfnamefont {C.}~\bibnamefont {Hallas}}, \bibinfo {author}
  {\bibfnamefont {N.~B.}\ \bibnamefont {Vilas}}, \bibinfo {author}
  {\bibfnamefont {A.}~\bibnamefont {Nasir}}, \bibinfo {author} {\bibfnamefont
  {Z.~D.}\ \bibnamefont {Lasner}}, \ and\ \bibinfo {author} {\bibfnamefont
  {J.~M.}\ \bibnamefont {Doyle}},\ }\href {https://arxiv.org/abs/2109.03067v1}
  {\bibfield  {journal} {\bibinfo  {journal} {Physical Review Letters}\ ,\
  \bibinfo {pages} {263002}} (\bibinfo {year} {2021})}\BibitemShut {NoStop}%
\bibitem [{\citenamefont {Mitra}\ \emph {et~al.}(2020)\citenamefont {Mitra},
  \citenamefont {Vilas}, \citenamefont {Hallas}, \citenamefont {Anderegg},
  \citenamefont {Augenbraun}, \citenamefont {Baum}, \citenamefont {Miller},
  \citenamefont {Raval},\ and\ \citenamefont {Doyle}}]{Mitra2020}%
  \BibitemOpen
  \bibfield  {author} {\bibinfo {author} {\bibfnamefont {D.}~\bibnamefont
  {Mitra}}, \bibinfo {author} {\bibfnamefont {N.~B.}\ \bibnamefont {Vilas}},
  \bibinfo {author} {\bibfnamefont {C.}~\bibnamefont {Hallas}}, \bibinfo
  {author} {\bibfnamefont {L.}~\bibnamefont {Anderegg}}, \bibinfo {author}
  {\bibfnamefont {B.~L.}\ \bibnamefont {Augenbraun}}, \bibinfo {author}
  {\bibfnamefont {L.}~\bibnamefont {Baum}}, \bibinfo {author} {\bibfnamefont
  {C.}~\bibnamefont {Miller}}, \bibinfo {author} {\bibfnamefont
  {S.}~\bibnamefont {Raval}}, \ and\ \bibinfo {author} {\bibfnamefont {J.~M.}\
  \bibnamefont {Doyle}},\ }\href {\doibase 10.1126/science.abc5357} {\bibfield
  {journal} {\bibinfo  {journal} {Science}\ }\textbf {\bibinfo {volume}
  {369}},\ \bibinfo {pages} {1366} (\bibinfo {year} {2020})}\BibitemShut
  {NoStop}%
\bibitem [{\citenamefont {Vilas}\ \emph {et~al.}(2022)\citenamefont {Vilas},
  \citenamefont {Hallas}, \citenamefont {Anderegg}, \citenamefont {Robichaud},
  \citenamefont {Winnicki}, \citenamefont {Mitra},\ and\ \citenamefont
  {Doyle}}]{Vilas2022}%
  \BibitemOpen
  \bibfield  {author} {\bibinfo {author} {\bibfnamefont {N.~B.}\ \bibnamefont
  {Vilas}}, \bibinfo {author} {\bibfnamefont {C.}~\bibnamefont {Hallas}},
  \bibinfo {author} {\bibfnamefont {L.}~\bibnamefont {Anderegg}}, \bibinfo
  {author} {\bibfnamefont {P.}~\bibnamefont {Robichaud}}, \bibinfo {author}
  {\bibfnamefont {A.}~\bibnamefont {Winnicki}}, \bibinfo {author}
  {\bibfnamefont {D.}~\bibnamefont {Mitra}}, \ and\ \bibinfo {author}
  {\bibfnamefont {J.~M.}\ \bibnamefont {Doyle}},\ }\href {\doibase
  10.1038/s41586-022-04620-5} {\bibfield  {journal} {\bibinfo  {journal}
  {Nature}\ }\textbf {\bibinfo {volume} {606}},\ \bibinfo {pages} {70}
  (\bibinfo {year} {2022})}\BibitemShut {NoStop}%
\bibitem [{\citenamefont {Kozyryev}\ \emph {et~al.}(2017)\citenamefont
  {Kozyryev}, \citenamefont {Baum}, \citenamefont {Matsuda}, \citenamefont
  {Augenbraun}, \citenamefont {Anderegg}, \citenamefont {Sedlack},\ and\
  \citenamefont {Doyle}}]{Kozyryev2017}%
  \BibitemOpen
  \bibfield  {author} {\bibinfo {author} {\bibfnamefont {I.}~\bibnamefont
  {Kozyryev}}, \bibinfo {author} {\bibfnamefont {L.}~\bibnamefont {Baum}},
  \bibinfo {author} {\bibfnamefont {K.}~\bibnamefont {Matsuda}}, \bibinfo
  {author} {\bibfnamefont {B.~L.}\ \bibnamefont {Augenbraun}}, \bibinfo
  {author} {\bibfnamefont {L.}~\bibnamefont {Anderegg}}, \bibinfo {author}
  {\bibfnamefont {A.~P.}\ \bibnamefont {Sedlack}}, \ and\ \bibinfo {author}
  {\bibfnamefont {J.~M.}\ \bibnamefont {Doyle}},\ }\href {\doibase
  10.1103/PhysRevLett.118.173201} {\bibfield  {journal} {\bibinfo  {journal}
  {Physical Review Letters}\ }\textbf {\bibinfo {volume} {118}},\ \bibinfo
  {pages} {173201} (\bibinfo {year} {2017})}\BibitemShut {NoStop}%
\bibitem [{\citenamefont {Augenbraun}\ \emph {et~al.}(2020)\citenamefont
  {Augenbraun}, \citenamefont {Lasner}, \citenamefont {Frenett}, \citenamefont
  {Sawaoka}, \citenamefont {Miller}, \citenamefont {Steimle},\ and\
  \citenamefont {Doyle}}]{Augenbraun2020}%
  \BibitemOpen
  \bibfield  {author} {\bibinfo {author} {\bibfnamefont {B.~L.}\ \bibnamefont
  {Augenbraun}}, \bibinfo {author} {\bibfnamefont {Z.~D.}\ \bibnamefont
  {Lasner}}, \bibinfo {author} {\bibfnamefont {A.}~\bibnamefont {Frenett}},
  \bibinfo {author} {\bibfnamefont {H.}~\bibnamefont {Sawaoka}}, \bibinfo
  {author} {\bibfnamefont {C.}~\bibnamefont {Miller}}, \bibinfo {author}
  {\bibfnamefont {T.~C.}\ \bibnamefont {Steimle}}, \ and\ \bibinfo {author}
  {\bibfnamefont {J.~M.}\ \bibnamefont {Doyle}},\ }\href {\doibase
  10.1088/1367-2630/ab687b} {\bibfield  {journal} {\bibinfo  {journal} {New
  Journal of Physics}\ }\textbf {\bibinfo {volume} {22}},\ \bibinfo {pages}
  {22003} (\bibinfo {year} {2020})}\BibitemShut {NoStop}%
\bibitem [{\citenamefont {Augenbraun}\ \emph {et~al.}(2023)\citenamefont
  {Augenbraun}, \citenamefont {Anderegg}, \citenamefont {Hallas}, \citenamefont
  {Lasner}, \citenamefont {Vilas},\ and\ \citenamefont
  {Doyle}}]{Augenbraun2023}%
  \BibitemOpen
  \bibfield  {author} {\bibinfo {author} {\bibfnamefont {B.~L.}\ \bibnamefont
  {Augenbraun}}, \bibinfo {author} {\bibfnamefont {L.}~\bibnamefont
  {Anderegg}}, \bibinfo {author} {\bibfnamefont {C.}~\bibnamefont {Hallas}},
  \bibinfo {author} {\bibfnamefont {Z.~D.}\ \bibnamefont {Lasner}}, \bibinfo
  {author} {\bibfnamefont {N.~B.}\ \bibnamefont {Vilas}}, \ and\ \bibinfo
  {author} {\bibfnamefont {J.~M.}\ \bibnamefont {Doyle}},\ }\href {\doibase
  https://doi.org/10.1016/bs.aamop.2023.04.005} {\bibfield  {journal} {\bibinfo
   {journal} {Advances In Atomic, Molecular, and Optical Physics}\ }\textbf
  {\bibinfo {volume} {72}},\ \bibinfo {pages} {89} (\bibinfo {year}
  {2023})}\BibitemShut {NoStop}%
\bibitem [{\citenamefont {Barry}\ \emph {et~al.}(2014)\citenamefont {Barry},
  \citenamefont {McCarron}, \citenamefont {Norrgard}, \citenamefont
  {Steinecker},\ and\ \citenamefont {DeMille}}]{Barry2014}%
  \BibitemOpen
  \bibfield  {author} {\bibinfo {author} {\bibfnamefont {J.~F.}\ \bibnamefont
  {Barry}}, \bibinfo {author} {\bibfnamefont {D.~J.}\ \bibnamefont {McCarron}},
  \bibinfo {author} {\bibfnamefont {E.~B.}\ \bibnamefont {Norrgard}}, \bibinfo
  {author} {\bibfnamefont {M.~H.}\ \bibnamefont {Steinecker}}, \ and\ \bibinfo
  {author} {\bibfnamefont {D.}~\bibnamefont {DeMille}},\ }\href {\doibase
  10.1038/nature13634} {\bibfield  {journal} {\bibinfo  {journal} {Nature}\
  }\textbf {\bibinfo {volume} {512}},\ \bibinfo {pages} {286} (\bibinfo {year}
  {2014})}\BibitemShut {NoStop}%
\bibitem [{\citenamefont {McCarron}\ \emph {et~al.}(2018)\citenamefont
  {McCarron}, \citenamefont {Steinecker}, \citenamefont {Zhu},\ and\
  \citenamefont {DeMille}}]{McCarron2018}%
  \BibitemOpen
  \bibfield  {author} {\bibinfo {author} {\bibfnamefont {D.~J.}\ \bibnamefont
  {McCarron}}, \bibinfo {author} {\bibfnamefont {M.~H.}\ \bibnamefont
  {Steinecker}}, \bibinfo {author} {\bibfnamefont {Y.}~\bibnamefont {Zhu}}, \
  and\ \bibinfo {author} {\bibfnamefont {D.}~\bibnamefont {DeMille}},\ }\href
  {\doibase 10.1103/PhysRevLett.121.013202} {\bibfield  {journal} {\bibinfo
  {journal} {Phys. Rev. Lett.}\ }\textbf {\bibinfo {volume} {121}},\ \bibinfo
  {pages} {13202} (\bibinfo {year} {2018})}\BibitemShut {NoStop}%
\bibitem [{\citenamefont {Anderegg}\ \emph {et~al.}(2017)\citenamefont
  {Anderegg}, \citenamefont {Augenbraun}, \citenamefont {Chae}, \citenamefont
  {Hemmerling}, \citenamefont {Hutzler}, \citenamefont {Ravi}, \citenamefont
  {Collopy}, \citenamefont {Ye}, \citenamefont {Ketterle},\ and\ \citenamefont
  {Doyle}}]{Anderegg2017}%
  \BibitemOpen
  \bibfield  {author} {\bibinfo {author} {\bibfnamefont {L.}~\bibnamefont
  {Anderegg}}, \bibinfo {author} {\bibfnamefont {B.~L.}\ \bibnamefont
  {Augenbraun}}, \bibinfo {author} {\bibfnamefont {E.}~\bibnamefont {Chae}},
  \bibinfo {author} {\bibfnamefont {B.}~\bibnamefont {Hemmerling}}, \bibinfo
  {author} {\bibfnamefont {N.~R.}\ \bibnamefont {Hutzler}}, \bibinfo {author}
  {\bibfnamefont {A.}~\bibnamefont {Ravi}}, \bibinfo {author} {\bibfnamefont
  {A.}~\bibnamefont {Collopy}}, \bibinfo {author} {\bibfnamefont
  {J.}~\bibnamefont {Ye}}, \bibinfo {author} {\bibfnamefont {W.}~\bibnamefont
  {Ketterle}}, \ and\ \bibinfo {author} {\bibfnamefont {J.~M.}\ \bibnamefont
  {Doyle}},\ }\href {\doibase 10.1103/PhysRevLett.119.103201} {\bibfield
  {journal} {\bibinfo  {journal} {Physical Review Letters}\ }\textbf {\bibinfo
  {volume} {119}},\ \bibinfo {pages} {103201} (\bibinfo {year}
  {2017})}\BibitemShut {NoStop}%
\bibitem [{\citenamefont {Truppe}\ \emph {et~al.}(2017)\citenamefont {Truppe},
  \citenamefont {Williams}, \citenamefont {Hambach}, \citenamefont {Caldwell},
  \citenamefont {Fitch}, \citenamefont {Hinds}, \citenamefont {Sauer},\ and\
  \citenamefont {Tarbutt}}]{Truppe2017}%
  \BibitemOpen
  \bibfield  {author} {\bibinfo {author} {\bibfnamefont {S.}~\bibnamefont
  {Truppe}}, \bibinfo {author} {\bibfnamefont {H.~J.}\ \bibnamefont
  {Williams}}, \bibinfo {author} {\bibfnamefont {M.}~\bibnamefont {Hambach}},
  \bibinfo {author} {\bibfnamefont {L.}~\bibnamefont {Caldwell}}, \bibinfo
  {author} {\bibfnamefont {N.~J.}\ \bibnamefont {Fitch}}, \bibinfo {author}
  {\bibfnamefont {E.~A.}\ \bibnamefont {Hinds}}, \bibinfo {author}
  {\bibfnamefont {B.~E.}\ \bibnamefont {Sauer}}, \ and\ \bibinfo {author}
  {\bibfnamefont {M.~R.}\ \bibnamefont {Tarbutt}},\ }\href {\doibase
  10.1038/nphys4241} {\bibfield  {journal} {\bibinfo  {journal} {Nature
  Physics}\ }\textbf {\bibinfo {volume} {13}},\ \bibinfo {pages} {1173}
  (\bibinfo {year} {2017})}\BibitemShut {NoStop}%
\bibitem [{\citenamefont {Williams}\ \emph {et~al.}(2017)\citenamefont
  {Williams}, \citenamefont {Truppe}, \citenamefont {Hambach}, \citenamefont
  {Caldwell}, \citenamefont {Fitch}, \citenamefont {Hinds}, \citenamefont
  {Sauer},\ and\ \citenamefont {Tarbutt}}]{Williams2017}%
  \BibitemOpen
  \bibfield  {author} {\bibinfo {author} {\bibfnamefont {H.~J.}\ \bibnamefont
  {Williams}}, \bibinfo {author} {\bibfnamefont {S.}~\bibnamefont {Truppe}},
  \bibinfo {author} {\bibfnamefont {M.}~\bibnamefont {Hambach}}, \bibinfo
  {author} {\bibfnamefont {L.}~\bibnamefont {Caldwell}}, \bibinfo {author}
  {\bibfnamefont {N.~J.}\ \bibnamefont {Fitch}}, \bibinfo {author}
  {\bibfnamefont {E.~A.}\ \bibnamefont {Hinds}}, \bibinfo {author}
  {\bibfnamefont {B.~E.}\ \bibnamefont {Sauer}}, \ and\ \bibinfo {author}
  {\bibfnamefont {M.~R.}\ \bibnamefont {Tarbutt}},\ }\href {\doibase
  10.1088/1367-2630/aa8e52} {\bibfield  {journal} {\bibinfo  {journal} {New
  Journal of Physics}\ }\textbf {\bibinfo {volume} {19}},\ \bibinfo {pages}
  {113035} (\bibinfo {year} {2017})}\BibitemShut {NoStop}%
\bibitem [{\citenamefont {Collopy}\ \emph {et~al.}(2018)\citenamefont
  {Collopy}, \citenamefont {Ding}, \citenamefont {Wu}, \citenamefont
  {Finneran}, \citenamefont {Anderegg}, \citenamefont {Augenbraun},
  \citenamefont {Doyle},\ and\ \citenamefont {Ye}}]{Collopy2018}%
  \BibitemOpen
  \bibfield  {author} {\bibinfo {author} {\bibfnamefont {A.~L.}\ \bibnamefont
  {Collopy}}, \bibinfo {author} {\bibfnamefont {S.}~\bibnamefont {Ding}},
  \bibinfo {author} {\bibfnamefont {Y.}~\bibnamefont {Wu}}, \bibinfo {author}
  {\bibfnamefont {I.~A.}\ \bibnamefont {Finneran}}, \bibinfo {author}
  {\bibfnamefont {L.}~\bibnamefont {Anderegg}}, \bibinfo {author}
  {\bibfnamefont {B.~L.}\ \bibnamefont {Augenbraun}}, \bibinfo {author}
  {\bibfnamefont {J.~M.}\ \bibnamefont {Doyle}}, \ and\ \bibinfo {author}
  {\bibfnamefont {J.}~\bibnamefont {Ye}},\ }\href {\doibase
  10.1103/PhysRevLett.121.213201} {\bibfield  {journal} {\bibinfo  {journal}
  {Physical Review Letters}\ }\textbf {\bibinfo {volume} {121}},\ \bibinfo
  {pages} {213201} (\bibinfo {year} {2018})}\BibitemShut {NoStop}%
\bibitem [{\citenamefont {Clayburn}\ \emph {et~al.}(2020)\citenamefont
  {Clayburn}, \citenamefont {Wright}, \citenamefont {Norrgard}, \citenamefont
  {DeMille},\ and\ \citenamefont {Hunter}}]{Clayburn2020}%
  \BibitemOpen
  \bibfield  {author} {\bibinfo {author} {\bibfnamefont {N.~B.}\ \bibnamefont
  {Clayburn}}, \bibinfo {author} {\bibfnamefont {T.~H.}\ \bibnamefont
  {Wright}}, \bibinfo {author} {\bibfnamefont {E.~B.}\ \bibnamefont
  {Norrgard}}, \bibinfo {author} {\bibfnamefont {D.}~\bibnamefont {DeMille}}, \
  and\ \bibinfo {author} {\bibfnamefont {L.~R.}\ \bibnamefont {Hunter}},\
  }\href {\doibase 10.1103/PhysRevA.102.052802} {\bibfield  {journal} {\bibinfo
   {journal} {Physical Review A}\ }\textbf {\bibinfo {volume} {102}},\ \bibinfo
  {pages} {052802} (\bibinfo {year} {2020})}\BibitemShut {NoStop}%
\bibitem [{\citenamefont {Grasdijk}\ \emph {et~al.}(2021)\citenamefont
  {Grasdijk}, \citenamefont {Timgren}, \citenamefont {Kastelic}, \citenamefont
  {Wright}, \citenamefont {Lamoreaux}, \citenamefont {DeMille}, \citenamefont
  {Wenz}, \citenamefont {Aitken}, \citenamefont {Zelevinsky}, \citenamefont
  {Winick},\ and\ \citenamefont {Kawall}}]{Grasdijk2021}%
  \BibitemOpen
  \bibfield  {author} {\bibinfo {author} {\bibfnamefont {O.}~\bibnamefont
  {Grasdijk}}, \bibinfo {author} {\bibfnamefont {O.}~\bibnamefont {Timgren}},
  \bibinfo {author} {\bibfnamefont {J.}~\bibnamefont {Kastelic}}, \bibinfo
  {author} {\bibfnamefont {T.}~\bibnamefont {Wright}}, \bibinfo {author}
  {\bibfnamefont {S.}~\bibnamefont {Lamoreaux}}, \bibinfo {author}
  {\bibfnamefont {D.}~\bibnamefont {DeMille}}, \bibinfo {author} {\bibfnamefont
  {K.}~\bibnamefont {Wenz}}, \bibinfo {author} {\bibfnamefont {M.}~\bibnamefont
  {Aitken}}, \bibinfo {author} {\bibfnamefont {T.}~\bibnamefont {Zelevinsky}},
  \bibinfo {author} {\bibfnamefont {T.}~\bibnamefont {Winick}}, \ and\ \bibinfo
  {author} {\bibfnamefont {D.}~\bibnamefont {Kawall}},\ }\href {\doibase
  10.1088/2058-9565/abdca3} {\bibfield  {journal} {\bibinfo  {journal} {Quantum
  Science and Technology}\ }\textbf {\bibinfo {volume} {6}},\ \bibinfo {pages}
  {044007} (\bibinfo {year} {2021})}\BibitemShut {NoStop}%
\bibitem [{\citenamefont {Wan}\ \emph {et~al.}(2016)\citenamefont {Wan},
  \citenamefont {Yuan}, \citenamefont {Jin}, \citenamefont {Wang},
  \citenamefont {Yang}, \citenamefont {Yu},\ and\ \citenamefont
  {Shao}}]{Wan2016}%
  \BibitemOpen
  \bibfield  {author} {\bibinfo {author} {\bibfnamefont {M.}~\bibnamefont
  {Wan}}, \bibinfo {author} {\bibfnamefont {D.}~\bibnamefont {Yuan}}, \bibinfo
  {author} {\bibfnamefont {C.}~\bibnamefont {Jin}}, \bibinfo {author}
  {\bibfnamefont {F.}~\bibnamefont {Wang}}, \bibinfo {author} {\bibfnamefont
  {Y.}~\bibnamefont {Yang}}, \bibinfo {author} {\bibfnamefont {Y.}~\bibnamefont
  {Yu}}, \ and\ \bibinfo {author} {\bibfnamefont {J.}~\bibnamefont {Shao}},\
  }\href {\doibase 10.1063/1.4955498} {\bibfield  {journal} {\bibinfo
  {journal} {The Journal of Chemical Physics}\ }\textbf {\bibinfo {volume}
  {145}},\ \bibinfo {pages} {024309} (\bibinfo {year} {2016})}\BibitemShut
  {NoStop}%
\bibitem [{\citenamefont {Daniel}\ \emph {et~al.}(2021)\citenamefont {Daniel},
  \citenamefont {Wang}, \citenamefont {Rodriguez}, \citenamefont {Lewis},
  \citenamefont {Teplukhin}, \citenamefont {Kendrick}, \citenamefont
  {Bardeen},\ and\ \citenamefont {Hemmerling}}]{Daniel2021}%
  \BibitemOpen
  \bibfield  {author} {\bibinfo {author} {\bibfnamefont {J.~R.}\ \bibnamefont
  {Daniel}}, \bibinfo {author} {\bibfnamefont {C.}~\bibnamefont {Wang}},
  \bibinfo {author} {\bibfnamefont {K.}~\bibnamefont {Rodriguez}}, \bibinfo
  {author} {\bibfnamefont {T.}~\bibnamefont {Lewis}}, \bibinfo {author}
  {\bibfnamefont {A.}~\bibnamefont {Teplukhin}}, \bibinfo {author}
  {\bibfnamefont {B.}~\bibnamefont {Kendrick}}, \bibinfo {author}
  {\bibfnamefont {C.}~\bibnamefont {Bardeen}}, \ and\ \bibinfo {author}
  {\bibfnamefont {B.}~\bibnamefont {Hemmerling}},\ }\href {\doibase
  10.1103/PhysRevA.104.012801} {\bibfield  {journal} {\bibinfo  {journal}
  {Physical Review A}\ }\textbf {\bibinfo {volume} {104}},\ \bibinfo {pages}
  {012801} (\bibinfo {year} {2021})}\BibitemShut {NoStop}%
\bibitem [{\citenamefont {Rogowski}\ and\ \citenamefont
  {Fontijn}(1987)}]{Rogowski1987}%
  \BibitemOpen
  \bibfield  {author} {\bibinfo {author} {\bibfnamefont {D.~F.}\ \bibnamefont
  {Rogowski}}\ and\ \bibinfo {author} {\bibfnamefont {A.}~\bibnamefont
  {Fontijn}},\ }\href {\doibase 10.1016/0009-2614(87)80207-5} {\bibfield
  {journal} {\bibinfo  {journal} {Chemical Physics Letters}\ }\textbf {\bibinfo
  {volume} {137}},\ \bibinfo {pages} {219} (\bibinfo {year}
  {1987})}\BibitemShut {NoStop}%
\bibitem [{\citenamefont {Langhoff}\ \emph
  {et~al.}(1988{\natexlab{a}})\citenamefont {Langhoff}, \citenamefont
  {Bauschlicher},\ and\ \citenamefont {Taylor}}]{Langhoff1988}%
  \BibitemOpen
  \bibfield  {author} {\bibinfo {author} {\bibfnamefont {S.~R.}\ \bibnamefont
  {Langhoff}}, \bibinfo {author} {\bibfnamefont {C.~W.}\ \bibnamefont
  {Bauschlicher}}, \ and\ \bibinfo {author} {\bibfnamefont {P.~R.}\
  \bibnamefont {Taylor}},\ }\href {\doibase 10.1063/1.454531} {\bibfield
  {journal} {\bibinfo  {journal} {The Journal of Chemical Physics}\ }\textbf
  {\bibinfo {volume} {88}},\ \bibinfo {pages} {5715} (\bibinfo {year}
  {1988}{\natexlab{a}})}\BibitemShut {NoStop}%
\bibitem [{\citenamefont {Ren}\ \emph {et~al.}(2021)\citenamefont {Ren},
  \citenamefont {Xiao}, \citenamefont {Liu},\ and\ \citenamefont
  {Yan}}]{Ren2021}%
  \BibitemOpen
  \bibfield  {author} {\bibinfo {author} {\bibfnamefont {X.~Y.}\ \bibnamefont
  {Ren}}, \bibinfo {author} {\bibfnamefont {Z.~Y.}\ \bibnamefont {Xiao}},
  \bibinfo {author} {\bibfnamefont {Y.}~\bibnamefont {Liu}}, \ and\ \bibinfo
  {author} {\bibfnamefont {B.}~\bibnamefont {Yan}},\ }\href {\doibase
  10.1088/1674-1056/abd46a} {\bibfield  {journal} {\bibinfo  {journal} {Chinese
  Physics B}\ }\textbf {\bibinfo {volume} {30}},\ \bibinfo {pages} {053101}
  (\bibinfo {year} {2021})}\BibitemShut {NoStop}%
\bibitem [{\citenamefont {Mes}\ \emph {et~al.}(2003)\citenamefont {Mes},
  \citenamefont {van Duijn}, \citenamefont {Zinkstok}, \citenamefont {Witte},\
  and\ \citenamefont {Hogervorst}}]{Mes2003b}%
  \BibitemOpen
  \bibfield  {author} {\bibinfo {author} {\bibfnamefont {J.}~\bibnamefont
  {Mes}}, \bibinfo {author} {\bibfnamefont {E.~J.}\ \bibnamefont {van Duijn}},
  \bibinfo {author} {\bibfnamefont {R.}~\bibnamefont {Zinkstok}}, \bibinfo
  {author} {\bibfnamefont {S.}~\bibnamefont {Witte}}, \ and\ \bibinfo {author}
  {\bibfnamefont {W.}~\bibnamefont {Hogervorst}},\ }\href {\doibase
  10.1063/1.1584515} {\bibfield  {journal} {\bibinfo  {journal} {Applied
  Physics Letters}\ }\textbf {\bibinfo {volume} {82}},\ \bibinfo {pages} {4423}
  (\bibinfo {year} {2003})}\BibitemShut {NoStop}%
\bibitem [{\citenamefont {Ostroumov}\ and\ \citenamefont
  {Seelert}(2008)}]{Ostroumov2008}%
  \BibitemOpen
  \bibfield  {author} {\bibinfo {author} {\bibfnamefont {V.}~\bibnamefont
  {Ostroumov}}\ and\ \bibinfo {author} {\bibfnamefont {W.}~\bibnamefont
  {Seelert}},\ }\href {\doibase 10.1117/12.767511} {\bibfield  {journal}
  {\bibinfo  {journal} {Solid State Lasers XVII: Technology and Devices}\
  }\textbf {\bibinfo {volume} {6871}},\ \bibinfo {pages} {68711K} (\bibinfo
  {year} {2008})}\BibitemShut {NoStop}%
\bibitem [{\citenamefont {Burkley}\ \emph {et~al.}(2019)\citenamefont
  {Burkley}, \citenamefont {Brandt}, \citenamefont {Rasor}, \citenamefont
  {Cooper},\ and\ \citenamefont {Yost}}]{Burkley2019}%
  \BibitemOpen
  \bibfield  {author} {\bibinfo {author} {\bibfnamefont {Z.}~\bibnamefont
  {Burkley}}, \bibinfo {author} {\bibfnamefont {A.~D.}\ \bibnamefont {Brandt}},
  \bibinfo {author} {\bibfnamefont {C.}~\bibnamefont {Rasor}}, \bibinfo
  {author} {\bibfnamefont {S.~F.}\ \bibnamefont {Cooper}}, \ and\ \bibinfo
  {author} {\bibfnamefont {D.~C.}\ \bibnamefont {Yost}},\ }\href {\doibase
  10.1364/AO.58.001657} {\bibfield  {journal} {\bibinfo  {journal} {Applied
  Optics}\ ,\ \bibinfo {pages} {1657}} (\bibinfo {year} {2019})}\BibitemShut
  {NoStop}%
\bibitem [{\citenamefont {Burkley}\ \emph {et~al.}(2021)\citenamefont
  {Burkley}, \citenamefont {Borges}, \citenamefont {Ohayon}, \citenamefont
  {Golovizin}, \citenamefont {Zhang},\ and\ \citenamefont
  {Crivelli}}]{Burkley2021}%
  \BibitemOpen
  \bibfield  {author} {\bibinfo {author} {\bibfnamefont {Z.}~\bibnamefont
  {Burkley}}, \bibinfo {author} {\bibfnamefont {L.~d.~S.}\ \bibnamefont
  {Borges}}, \bibinfo {author} {\bibfnamefont {B.}~\bibnamefont {Ohayon}},
  \bibinfo {author} {\bibfnamefont {A.}~\bibnamefont {Golovizin}}, \bibinfo
  {author} {\bibfnamefont {J.}~\bibnamefont {Zhang}}, \ and\ \bibinfo {author}
  {\bibfnamefont {P.}~\bibnamefont {Crivelli}},\ }\href {\doibase
  10.1364/OE.432552} {\bibfield  {journal} {\bibinfo  {journal} {Optics
  Express}\ }\textbf {\bibinfo {volume} {29}},\ \bibinfo {pages} {27450}
  (\bibinfo {year} {2021})}\BibitemShut {NoStop}%
\bibitem [{\citenamefont {McCarron}\ \emph {et~al.}(2021)\citenamefont
  {McCarron}, \citenamefont {Shaw},\ and\ \citenamefont
  {Hannig}}]{McCarron2021}%
  \BibitemOpen
  \bibfield  {author} {\bibinfo {author} {\bibfnamefont {D.~J.}\ \bibnamefont
  {McCarron}}, \bibinfo {author} {\bibfnamefont {J.~C.}\ \bibnamefont {Shaw}},
  \ and\ \bibinfo {author} {\bibfnamefont {S.}~\bibnamefont {Hannig}},\ }\href
  {\doibase 10.1364/OE.441741} {\bibfield  {journal} {\bibinfo  {journal}
  {Optics Express}\ }\textbf {\bibinfo {volume} {29}},\ \bibinfo {pages}
  {37140} (\bibinfo {year} {2021})}\BibitemShut {NoStop}%
\bibitem [{\citenamefont {Jevons}(1913)}]{Jevons1913}%
  \BibitemOpen
  \bibfield  {author} {\bibinfo {author} {\bibfnamefont {W.}~\bibnamefont
  {Jevons}},\ }\href {\doibase 10.1098/RSPA.1913.0077} {\bibfield  {journal}
  {\bibinfo  {journal} {Proceedings of the Royal Society of London. Series A,
  Containing Papers of a Mathematical and Physical Character}\ }\textbf
  {\bibinfo {volume} {89}},\ \bibinfo {pages} {187} (\bibinfo {year}
  {1913})}\BibitemShut {NoStop}%
\bibitem [{\citenamefont {Jevons}(1924)}]{Jevons1924}%
  \BibitemOpen
  \bibfield  {author} {\bibinfo {author} {\bibfnamefont {W.}~\bibnamefont
  {Jevons}},\ }\href {\doibase 10.1098/RSPA.1924.0061} {\bibfield  {journal}
  {\bibinfo  {journal} {Proceedings of the Royal Society of London. Series A,
  Containing Papers of a Mathematical and Physical Character}\ }\textbf
  {\bibinfo {volume} {106}},\ \bibinfo {pages} {174} (\bibinfo {year}
  {1924})}\BibitemShut {NoStop}%
\bibitem [{\citenamefont {Bhaduri}\ and\ \citenamefont
  {Fowler}(1934)}]{Bhaduri1934}%
  \BibitemOpen
  \bibfield  {author} {\bibinfo {author} {\bibfnamefont {B.~N.}\ \bibnamefont
  {Bhaduri}}\ and\ \bibinfo {author} {\bibfnamefont {A.}~\bibnamefont
  {Fowler}},\ }\href {\doibase 10.1098/rspa.1934.0099} {\bibfield  {journal}
  {\bibinfo  {journal} {Proceedings of the Royal Society of London A}\ }\textbf
  {\bibinfo {volume} {145}},\ \bibinfo {pages} {321} (\bibinfo {year}
  {1934})}\BibitemShut {NoStop}%
\bibitem [{\citenamefont {Holst}(1935)}]{Holst1935}%
  \BibitemOpen
  \bibfield  {author} {\bibinfo {author} {\bibfnamefont {W.}~\bibnamefont
  {Holst}},\ }\href {\doibase 10.1007/BF01341191} {\bibfield  {journal}
  {\bibinfo  {journal} {Zeitschrift für Physik}\ }\textbf {\bibinfo {volume}
  {93}},\ \bibinfo {pages} {55} (\bibinfo {year} {1935})}\BibitemShut {NoStop}%
\bibitem [{\citenamefont {Mahanti}(1934)}]{Mahanti1934}%
  \BibitemOpen
  \bibfield  {author} {\bibinfo {author} {\bibfnamefont {P.~C.}\ \bibnamefont
  {Mahanti}},\ }\href {\doibase 10.1007/BF01351785} {\bibfield  {journal}
  {\bibinfo  {journal} {Zeitschrift für Physik}\ }\textbf {\bibinfo {volume}
  {88}},\ \bibinfo {pages} {550} (\bibinfo {year} {1934})}\BibitemShut
  {NoStop}%
\bibitem [{\citenamefont {Miescher}(1936)}]{Miescher1936}%
  \BibitemOpen
  \bibfield  {author} {\bibinfo {author} {\bibfnamefont {E.}~\bibnamefont
  {Miescher}},\ }\href {\doibase 10.5169/SEALS-110658} {\bibfield  {journal}
  {\bibinfo  {journal} {Helvetica Physica Acta}\ }\textbf {\bibinfo {volume}
  {9}},\ \bibinfo {pages} {693} (\bibinfo {year} {1936})}\BibitemShut {NoStop}%
\bibitem [{\citenamefont {Sharma}(1951)}]{Sharma1951}%
  \BibitemOpen
  \bibfield  {author} {\bibinfo {author} {\bibfnamefont {D.}~\bibnamefont
  {Sharma}},\ }\href {\doibase 10.1086/145389} {\bibfield  {journal} {\bibinfo
  {journal} {The Astrophysical Journal}\ }\textbf {\bibinfo {volume} {113}},\
  \bibinfo {pages} {210} (\bibinfo {year} {1951})}\BibitemShut {NoStop}%
\bibitem [{\citenamefont {Barrow}(1954)}]{Barrow1954}%
  \BibitemOpen
  \bibfield  {author} {\bibinfo {author} {\bibfnamefont {R.~F.}\ \bibnamefont
  {Barrow}},\ }\href {\doibase 10.1063/1.1740123} {\bibfield  {journal}
  {\bibinfo  {journal} {The Journal of Chemical Physics}\ }\textbf {\bibinfo
  {volume} {22}},\ \bibinfo {pages} {573} (\bibinfo {year} {1954})}\BibitemShut
  {NoStop}%
\bibitem [{\citenamefont {Reddy}\ and\ \citenamefont {Rao}(1957)}]{Reddy1957}%
  \BibitemOpen
  \bibfield  {author} {\bibinfo {author} {\bibfnamefont {S.~P.}\ \bibnamefont
  {Reddy}}\ and\ \bibinfo {author} {\bibfnamefont {P.~T.}\ \bibnamefont
  {Rao}},\ }\href {\doibase 10.1139/p57-100} {\bibfield  {journal} {\bibinfo
  {journal} {Canadian Journal of Physics}\ }\textbf {\bibinfo {volume} {35}},\
  \bibinfo {pages} {912} (\bibinfo {year} {1957})}\BibitemShut {NoStop}%
\bibitem [{\citenamefont {Barrow}(1960)}]{Barrow1960}%
  \BibitemOpen
  \bibfield  {author} {\bibinfo {author} {\bibfnamefont {R.~F.}\ \bibnamefont
  {Barrow}},\ }\href {\doibase 10.1039/tf9605600952} {\bibfield  {journal}
  {\bibinfo  {journal} {Transactions of the Faraday Society}\ }\textbf
  {\bibinfo {volume} {56}},\ \bibinfo {pages} {952} (\bibinfo {year}
  {1960})}\BibitemShut {NoStop}%
\bibitem [{\citenamefont {Lide}(1965)}]{Lide1965}%
  \BibitemOpen
  \bibfield  {author} {\bibinfo {author} {\bibfnamefont {D.~R.}\ \bibnamefont
  {Lide}},\ }\href {\doibase 10.1063/1.1696035} {\bibfield  {journal} {\bibinfo
   {journal} {The Journal of Chemical Physics}\ }\textbf {\bibinfo {volume}
  {42}},\ \bibinfo {pages} {1013} (\bibinfo {year} {1965})}\BibitemShut
  {NoStop}%
\bibitem [{\citenamefont {Lide}(1967)}]{Lide1967}%
  \BibitemOpen
  \bibfield  {author} {\bibinfo {author} {\bibfnamefont {D.~R.}\ \bibnamefont
  {Lide}},\ }\href {\doibase 10.1063/1.1840811} {\bibfield  {journal} {\bibinfo
   {journal} {The Journal of Chemical Physics}\ }\textbf {\bibinfo {volume}
  {46}},\ \bibinfo {pages} {1224} (\bibinfo {year} {1967})}\BibitemShut
  {NoStop}%
\bibitem [{\citenamefont {Wyse}\ and\ \citenamefont {Gordy}(1972)}]{Wyse1972}%
  \BibitemOpen
  \bibfield  {author} {\bibinfo {author} {\bibfnamefont {F.~C.}\ \bibnamefont
  {Wyse}}\ and\ \bibinfo {author} {\bibfnamefont {W.}~\bibnamefont {Gordy}},\
  }\href {\doibase 10.1063/1.1677509} {\bibfield  {journal} {\bibinfo
  {journal} {The Journal of Chemical Physics}\ }\textbf {\bibinfo {volume}
  {56}},\ \bibinfo {pages} {2130} (\bibinfo {year} {1972})}\BibitemShut
  {NoStop}%
\bibitem [{\citenamefont {Hoeft}\ \emph {et~al.}(1973)\citenamefont {Hoeft},
  \citenamefont {Törring},\ and\ \citenamefont {Tiemann}}]{Hoeft1973}%
  \BibitemOpen
  \bibfield  {author} {\bibinfo {author} {\bibfnamefont {J.}~\bibnamefont
  {Hoeft}}, \bibinfo {author} {\bibfnamefont {T.}~\bibnamefont {Törring}}, \
  and\ \bibinfo {author} {\bibfnamefont {E.}~\bibnamefont {Tiemann}},\ }\href
  {\doibase 10.1515/zna-1973-0705} {\bibfield  {journal} {\bibinfo  {journal}
  {Zeitschrift für Naturforschung A}\ }\textbf {\bibinfo {volume} {28}},\
  \bibinfo {pages} {1066} (\bibinfo {year} {1973})}\BibitemShut {NoStop}%
\bibitem [{\citenamefont {Lovas}\ and\ \citenamefont
  {Tiemann}(1974)}]{Lovas1974}%
  \BibitemOpen
  \bibfield  {author} {\bibinfo {author} {\bibfnamefont {F.~J.}\ \bibnamefont
  {Lovas}}\ and\ \bibinfo {author} {\bibfnamefont {E.}~\bibnamefont
  {Tiemann}},\ }\href {\doibase 10.1063/1.3253146} {\bibfield  {journal}
  {\bibinfo  {journal} {Journal of Physical and Chemical Reference Data}\
  }\textbf {\bibinfo {volume} {3}},\ \bibinfo {pages} {609} (\bibinfo {year}
  {1974})}\BibitemShut {NoStop}%
\bibitem [{\citenamefont {Schnöckel}(1976)}]{Schnoeckel1976}%
  \BibitemOpen
  \bibfield  {author} {\bibinfo {author} {\bibfnamefont {H.}~\bibnamefont
  {Schnöckel}},\ }\href {\doibase 10.1515/znb-1976-0925} {\bibfield  {journal}
  {\bibinfo  {journal} {Zeitschrift für Naturforschung B}\ }\textbf {\bibinfo
  {volume} {31}},\ \bibinfo {pages} {1291} (\bibinfo {year}
  {1976})}\BibitemShut {NoStop}%
\bibitem [{\citenamefont {Tsunoda}\ \emph {et~al.}(1978)\citenamefont
  {Tsunoda}, \citenamefont {Fujiwara},\ and\ \citenamefont
  {Fuwa}}]{Tsunoda1978}%
  \BibitemOpen
  \bibfield  {author} {\bibinfo {author} {\bibfnamefont {K.~I.}\ \bibnamefont
  {Tsunoda}}, \bibinfo {author} {\bibfnamefont {K.}~\bibnamefont {Fujiwara}}, \
  and\ \bibinfo {author} {\bibfnamefont {K.}~\bibnamefont {Fuwa}},\ }\href
  {\doibase 10.1021/ac50029a011} {\bibfield  {journal} {\bibinfo  {journal}
  {Analytical Chemistry}\ }\textbf {\bibinfo {volume} {50}},\ \bibinfo {pages}
  {861} (\bibinfo {year} {1978})}\BibitemShut {NoStop}%
\bibitem [{\citenamefont {Ram}\ \emph {et~al.}(1982)\citenamefont {Ram},
  \citenamefont {Rai}, \citenamefont {Upadhya},\ and\ \citenamefont
  {Rai}}]{Ram1982}%
  \BibitemOpen
  \bibfield  {author} {\bibinfo {author} {\bibfnamefont {R.~S.}\ \bibnamefont
  {Ram}}, \bibinfo {author} {\bibfnamefont {S.~B.}\ \bibnamefont {Rai}},
  \bibinfo {author} {\bibfnamefont {K.~N.}\ \bibnamefont {Upadhya}}, \ and\
  \bibinfo {author} {\bibfnamefont {D.~K.}\ \bibnamefont {Rai}},\ }\href
  {\doibase 10.1088/0031-8949/26/5/007} {\bibfield  {journal} {\bibinfo
  {journal} {Physica Scripta}\ }\textbf {\bibinfo {volume} {26}},\ \bibinfo
  {pages} {383} (\bibinfo {year} {1982})}\BibitemShut {NoStop}%
\bibitem [{\citenamefont {Mahieu}\ \emph
  {et~al.}(1989{\natexlab{a}})\citenamefont {Mahieu}, \citenamefont {Dubois},\
  and\ \citenamefont {Bredohl}}]{Mahieu1989}%
  \BibitemOpen
  \bibfield  {author} {\bibinfo {author} {\bibfnamefont {E.}~\bibnamefont
  {Mahieu}}, \bibinfo {author} {\bibfnamefont {I.}~\bibnamefont {Dubois}}, \
  and\ \bibinfo {author} {\bibfnamefont {H.}~\bibnamefont {Bredohl}},\ }\href
  {\doibase 10.1016/0022-2852(89)90319-6} {\bibfield  {journal} {\bibinfo
  {journal} {Journal of Molecular Spectroscopy}\ }\textbf {\bibinfo {volume}
  {134}},\ \bibinfo {pages} {317} (\bibinfo {year}
  {1989}{\natexlab{a}})}\BibitemShut {NoStop}%
\bibitem [{\citenamefont {Mahieu}\ \emph
  {et~al.}(1989{\natexlab{b}})\citenamefont {Mahieu}, \citenamefont {Dubois},\
  and\ \citenamefont {Bredohl}}]{Mahieu1989a}%
  \BibitemOpen
  \bibfield  {author} {\bibinfo {author} {\bibfnamefont {E.}~\bibnamefont
  {Mahieu}}, \bibinfo {author} {\bibfnamefont {I.}~\bibnamefont {Dubois}}, \
  and\ \bibinfo {author} {\bibfnamefont {H.}~\bibnamefont {Bredohl}},\ }\href
  {\doibase 10.1016/0022-2852(89)90116-1} {\bibfield  {journal} {\bibinfo
  {journal} {Journal of Molecular Spectroscopy}\ }\textbf {\bibinfo {volume}
  {138}},\ \bibinfo {pages} {264} (\bibinfo {year}
  {1989}{\natexlab{b}})}\BibitemShut {NoStop}%
\bibitem [{\citenamefont {Dearden}\ \emph {et~al.}(1993)\citenamefont
  {Dearden}, \citenamefont {Johnson},\ and\ \citenamefont
  {Hudgens}}]{Dearden1993}%
  \BibitemOpen
  \bibfield  {author} {\bibinfo {author} {\bibfnamefont {D.~V.}\ \bibnamefont
  {Dearden}}, \bibinfo {author} {\bibfnamefont {R.~D.}\ \bibnamefont
  {Johnson}}, \ and\ \bibinfo {author} {\bibfnamefont {J.~W.}\ \bibnamefont
  {Hudgens}},\ }\href {\doibase 10.1063/1.465682} {\bibfield  {journal}
  {\bibinfo  {journal} {The Journal of Chemical Physics}\ }\textbf {\bibinfo
  {volume} {99}},\ \bibinfo {pages} {7521} (\bibinfo {year}
  {1993})}\BibitemShut {NoStop}%
\bibitem [{\citenamefont {Ogilvie}\ and\ \citenamefont
  {Liao}(1994)}]{Ogilvie1994}%
  \BibitemOpen
  \bibfield  {author} {\bibinfo {author} {\bibfnamefont {J.~F.}\ \bibnamefont
  {Ogilvie}}\ and\ \bibinfo {author} {\bibfnamefont {S.~C.}\ \bibnamefont
  {Liao}},\ }\href {https://link.springer.com/article/10.1007/BF03156408}
  {\bibfield  {journal} {\bibinfo  {journal} {Acta Physica Hungarica}\ }\textbf
  {\bibinfo {volume} {74}},\ \bibinfo {pages} {365} (\bibinfo {year}
  {1994})}\BibitemShut {NoStop}%
\bibitem [{\citenamefont {Hedderich}\ \emph {et~al.}(1993)\citenamefont
  {Hedderich}, \citenamefont {Dulick},\ and\ \citenamefont
  {Bernath}}]{Hedderich1993}%
  \BibitemOpen
  \bibfield  {author} {\bibinfo {author} {\bibfnamefont {H.~G.}\ \bibnamefont
  {Hedderich}}, \bibinfo {author} {\bibfnamefont {M.}~\bibnamefont {Dulick}}, \
  and\ \bibinfo {author} {\bibfnamefont {P.~F.}\ \bibnamefont {Bernath}},\
  }\href {\doibase 10.1063/1.465611} {\bibfield  {journal} {\bibinfo  {journal}
  {The Journal of Chemical Physics}\ }\textbf {\bibinfo {volume} {99}},\
  \bibinfo {pages} {8363} (\bibinfo {year} {1993})}\BibitemShut {NoStop}%
\bibitem [{\citenamefont {Hensel}\ \emph {et~al.}(1993)\citenamefont {Hensel},
  \citenamefont {Styger}, \citenamefont {Jäger}, \citenamefont {Merer},\ and\
  \citenamefont {Gerry}}]{Hensel1993}%
  \BibitemOpen
  \bibfield  {author} {\bibinfo {author} {\bibfnamefont {K.~D.}\ \bibnamefont
  {Hensel}}, \bibinfo {author} {\bibfnamefont {C.}~\bibnamefont {Styger}},
  \bibinfo {author} {\bibfnamefont {W.}~\bibnamefont {Jäger}}, \bibinfo
  {author} {\bibfnamefont {A.~J.}\ \bibnamefont {Merer}}, \ and\ \bibinfo
  {author} {\bibfnamefont {M.~C.~L.}\ \bibnamefont {Gerry}},\ }\href {\doibase
  10.1063/1.465141} {\bibfield  {journal} {\bibinfo  {journal} {The Journal of
  Chemical Physics}\ }\textbf {\bibinfo {volume} {99}},\ \bibinfo {pages}
  {3320} (\bibinfo {year} {1993})}\BibitemShut {NoStop}%
\bibitem [{\citenamefont {Saksena}\ \emph {et~al.}(1998)\citenamefont
  {Saksena}, \citenamefont {Dixit},\ and\ \citenamefont {Singh}}]{Saksena1998}%
  \BibitemOpen
  \bibfield  {author} {\bibinfo {author} {\bibfnamefont {M.}~\bibnamefont
  {Saksena}}, \bibinfo {author} {\bibfnamefont {V.}~\bibnamefont {Dixit}}, \
  and\ \bibinfo {author} {\bibfnamefont {M.}~\bibnamefont {Singh}},\ }\href
  {\doibase 10.1006/JMSP.1997.7442} {\bibfield  {journal} {\bibinfo  {journal}
  {Journal of Molecular Spectroscopy}\ }\textbf {\bibinfo {volume} {187}},\
  \bibinfo {pages} {1} (\bibinfo {year} {1998})}\BibitemShut {NoStop}%
\bibitem [{\citenamefont {Parvinen}\ and\ \citenamefont
  {Lajunen}(1998)}]{Parvinen1998}%
  \BibitemOpen
  \bibfield  {author} {\bibinfo {author} {\bibfnamefont {P.}~\bibnamefont
  {Parvinen}}\ and\ \bibinfo {author} {\bibfnamefont {L.~H.}\ \bibnamefont
  {Lajunen}},\ }\href {\doibase 10.1080/00387019808007452} {\bibfield
  {journal} {\bibinfo  {journal} {Spectroscopy Letters}\ }\textbf {\bibinfo
  {volume} {31}},\ \bibinfo {pages} {1761} (\bibinfo {year}
  {1998})}\BibitemShut {NoStop}%
\bibitem [{\citenamefont {Brites}\ \emph {et~al.}(2008)\citenamefont {Brites},
  \citenamefont {Hammoutene},\ and\ \citenamefont {Hochlaf}}]{Brites2008b}%
  \BibitemOpen
  \bibfield  {author} {\bibinfo {author} {\bibfnamefont {V.}~\bibnamefont
  {Brites}}, \bibinfo {author} {\bibfnamefont {D.}~\bibnamefont {Hammoutene}},
  \ and\ \bibinfo {author} {\bibfnamefont {M.}~\bibnamefont {Hochlaf}},\ }\href
  {\doibase 10.1021/jp805508f} {\bibfield  {journal} {\bibinfo  {journal} {The
  Journal of Physical Chemistry A}\ }\textbf {\bibinfo {volume} {112}},\
  \bibinfo {pages} {13419} (\bibinfo {year} {2008})}\BibitemShut {NoStop}%
\bibitem [{\citenamefont {Pamboundom}\ \emph {et~al.}(2016)\citenamefont
  {Pamboundom}, \citenamefont {Tchakoua},\ and\ \citenamefont
  {Nsangou}}]{Pamboundom2016}%
  \BibitemOpen
  \bibfield  {author} {\bibinfo {author} {\bibfnamefont {M.}~\bibnamefont
  {Pamboundom}}, \bibinfo {author} {\bibfnamefont {T.}~\bibnamefont
  {Tchakoua}}, \ and\ \bibinfo {author} {\bibfnamefont {M.}~\bibnamefont
  {Nsangou}},\ }\href {\doibase 10.1007/s10509-016-2735-y} {\bibfield
  {journal} {\bibinfo  {journal} {Astrophysics and Space Science}\ }\textbf
  {\bibinfo {volume} {361}},\ \bibinfo {pages} {150} (\bibinfo {year}
  {2016})}\BibitemShut {NoStop}%
\bibitem [{\citenamefont {Preston}\ \emph {et~al.}(2022)\citenamefont
  {Preston}, \citenamefont {Jackson},\ and\ \citenamefont
  {Mawhorter}}]{Preston2022}%
  \BibitemOpen
  \bibfield  {author} {\bibinfo {author} {\bibfnamefont {A.}~\bibnamefont
  {Preston}}, \bibinfo {author} {\bibfnamefont {S.}~\bibnamefont {Jackson}}, \
  and\ \bibinfo {author} {\bibfnamefont {R.}~\bibnamefont {Mawhorter}},\ }\href
  {\doibase 10.1016/j.cplett.2022.140089} {\bibfield  {journal} {\bibinfo
  {journal} {Chemical Physics Letters}\ }\textbf {\bibinfo {volume} {807}},\
  \bibinfo {pages} {140089} (\bibinfo {year} {2022})}\BibitemShut {NoStop}%
\bibitem [{\citenamefont {Langhoff}\ \emph
  {et~al.}(1988{\natexlab{b}})\citenamefont {Langhoff}, \citenamefont
  {Bauschlicher},\ and\ \citenamefont {Taylor}}]{Langhoff1988a}%
  \BibitemOpen
  \bibfield  {author} {\bibinfo {author} {\bibfnamefont {S.~R.}\ \bibnamefont
  {Langhoff}}, \bibinfo {author} {\bibfnamefont {C.~W.}\ \bibnamefont
  {Bauschlicher}}, \ and\ \bibinfo {author} {\bibfnamefont {P.~R.}\
  \bibnamefont {Taylor}},\ }\href {\doibase 10.1063/1.455757} {\bibfield
  {journal} {\bibinfo  {journal} {The Journal of Chemical Physics}\ }\textbf
  {\bibinfo {volume} {89}},\ \bibinfo {pages} {7650} (\bibinfo {year}
  {1988}{\natexlab{b}})}\BibitemShut {NoStop}%
\bibitem [{\citenamefont {Andreazza}\ \emph {et~al.}(2018)\citenamefont
  {Andreazza}, \citenamefont {de~Almeida},\ and\ \citenamefont
  {Vichietti}}]{Andreazza2018}%
  \BibitemOpen
  \bibfield  {author} {\bibinfo {author} {\bibfnamefont {C.~M.}\ \bibnamefont
  {Andreazza}}, \bibinfo {author} {\bibfnamefont {A.~A.}\ \bibnamefont
  {de~Almeida}}, \ and\ \bibinfo {author} {\bibfnamefont {R.~M.}\ \bibnamefont
  {Vichietti}},\ }\href {\doibase 10.1093/mnras/sty753} {\bibfield  {journal}
  {\bibinfo  {journal} {Monthly Notices of the Royal Astronomical Society}\
  }\textbf {\bibinfo {volume} {477}},\ \bibinfo {pages} {548} (\bibinfo {year}
  {2018})}\BibitemShut {NoStop}%
\bibitem [{\citenamefont {Yousefi}\ and\ \citenamefont
  {Bernath}(2018)}]{Yousefi2018}%
  \BibitemOpen
  \bibfield  {author} {\bibinfo {author} {\bibfnamefont {M.}~\bibnamefont
  {Yousefi}}\ and\ \bibinfo {author} {\bibfnamefont {P.~F.}\ \bibnamefont
  {Bernath}},\ }\href {\doibase 10.3847/1538-4365/aacc6a} {\bibfield  {journal}
  {\bibinfo  {journal} {The Astrophysical Journal Supplement Series}\ }\textbf
  {\bibinfo {volume} {237}},\ \bibinfo {pages} {8} (\bibinfo {year}
  {2018})}\BibitemShut {NoStop}%
\bibitem [{\citenamefont {Aerts}\ and\ \citenamefont
  {Brown}(2019)}]{Aerts2019}%
  \BibitemOpen
  \bibfield  {author} {\bibinfo {author} {\bibfnamefont {A.}~\bibnamefont
  {Aerts}}\ and\ \bibinfo {author} {\bibfnamefont {A.}~\bibnamefont {Brown}},\
  }\href {\doibase 10.1063/1.5097151} {\bibfield  {journal} {\bibinfo
  {journal} {The Journal of Chemical Physics}\ }\textbf {\bibinfo {volume}
  {150}},\ \bibinfo {pages} {224302} (\bibinfo {year} {2019})}\BibitemShut
  {NoStop}%
\bibitem [{\citenamefont {Xu}\ \emph {et~al.}(2020)\citenamefont {Xu},
  \citenamefont {Zhang},\ and\ \citenamefont {Zhang}}]{Xu2020}%
  \BibitemOpen
  \bibfield  {author} {\bibinfo {author} {\bibfnamefont {J.~G.}\ \bibnamefont
  {Xu}}, \bibinfo {author} {\bibfnamefont {C.~Y.}\ \bibnamefont {Zhang}}, \
  and\ \bibinfo {author} {\bibfnamefont {Y.~G.}\ \bibnamefont {Zhang}},\ }\href
  {\doibase 10.1088/1674-1056/ab6c46} {\bibfield  {journal} {\bibinfo
  {journal} {Chinese Physics B}\ }\textbf {\bibinfo {volume} {29}},\ \bibinfo
  {pages} {033102} (\bibinfo {year} {2020})}\BibitemShut {NoStop}%
\bibitem [{\citenamefont {Zhang}\ \emph {et~al.}(2021)\citenamefont {Zhang},
  \citenamefont {Li},\ and\ \citenamefont {Ma}}]{Zhang2021b}%
  \BibitemOpen
  \bibfield  {author} {\bibinfo {author} {\bibfnamefont {J.}~\bibnamefont
  {Zhang}}, \bibinfo {author} {\bibfnamefont {H.}~\bibnamefont {Li}}, \ and\
  \bibinfo {author} {\bibfnamefont {Y.}~\bibnamefont {Ma}},\ }\href {\doibase
  10.1016/j.comptc.2021.113307} {\bibfield  {journal} {\bibinfo  {journal}
  {Computational and Theoretical Chemistry}\ }\textbf {\bibinfo {volume}
  {1202}},\ \bibinfo {pages} {113307} (\bibinfo {year} {2021})}\BibitemShut
  {NoStop}%
\bibitem [{\citenamefont {Qin}\ \emph {et~al.}(2021)\citenamefont {Qin},
  \citenamefont {Bai},\ and\ \citenamefont {Liu}}]{Qin2021}%
  \BibitemOpen
  \bibfield  {author} {\bibinfo {author} {\bibfnamefont {Z.}~\bibnamefont
  {Qin}}, \bibinfo {author} {\bibfnamefont {T.}~\bibnamefont {Bai}}, \ and\
  \bibinfo {author} {\bibfnamefont {L.}~\bibnamefont {Liu}},\ }\href {\doibase
  10.1093/MNRAS/STAB2655} {\bibfield  {journal} {\bibinfo  {journal} {Monthly
  Notices of the Royal Astronomical Society}\ }\textbf {\bibinfo {volume}
  {508}},\ \bibinfo {pages} {2848} (\bibinfo {year} {2021})}\BibitemShut
  {NoStop}%
\bibitem [{\citenamefont {Bala}\ \emph {et~al.}(2023)\citenamefont {Bala},
  \citenamefont {Prasannaa}, \citenamefont {Chakravarti}, \citenamefont
  {Mukherjee},\ and\ \citenamefont {Das}}]{Bala2023}%
  \BibitemOpen
  \bibfield  {author} {\bibinfo {author} {\bibfnamefont {R.}~\bibnamefont
  {Bala}}, \bibinfo {author} {\bibfnamefont {V.~S.}\ \bibnamefont {Prasannaa}},
  \bibinfo {author} {\bibfnamefont {D.}~\bibnamefont {Chakravarti}}, \bibinfo
  {author} {\bibfnamefont {D.}~\bibnamefont {Mukherjee}}, \ and\ \bibinfo
  {author} {\bibfnamefont {B.~P.}\ \bibnamefont {Das}},\ }\href {\doibase
  10.48550/arXiv.2303.08681} {\bibfield  {journal} {\bibinfo  {journal}
  {ArXiv}\ } (\bibinfo {year} {2023}),\ 10.48550/arXiv.2303.08681}\BibitemShut
  {NoStop}%
\bibitem [{\citenamefont {Kaur}\ \emph {et~al.}(2023)\citenamefont {Kaur},
  \citenamefont {Bharadvaja},\ and\ \citenamefont {Baluja}}]{Kaur2023}%
  \BibitemOpen
  \bibfield  {author} {\bibinfo {author} {\bibfnamefont {S.}~\bibnamefont
  {Kaur}}, \bibinfo {author} {\bibfnamefont {A.}~\bibnamefont {Bharadvaja}}, \
  and\ \bibinfo {author} {\bibfnamefont {K.~L.}\ \bibnamefont {Baluja}},\
  }\href {\doibase 10.1140/epjd/s10053-023-00707-4} {\bibfield  {journal}
  {\bibinfo  {journal} {The European Physical Journal D}\ }\textbf {\bibinfo
  {volume} {77}},\ \bibinfo {pages} {142} (\bibinfo {year} {2023})}\BibitemShut
  {NoStop}%
\bibitem [{\citenamefont {Cernicharo}\ and\ \citenamefont
  {Guelin}(1987)}]{Cernicharo1987}%
  \BibitemOpen
  \bibfield  {author} {\bibinfo {author} {\bibfnamefont {J.}~\bibnamefont
  {Cernicharo}}\ and\ \bibinfo {author} {\bibfnamefont {M.}~\bibnamefont
  {Guelin}},\ }\href
  {https://ui.adsabs.harvard.edu/abs/1987A&A...183L..10C/abstract} {\bibfield
  {journal} {\bibinfo  {journal} {Astronomy and Astrophysics}\ }\textbf
  {\bibinfo {volume} {183}},\ \bibinfo {pages} {L10} (\bibinfo {year}
  {1987})}\BibitemShut {NoStop}%
\bibitem [{\citenamefont {Ford}\ \emph {et~al.}(2004)\citenamefont {Ford},
  \citenamefont {Neufeld}, \citenamefont {Schilke},\ and\ \citenamefont
  {Melnick}}]{Ford2004}%
  \BibitemOpen
  \bibfield  {author} {\bibinfo {author} {\bibfnamefont {K.~E.~S.}\
  \bibnamefont {Ford}}, \bibinfo {author} {\bibfnamefont {D.~A.}\ \bibnamefont
  {Neufeld}}, \bibinfo {author} {\bibfnamefont {P.}~\bibnamefont {Schilke}}, \
  and\ \bibinfo {author} {\bibfnamefont {G.~J.}\ \bibnamefont {Melnick}},\
  }\href {\doibase 10.1086/423886} {\bibfield  {journal} {\bibinfo  {journal}
  {The Astrophysical Journal}\ }\textbf {\bibinfo {volume} {614}},\ \bibinfo
  {pages} {990} (\bibinfo {year} {2004})}\BibitemShut {NoStop}%
\bibitem [{\citenamefont {Agúndez}\ \emph {et~al.}(2012)\citenamefont
  {Agúndez}, \citenamefont {Fonfría}, \citenamefont {Cernicharo},
  \citenamefont {Kahane}, \citenamefont {Daniel},\ and\ \citenamefont
  {Guélin}}]{Agundez2012}%
  \BibitemOpen
  \bibfield  {author} {\bibinfo {author} {\bibfnamefont {M.}~\bibnamefont
  {Agúndez}}, \bibinfo {author} {\bibfnamefont {J.~P.}\ \bibnamefont
  {Fonfría}}, \bibinfo {author} {\bibfnamefont {J.}~\bibnamefont
  {Cernicharo}}, \bibinfo {author} {\bibfnamefont {C.}~\bibnamefont {Kahane}},
  \bibinfo {author} {\bibfnamefont {F.}~\bibnamefont {Daniel}}, \ and\ \bibinfo
  {author} {\bibfnamefont {M.}~\bibnamefont {Guélin}},\ }\href {\doibase
  10.1051/0004-6361/201218963} {\bibfield  {journal} {\bibinfo  {journal}
  {Astronomy and Astrophysics}\ }\textbf {\bibinfo {volume} {543}},\ \bibinfo
  {pages} {A48} (\bibinfo {year} {2012})}\BibitemShut {NoStop}%
\bibitem [{\citenamefont {Kamiński}\ \emph {et~al.}(2016)\citenamefont
  {Kamiński}, \citenamefont {Wong}, \citenamefont {Schmidt}, \citenamefont
  {Müller}, \citenamefont {Gottlieb}, \citenamefont {Cherchneff},
  \citenamefont {Menten}, \citenamefont {Keller}, \citenamefont {Brünken},
  \citenamefont {Winters},\ and\ \citenamefont {Patel}}]{Kaminski2016}%
  \BibitemOpen
  \bibfield  {author} {\bibinfo {author} {\bibfnamefont {T.}~\bibnamefont
  {Kamiński}}, \bibinfo {author} {\bibfnamefont {K.~T.}\ \bibnamefont {Wong}},
  \bibinfo {author} {\bibfnamefont {M.~R.}\ \bibnamefont {Schmidt}}, \bibinfo
  {author} {\bibfnamefont {H.~S.}\ \bibnamefont {Müller}}, \bibinfo {author}
  {\bibfnamefont {C.~A.}\ \bibnamefont {Gottlieb}}, \bibinfo {author}
  {\bibfnamefont {I.}~\bibnamefont {Cherchneff}}, \bibinfo {author}
  {\bibfnamefont {K.~M.}\ \bibnamefont {Menten}}, \bibinfo {author}
  {\bibfnamefont {D.}~\bibnamefont {Keller}}, \bibinfo {author} {\bibfnamefont
  {S.}~\bibnamefont {Brünken}}, \bibinfo {author} {\bibfnamefont {J.~M.}\
  \bibnamefont {Winters}}, \ and\ \bibinfo {author} {\bibfnamefont {N.~A.}\
  \bibnamefont {Patel}},\ }\href
  {http://cdsarc.u-strasbg.fr/viz-bin/qcat?J/A+A/592/A42} {\bibfield  {journal}
  {\bibinfo  {journal} {Astronomy and Astrophysics}\ }\textbf {\bibinfo
  {volume} {592}},\ \bibinfo {pages} {42} (\bibinfo {year} {2016})}\BibitemShut
  {NoStop}%
\bibitem [{\citenamefont {Decin}\ \emph {et~al.}(2017)\citenamefont {Decin},
  \citenamefont {Richards}, \citenamefont {Waters}, \citenamefont {Danilovich},
  \citenamefont {Gobrecht}, \citenamefont {Khouri}, \citenamefont {Homan},
  \citenamefont {Bakker}, \citenamefont {de~Sande}, \citenamefont {Nuth},\ and\
  \citenamefont {Beck}}]{Decin2017}%
  \BibitemOpen
  \bibfield  {author} {\bibinfo {author} {\bibfnamefont {L.}~\bibnamefont
  {Decin}}, \bibinfo {author} {\bibfnamefont {A.~M.~S.}\ \bibnamefont
  {Richards}}, \bibinfo {author} {\bibfnamefont {L.~B. F.~M.}\ \bibnamefont
  {Waters}}, \bibinfo {author} {\bibfnamefont {T.}~\bibnamefont {Danilovich}},
  \bibinfo {author} {\bibfnamefont {D.}~\bibnamefont {Gobrecht}}, \bibinfo
  {author} {\bibfnamefont {T.}~\bibnamefont {Khouri}}, \bibinfo {author}
  {\bibfnamefont {W.}~\bibnamefont {Homan}}, \bibinfo {author} {\bibfnamefont
  {J.~M.}\ \bibnamefont {Bakker}}, \bibinfo {author} {\bibfnamefont {M.~V.}\
  \bibnamefont {de~Sande}}, \bibinfo {author} {\bibfnamefont {J.~A.}\
  \bibnamefont {Nuth}}, \ and\ \bibinfo {author} {\bibfnamefont {E.~D.}\
  \bibnamefont {Beck}},\ }\href {\doibase 10.1051/0004-6361/201730782}
  {\bibfield  {journal} {\bibinfo  {journal} {Astronomy and Astrophysics}\
  }\textbf {\bibinfo {volume} {608}},\ \bibinfo {pages} {A55} (\bibinfo {year}
  {2017})}\BibitemShut {NoStop}%
\bibitem [{\citenamefont {Bernath}(2020)}]{Bernath2020}%
  \BibitemOpen
  \bibfield  {author} {\bibinfo {author} {\bibfnamefont {P.~F.}\ \bibnamefont
  {Bernath}},\ }\href {\doibase 10.1016/j.jqsrt.2019.106687} {\bibfield
  {journal} {\bibinfo  {journal} {Journal of Quantitative Spectroscopy and
  Radiative Transfer}\ }\textbf {\bibinfo {volume} {240}},\ \bibinfo {pages}
  {106687} (\bibinfo {year} {2020})}\BibitemShut {NoStop}%
\bibitem [{\citenamefont {Yurchenko}\ \emph {et~al.}(2023)\citenamefont
  {Yurchenko}, \citenamefont {Nogué}, \citenamefont {Azzam},\ and\
  \citenamefont {Tennyson}}]{Yurchenko2023}%
  \BibitemOpen
  \bibfield  {author} {\bibinfo {author} {\bibfnamefont {S.~N.}\ \bibnamefont
  {Yurchenko}}, \bibinfo {author} {\bibfnamefont {E.}~\bibnamefont {Nogué}},
  \bibinfo {author} {\bibfnamefont {A.~A.~A.}\ \bibnamefont {Azzam}}, \ and\
  \bibinfo {author} {\bibfnamefont {J.}~\bibnamefont {Tennyson}},\ }\href
  {\doibase 10.1093/mnras/stac3757} {\bibfield  {journal} {\bibinfo  {journal}
  {Monthly Notices of the Royal Astronomical Society}\ }\textbf {\bibinfo
  {volume} {520}},\ \bibinfo {pages} {5183} (\bibinfo {year}
  {2023})}\BibitemShut {NoStop}%
\bibitem [{\citenamefont {Tennyson}\ \emph {et~al.}(2020)\citenamefont
  {Tennyson}, \citenamefont {Yurchenko}, \citenamefont {Al-Refaie},
  \citenamefont {Clark}, \citenamefont {Chubb}, \citenamefont {Conway},
  \citenamefont {Dewan}, \citenamefont {Gorman}, \citenamefont {Hill},
  \citenamefont {Lynas-Gray}, \citenamefont {Mellor}, \citenamefont
  {McKemmish}, \citenamefont {Owens}, \citenamefont {Polyansky}, \citenamefont
  {Semenov}, \citenamefont {Somogyi}, \citenamefont {Tinetti}, \citenamefont
  {Upadhyay}, \citenamefont {Waldmann}, \citenamefont {Wang}, \citenamefont
  {Wright},\ and\ \citenamefont {Yurchenko}}]{Tennyson2020}%
  \BibitemOpen
  \bibfield  {author} {\bibinfo {author} {\bibfnamefont {J.}~\bibnamefont
  {Tennyson}}, \bibinfo {author} {\bibfnamefont {S.~N.}\ \bibnamefont
  {Yurchenko}}, \bibinfo {author} {\bibfnamefont {A.~F.}\ \bibnamefont
  {Al-Refaie}}, \bibinfo {author} {\bibfnamefont {V.~H.}\ \bibnamefont
  {Clark}}, \bibinfo {author} {\bibfnamefont {K.~L.}\ \bibnamefont {Chubb}},
  \bibinfo {author} {\bibfnamefont {E.~K.}\ \bibnamefont {Conway}}, \bibinfo
  {author} {\bibfnamefont {A.}~\bibnamefont {Dewan}}, \bibinfo {author}
  {\bibfnamefont {M.~N.}\ \bibnamefont {Gorman}}, \bibinfo {author}
  {\bibfnamefont {C.}~\bibnamefont {Hill}}, \bibinfo {author} {\bibfnamefont
  {A.~E.}\ \bibnamefont {Lynas-Gray}}, \bibinfo {author} {\bibfnamefont
  {T.}~\bibnamefont {Mellor}}, \bibinfo {author} {\bibfnamefont {L.~K.}\
  \bibnamefont {McKemmish}}, \bibinfo {author} {\bibfnamefont {A.}~\bibnamefont
  {Owens}}, \bibinfo {author} {\bibfnamefont {O.~L.}\ \bibnamefont
  {Polyansky}}, \bibinfo {author} {\bibfnamefont {M.}~\bibnamefont {Semenov}},
  \bibinfo {author} {\bibfnamefont {W.}~\bibnamefont {Somogyi}}, \bibinfo
  {author} {\bibfnamefont {G.}~\bibnamefont {Tinetti}}, \bibinfo {author}
  {\bibfnamefont {A.}~\bibnamefont {Upadhyay}}, \bibinfo {author}
  {\bibfnamefont {I.}~\bibnamefont {Waldmann}}, \bibinfo {author}
  {\bibfnamefont {Y.}~\bibnamefont {Wang}}, \bibinfo {author} {\bibfnamefont
  {S.}~\bibnamefont {Wright}}, \ and\ \bibinfo {author} {\bibfnamefont {O.~P.}\
  \bibnamefont {Yurchenko}},\ }\href {\doibase 10.1016/j.jqsrt.2020.107228}
  {\bibfield  {journal} {\bibinfo  {journal} {Journal of Quantitative
  Spectroscopy and Radiative Transfer}\ }\textbf {\bibinfo {volume} {255}},\
  \bibinfo {pages} {107228} (\bibinfo {year} {2020})}\BibitemShut {NoStop}%
\bibitem [{\citenamefont {Wang}\ \emph {et~al.}(2020)\citenamefont {Wang},
  \citenamefont {Tennyson},\ and\ \citenamefont {Yurchenko}}]{Wang2020}%
  \BibitemOpen
  \bibfield  {author} {\bibinfo {author} {\bibfnamefont {Y.}~\bibnamefont
  {Wang}}, \bibinfo {author} {\bibfnamefont {J.}~\bibnamefont {Tennyson}}, \
  and\ \bibinfo {author} {\bibfnamefont {S.}~\bibnamefont {Yurchenko}},\ }\href
  {\doibase 10.3390/atoms8010007} {\bibfield  {journal} {\bibinfo  {journal}
  {Atoms}\ }\textbf {\bibinfo {volume} {8}},\ \bibinfo {pages} {7} (\bibinfo
  {year} {2020})}\BibitemShut {NoStop}%
\bibitem [{\citenamefont {Yasuda}\ \emph {et~al.}(2009)\citenamefont {Yasuda},
  \citenamefont {Saegusa},\ and\ \citenamefont {Okabe}}]{Yasuda2009}%
  \BibitemOpen
  \bibfield  {author} {\bibinfo {author} {\bibfnamefont {K.}~\bibnamefont
  {Yasuda}}, \bibinfo {author} {\bibfnamefont {K.}~\bibnamefont {Saegusa}}, \
  and\ \bibinfo {author} {\bibfnamefont {T.~H.}\ \bibnamefont {Okabe}},\ }\href
  {\doibase 10.2320/matertrans.M2009260} {\bibfield  {journal} {\bibinfo
  {journal} {Materials Transactions}\ }\textbf {\bibinfo {volume} {50}},\
  \bibinfo {pages} {2873} (\bibinfo {year} {2009})}\BibitemShut {NoStop}%
\bibitem [{\citenamefont {Yasuda}\ \emph
  {et~al.}(2011{\natexlab{a}})\citenamefont {Yasuda}, \citenamefont {Saegusa},\
  and\ \citenamefont {Okabe}}]{Yasuda2011}%
  \BibitemOpen
  \bibfield  {author} {\bibinfo {author} {\bibfnamefont {K.}~\bibnamefont
  {Yasuda}}, \bibinfo {author} {\bibfnamefont {K.}~\bibnamefont {Saegusa}}, \
  and\ \bibinfo {author} {\bibfnamefont {T.~H.}\ \bibnamefont {Okabe}},\ }\href
  {\doibase 10.1007/s11663-010-9440-y} {\bibfield  {journal} {\bibinfo
  {journal} {Metallurgical and Materials Transactions B: Process Metallurgy and
  Materials Processing Science}\ }\textbf {\bibinfo {volume} {42}},\ \bibinfo
  {pages} {37} (\bibinfo {year} {2011}{\natexlab{a}})}\BibitemShut {NoStop}%
\bibitem [{\citenamefont {Yasuda}\ \emph
  {et~al.}(2011{\natexlab{b}})\citenamefont {Yasuda}, \citenamefont {Saegusa},\
  and\ \citenamefont {Okabe}}]{Yasuda2011a}%
  \BibitemOpen
  \bibfield  {author} {\bibinfo {author} {\bibfnamefont {K.}~\bibnamefont
  {Yasuda}}, \bibinfo {author} {\bibfnamefont {K.}~\bibnamefont {Saegusa}}, \
  and\ \bibinfo {author} {\bibfnamefont {T.~H.}\ \bibnamefont {Okabe}},\ }\href
  {\doibase 10.1515/HTMP.2011.063} {\bibfield  {journal} {\bibinfo  {journal}
  {High Temperature Materials and Processes}\ }\textbf {\bibinfo {volume}
  {30}},\ \bibinfo {pages} {411} (\bibinfo {year}
  {2011}{\natexlab{b}})}\BibitemShut {NoStop}%
\bibitem [{\citenamefont {McGregor}\ \emph {et~al.}(1992)\citenamefont
  {McGregor}, \citenamefont {Drakes}, \citenamefont {Beale},\ and\
  \citenamefont {Sherrell}}]{McGregor1992}%
  \BibitemOpen
  \bibfield  {author} {\bibinfo {author} {\bibfnamefont {W.~K.}\ \bibnamefont
  {McGregor}}, \bibinfo {author} {\bibfnamefont {J.~A.}\ \bibnamefont
  {Drakes}}, \bibinfo {author} {\bibfnamefont {K.~S.}\ \bibnamefont {Beale}}, \
  and\ \bibinfo {author} {\bibfnamefont {F.~G.}\ \bibnamefont {Sherrell}}\
  }(\bibinfo  {publisher} {American Institute of Aeronautics and Astronautics
  Inc, AIAA},\ \bibinfo {year} {1992})\BibitemShut {NoStop}%
\bibitem [{\citenamefont {McGregor}\ \emph {et~al.}(1993)\citenamefont
  {McGregor}, \citenamefont {Drakes}, \citenamefont {Beale},\ and\
  \citenamefont {Sherrell}}]{McGregor1993}%
  \BibitemOpen
  \bibfield  {author} {\bibinfo {author} {\bibfnamefont {W.~K.}\ \bibnamefont
  {McGregor}}, \bibinfo {author} {\bibfnamefont {J.~A.}\ \bibnamefont
  {Drakes}}, \bibinfo {author} {\bibfnamefont {K.~S.}\ \bibnamefont {Beale}}, \
  and\ \bibinfo {author} {\bibfnamefont {F.~G.}\ \bibnamefont {Sherrell}},\
  }\href {https://ui.adsabs.harvard.edu/abs/1993JTHT....7R.736M/abstract}
  {\bibfield  {journal} {\bibinfo  {journal} {J. Thermophys. Heat Transfer}\
  }\textbf {\bibinfo {volume} {7}},\ \bibinfo {pages} {736} (\bibinfo {year}
  {1993})}\BibitemShut {NoStop}%
\bibitem [{\citenamefont {Oliver}\ \emph {et~al.}(1992)\citenamefont {Oliver},
  \citenamefont {McGregor}, \citenamefont {Reed}, \citenamefont {Drakes},\ and\
  \citenamefont {Beale}}]{Oliver1992}%
  \BibitemOpen
  \bibfield  {author} {\bibinfo {author} {\bibfnamefont {S.~M.}\ \bibnamefont
  {Oliver}}, \bibinfo {author} {\bibfnamefont {W.~K.}\ \bibnamefont
  {McGregor}}, \bibinfo {author} {\bibfnamefont {R.~A.}\ \bibnamefont {Reed}},
  \bibinfo {author} {\bibfnamefont {J.~A.}\ \bibnamefont {Drakes}}, \ and\
  \bibinfo {author} {\bibfnamefont {K.}~\bibnamefont {Beale}},\ }\href
  {https://ui.adsabs.harvard.edu/abs/1992aedc.rept.....O/abstract} {\bibfield
  {journal} {\bibinfo  {journal} {Arnold Engineering Development Center}\ }
  (\bibinfo {year} {1992})}\BibitemShut {NoStop}%
\bibitem [{\citenamefont {Parvinen}\ and\ \citenamefont
  {Lajunen}(1999)}]{Parvinen1999}%
  \BibitemOpen
  \bibfield  {author} {\bibinfo {author} {\bibfnamefont {P.}~\bibnamefont
  {Parvinen}}\ and\ \bibinfo {author} {\bibfnamefont {L.~H.}\ \bibnamefont
  {Lajunen}},\ }\href {\doibase 10.1016/S0039-9140(99)00103-4} {\bibfield
  {journal} {\bibinfo  {journal} {Talanta}\ }\textbf {\bibinfo {volume} {50}},\
  \bibinfo {pages} {67} (\bibinfo {year} {1999})}\BibitemShut {NoStop}%
\bibitem [{\citenamefont {Tacke}\ and\ \citenamefont
  {Schnöckel}(1989)}]{Tacke1989}%
  \BibitemOpen
  \bibfield  {author} {\bibinfo {author} {\bibfnamefont {M.}~\bibnamefont
  {Tacke}}\ and\ \bibinfo {author} {\bibfnamefont {H.}~\bibnamefont
  {Schnöckel}},\ }\href {\doibase 10.1021/ic00313a039} {\bibfield  {journal}
  {\bibinfo  {journal} {Inorganic Chemistry}\ }\textbf {\bibinfo {volume}
  {28}},\ \bibinfo {pages} {2895} (\bibinfo {year} {1989})}\BibitemShut
  {NoStop}%
\bibitem [{\citenamefont {Shaw}(2022)}]{Thesis_Shaw_2022}%
  \BibitemOpen
  \bibfield  {author} {\bibinfo {author} {\bibfnamefont {J.~C.}\ \bibnamefont
  {Shaw}},\ }\emph {\bibinfo {title} {Prospects for laser cooling and trapping
  aluminum monochloride}},\ \href@noop {} {Ph.D. thesis},\ \bibinfo  {school}
  {University of Connecticut} (\bibinfo {year} {2022})\BibitemShut {NoStop}%
\bibitem [{\citenamefont {Hutzler}\ \emph {et~al.}(2012)\citenamefont
  {Hutzler}, \citenamefont {Lu},\ and\ \citenamefont {Doyle}}]{Hutzler2012}%
  \BibitemOpen
  \bibfield  {author} {\bibinfo {author} {\bibfnamefont {N.~R.}\ \bibnamefont
  {Hutzler}}, \bibinfo {author} {\bibfnamefont {H.-I.~I.}\ \bibnamefont {Lu}},
  \ and\ \bibinfo {author} {\bibfnamefont {J.~M.}\ \bibnamefont {Doyle}},\
  }\href {\doibase 10.1021/cr200362u} {\bibfield  {journal} {\bibinfo
  {journal} {Chemical Reviews}\ }\textbf {\bibinfo {volume} {112}},\ \bibinfo
  {pages} {4803} (\bibinfo {year} {2012})}\BibitemShut {NoStop}%
\bibitem [{\citenamefont {Barry}\ \emph {et~al.}(2011)\citenamefont {Barry},
  \citenamefont {Shuman},\ and\ \citenamefont {DeMille}}]{Barry2011}%
  \BibitemOpen
  \bibfield  {author} {\bibinfo {author} {\bibfnamefont {J.~F.}\ \bibnamefont
  {Barry}}, \bibinfo {author} {\bibfnamefont {E.~S.}\ \bibnamefont {Shuman}}, \
  and\ \bibinfo {author} {\bibfnamefont {D.}~\bibnamefont {DeMille}},\ }\href
  {\doibase 10.1039/C1CP20335E} {\bibfield  {journal} {\bibinfo  {journal}
  {Physical Chemistry Chemical Physics}\ }\textbf {\bibinfo {volume} {13}},\
  \bibinfo {pages} {18936} (\bibinfo {year} {2011})}\BibitemShut {NoStop}%
\bibitem [{\citenamefont {Lewis}\ \emph {et~al.}(2021)\citenamefont {Lewis},
  \citenamefont {Wang}, \citenamefont {Daniel}, \citenamefont {Dhital},
  \citenamefont {Bardeen},\ and\ \citenamefont {Hemmerling}}]{Lewis2021}%
  \BibitemOpen
  \bibfield  {author} {\bibinfo {author} {\bibfnamefont {T.~N.}\ \bibnamefont
  {Lewis}}, \bibinfo {author} {\bibfnamefont {C.}~\bibnamefont {Wang}},
  \bibinfo {author} {\bibfnamefont {J.~R.}\ \bibnamefont {Daniel}}, \bibinfo
  {author} {\bibfnamefont {M.}~\bibnamefont {Dhital}}, \bibinfo {author}
  {\bibfnamefont {C.~J.}\ \bibnamefont {Bardeen}}, \ and\ \bibinfo {author}
  {\bibfnamefont {B.}~\bibnamefont {Hemmerling}},\ }\href
  {https://arxiv.org/abs/2108.01158v1} {\bibfield  {journal} {\bibinfo
  {journal} {Physical Chemistry Chemical Physics}\ }\textbf {\bibinfo {volume}
  {23}},\ \bibinfo {pages} {22785} (\bibinfo {year} {2021})}\BibitemShut
  {NoStop}%
\bibitem [{\citenamefont {Shaw}\ and\ \citenamefont
  {McCarron}(2020)}]{Shaw2020}%
  \BibitemOpen
  \bibfield  {author} {\bibinfo {author} {\bibfnamefont {J.~C.}\ \bibnamefont
  {Shaw}}\ and\ \bibinfo {author} {\bibfnamefont {D.~J.}\ \bibnamefont
  {McCarron}},\ }\href {\doibase 10.1103/PhysRevA.102.041302} {\bibfield
  {journal} {\bibinfo  {journal} {Physical Review A}\ }\textbf {\bibinfo
  {volume} {102}},\ \bibinfo {pages} {041302} (\bibinfo {year}
  {2020})}\BibitemShut {NoStop}%
\bibitem [{\citenamefont {Brown}\ and\ \citenamefont
  {Carrington}(2003)}]{Brown2003}%
  \BibitemOpen
  \bibfield  {author} {\bibinfo {author} {\bibfnamefont {J.~M.}\ \bibnamefont
  {Brown}}\ and\ \bibinfo {author} {\bibfnamefont {A.}~\bibnamefont
  {Carrington}},\ }\href@noop {} {\emph {\bibinfo {title} {Rotational
  Spectroscopy of Diatomic Molecules}}}\ (\bibinfo  {publisher} {Cambridge
  University Press},\ \bibinfo {year} {2003})\BibitemShut {NoStop}%
\bibitem [{\citenamefont {Alonso}\ \emph {et~al.}(2004)\citenamefont {Alonso},
  \citenamefont {Svane}, \citenamefont {Rodríguez},\ and\ \citenamefont
  {Christensen}}]{Alonso2004}%
  \BibitemOpen
  \bibfield  {author} {\bibinfo {author} {\bibfnamefont {R.~E.}\ \bibnamefont
  {Alonso}}, \bibinfo {author} {\bibfnamefont {A.}~\bibnamefont {Svane}},
  \bibinfo {author} {\bibfnamefont {C.~O.}\ \bibnamefont {Rodríguez}}, \ and\
  \bibinfo {author} {\bibfnamefont {N.~E.}\ \bibnamefont {Christensen}},\
  }\href {\doibase 10.1103/PhysRevB.69.125101} {\bibfield  {journal} {\bibinfo
  {journal} {Physical Review B}\ }\textbf {\bibinfo {volume} {69}},\ \bibinfo
  {pages} {125101} (\bibinfo {year} {2004})}\BibitemShut {NoStop}%
\bibitem [{\citenamefont {Barry}(2013)}]{Thesis_Barry_2013}%
  \BibitemOpen
  \bibfield  {author} {\bibinfo {author} {\bibfnamefont {J.~F.}\ \bibnamefont
  {Barry}},\ }\emph {\bibinfo {title} {Laser cooling and slowing of a diatomic
  molecule}},\ \href@noop {} {Ph.D. thesis},\ \bibinfo  {school} {Yale
  University} (\bibinfo {year} {2013})\BibitemShut {NoStop}%
\bibitem [{\citenamefont {Tarbutt}\ and\ \citenamefont
  {Steimle}(2021)}]{Tarbutt2015}%
  \BibitemOpen
  \bibfield  {author} {\bibinfo {author} {\bibfnamefont {M.~R.}\ \bibnamefont
  {Tarbutt}}\ and\ \bibinfo {author} {\bibfnamefont {T.~C.}\ \bibnamefont
  {Steimle}},\ }\href {\doibase 10.1103/PhysRevA.92.053401} {\bibfield
  {journal} {\bibinfo  {journal} {Physical Review A}\ }\textbf {\bibinfo
  {volume} {92}},\ \bibinfo {pages} {053401} (\bibinfo {year}
  {2021})}\BibitemShut {NoStop}%
\bibitem [{\citenamefont {Hummon}\ \emph {et~al.}(2013)\citenamefont {Hummon},
  \citenamefont {Yeo}, \citenamefont {Stuhl}, \citenamefont {Collopy},
  \citenamefont {Xia},\ and\ \citenamefont {Ye}}]{Hummon2013}%
  \BibitemOpen
  \bibfield  {author} {\bibinfo {author} {\bibfnamefont {M.~T.}\ \bibnamefont
  {Hummon}}, \bibinfo {author} {\bibfnamefont {M.}~\bibnamefont {Yeo}},
  \bibinfo {author} {\bibfnamefont {B.~K.}\ \bibnamefont {Stuhl}}, \bibinfo
  {author} {\bibfnamefont {A.~L.}\ \bibnamefont {Collopy}}, \bibinfo {author}
  {\bibfnamefont {Y.}~\bibnamefont {Xia}}, \ and\ \bibinfo {author}
  {\bibfnamefont {J.}~\bibnamefont {Ye}},\ }\href {\doibase
  10.1103/PhysRevLett.110.143001} {\bibfield  {journal} {\bibinfo  {journal}
  {Physical Review Letters}\ }\textbf {\bibinfo {volume} {110}},\ \bibinfo
  {pages} {143001} (\bibinfo {year} {2013})}\BibitemShut {NoStop}%
\bibitem [{\citenamefont {Norrgard}\ \emph {et~al.}(2016)\citenamefont
  {Norrgard}, \citenamefont {McCarron}, \citenamefont {Steinecker},
  \citenamefont {Tarbutt},\ and\ \citenamefont {DeMille}}]{Norrgard2016}%
  \BibitemOpen
  \bibfield  {author} {\bibinfo {author} {\bibfnamefont {E.~B.}\ \bibnamefont
  {Norrgard}}, \bibinfo {author} {\bibfnamefont {D.~J.}\ \bibnamefont
  {McCarron}}, \bibinfo {author} {\bibfnamefont {M.~H.}\ \bibnamefont
  {Steinecker}}, \bibinfo {author} {\bibfnamefont {M.~R.}\ \bibnamefont
  {Tarbutt}}, \ and\ \bibinfo {author} {\bibfnamefont {D.}~\bibnamefont
  {DeMille}},\ }\href {\doibase 10.1103/PhysRevLett.116.063004} {\bibfield
  {journal} {\bibinfo  {journal} {Physical Review Letters}\ }\textbf {\bibinfo
  {volume} {116}},\ \bibinfo {pages} {063004} (\bibinfo {year}
  {2016})}\BibitemShut {NoStop}%
\bibitem [{\citenamefont {Devlin}\ and\ \citenamefont
  {Tarbutt}(2016)}]{Devlin2016}%
  \BibitemOpen
  \bibfield  {author} {\bibinfo {author} {\bibfnamefont {J.~A.}\ \bibnamefont
  {Devlin}}\ and\ \bibinfo {author} {\bibfnamefont {M.~R.}\ \bibnamefont
  {Tarbutt}},\ }\href {\doibase 10.1088/1367-2630/18/12/123017} {\bibfield
  {journal} {\bibinfo  {journal} {New Journal of Physics}\ }\textbf {\bibinfo
  {volume} {18}},\ \bibinfo {pages} {123017} (\bibinfo {year}
  {2016})}\BibitemShut {NoStop}%
\bibitem [{\citenamefont {Xu}\ \emph {et~al.}(2003)\citenamefont {Xu},
  \citenamefont {Loftus}, \citenamefont {Hall}, \citenamefont {Gallagher},\
  and\ \citenamefont {Ye}}]{Xu2003}%
  \BibitemOpen
  \bibfield  {author} {\bibinfo {author} {\bibfnamefont {X.}~\bibnamefont
  {Xu}}, \bibinfo {author} {\bibfnamefont {T.~H.}\ \bibnamefont {Loftus}},
  \bibinfo {author} {\bibfnamefont {J.~L.}\ \bibnamefont {Hall}}, \bibinfo
  {author} {\bibfnamefont {A.}~\bibnamefont {Gallagher}}, \ and\ \bibinfo
  {author} {\bibfnamefont {J.}~\bibnamefont {Ye}},\ }\href {\doibase
  10.1364/JOSAB.20.000968} {\bibfield  {journal} {\bibinfo  {journal} {Journal
  of the Optical Society of America B}\ }\textbf {\bibinfo {volume} {20}},\
  \bibinfo {pages} {968} (\bibinfo {year} {2003})}\BibitemShut {NoStop}%
\bibitem [{\citenamefont {Kaebert}\ \emph {et~al.}(2021)\citenamefont
  {Kaebert}, \citenamefont {Stepanova}, \citenamefont {Poll}, \citenamefont
  {Petzold}, \citenamefont {Xu}, \citenamefont {Siercke},\ and\ \citenamefont
  {Ospelkaus}}]{Kaebert2021}%
  \BibitemOpen
  \bibfield  {author} {\bibinfo {author} {\bibfnamefont {P.}~\bibnamefont
  {Kaebert}}, \bibinfo {author} {\bibfnamefont {M.}~\bibnamefont {Stepanova}},
  \bibinfo {author} {\bibfnamefont {T.}~\bibnamefont {Poll}}, \bibinfo {author}
  {\bibfnamefont {M.}~\bibnamefont {Petzold}}, \bibinfo {author} {\bibfnamefont
  {S.}~\bibnamefont {Xu}}, \bibinfo {author} {\bibfnamefont {M.}~\bibnamefont
  {Siercke}}, \ and\ \bibinfo {author} {\bibfnamefont {S.}~\bibnamefont
  {Ospelkaus}},\ }\href {\doibase 10.1088/1367-2630/ac1ed7} {\bibfield
  {journal} {\bibinfo  {journal} {New Journal of Physics}\ }\textbf {\bibinfo
  {volume} {23}},\ \bibinfo {pages} {093013} (\bibinfo {year}
  {2021})}\BibitemShut {NoStop}%
\bibitem [{\citenamefont {Haubrich}\ \emph {et~al.}(1996)\citenamefont
  {Haubrich}, \citenamefont {Schadwinkel}, \citenamefont {Strauch},
  \citenamefont {Ueberholz}, \citenamefont {Wynands},\ and\ \citenamefont
  {Meschede}}]{Haubrich1996}%
  \BibitemOpen
  \bibfield  {author} {\bibinfo {author} {\bibfnamefont {D.}~\bibnamefont
  {Haubrich}}, \bibinfo {author} {\bibfnamefont {H.}~\bibnamefont
  {Schadwinkel}}, \bibinfo {author} {\bibfnamefont {F.}~\bibnamefont
  {Strauch}}, \bibinfo {author} {\bibfnamefont {B.}~\bibnamefont {Ueberholz}},
  \bibinfo {author} {\bibfnamefont {R.}~\bibnamefont {Wynands}}, \ and\
  \bibinfo {author} {\bibfnamefont {D.}~\bibnamefont {Meschede}},\ }\href
  {\doibase 10.1209/epl/i1996-00512-5} {\bibfield  {journal} {\bibinfo
  {journal} {Europhysics Letters}\ }\textbf {\bibinfo {volume} {34}},\ \bibinfo
  {pages} {663} (\bibinfo {year} {1996})}\BibitemShut {NoStop}%
\bibitem [{\citenamefont {Willems}\ \emph {et~al.}(1997)\citenamefont
  {Willems}, \citenamefont {Boyd}, \citenamefont {Bliss},\ and\ \citenamefont
  {Libbrecht}}]{Willems1997}%
  \BibitemOpen
  \bibfield  {author} {\bibinfo {author} {\bibfnamefont {P.~A.}\ \bibnamefont
  {Willems}}, \bibinfo {author} {\bibfnamefont {R.~A.}\ \bibnamefont {Boyd}},
  \bibinfo {author} {\bibfnamefont {J.~L.}\ \bibnamefont {Bliss}}, \ and\
  \bibinfo {author} {\bibfnamefont {K.~G.}\ \bibnamefont {Libbrecht}},\ }\href
  {\doibase 10.1103/PhysRevLett.78.1660} {\bibfield  {journal} {\bibinfo
  {journal} {Physical Review Letters}\ }\textbf {\bibinfo {volume} {78}},\
  \bibinfo {pages} {1660} (\bibinfo {year} {1997})}\BibitemShut {NoStop}%
\bibitem [{\citenamefont {Yoon}\ \emph {et~al.}(2007)\citenamefont {Yoon},
  \citenamefont {Choi}, \citenamefont {Park}, \citenamefont {Ji}, \citenamefont
  {Lee},\ and\ \citenamefont {An}}]{Yoon2007}%
  \BibitemOpen
  \bibfield  {author} {\bibinfo {author} {\bibfnamefont {S.}~\bibnamefont
  {Yoon}}, \bibinfo {author} {\bibfnamefont {Y.}~\bibnamefont {Choi}}, \bibinfo
  {author} {\bibfnamefont {S.}~\bibnamefont {Park}}, \bibinfo {author}
  {\bibfnamefont {W.}~\bibnamefont {Ji}}, \bibinfo {author} {\bibfnamefont
  {J.-H.}\ \bibnamefont {Lee}}, \ and\ \bibinfo {author} {\bibfnamefont
  {K.}~\bibnamefont {An}},\ }\href {\doibase 10.1088/1742-6596/80/1/012046}
  {\bibfield  {journal} {\bibinfo  {journal} {Journal of Physics: Conference
  Series}\ }\textbf {\bibinfo {volume} {80}},\ \bibinfo {pages} {012046}
  (\bibinfo {year} {2007})}\BibitemShut {NoStop}%
\bibitem [{\citenamefont {Partlow}\ \emph {et~al.}(2004)\citenamefont
  {Partlow}, \citenamefont {Miao}, \citenamefont {Bochmann}, \citenamefont
  {Cashen},\ and\ \citenamefont {Metcalf}}]{Partlow2004}%
  \BibitemOpen
  \bibfield  {author} {\bibinfo {author} {\bibfnamefont {M.}~\bibnamefont
  {Partlow}}, \bibinfo {author} {\bibfnamefont {X.}~\bibnamefont {Miao}},
  \bibinfo {author} {\bibfnamefont {J.}~\bibnamefont {Bochmann}}, \bibinfo
  {author} {\bibfnamefont {M.}~\bibnamefont {Cashen}}, \ and\ \bibinfo {author}
  {\bibfnamefont {H.}~\bibnamefont {Metcalf}},\ }\href {\doibase
  10.1103/PhysRevLett.93.213004} {\bibfield  {journal} {\bibinfo  {journal}
  {Physical Review Letters}\ }\textbf {\bibinfo {volume} {93}},\ \bibinfo
  {pages} {213004} (\bibinfo {year} {2004})}\BibitemShut {NoStop}%
\bibitem [{\citenamefont {Chieda}\ and\ \citenamefont
  {Eyler}(2012)}]{Chieda2012}%
  \BibitemOpen
  \bibfield  {author} {\bibinfo {author} {\bibfnamefont {M.~A.}\ \bibnamefont
  {Chieda}}\ and\ \bibinfo {author} {\bibfnamefont {E.~E.}\ \bibnamefont
  {Eyler}},\ }\href {\doibase 10.1103/PhysRevA.86.053415} {\bibfield  {journal}
  {\bibinfo  {journal} {Physical Review A}\ }\textbf {\bibinfo {volume} {86}},\
  \bibinfo {pages} {053415} (\bibinfo {year} {2012})}\BibitemShut {NoStop}%
\bibitem [{\citenamefont {Yang}\ \emph
  {et~al.}(2016{\natexlab{b}})\citenamefont {Yang}, \citenamefont {Li},
  \citenamefont {Yin}, \citenamefont {Xu}, \citenamefont {Li}, \citenamefont
  {Xia},\ and\ \citenamefont {Yin}}]{Yang2016d}%
  \BibitemOpen
  \bibfield  {author} {\bibinfo {author} {\bibfnamefont {X.}~\bibnamefont
  {Yang}}, \bibinfo {author} {\bibfnamefont {C.}~\bibnamefont {Li}}, \bibinfo
  {author} {\bibfnamefont {Y.}~\bibnamefont {Yin}}, \bibinfo {author}
  {\bibfnamefont {S.}~\bibnamefont {Xu}}, \bibinfo {author} {\bibfnamefont
  {X.}~\bibnamefont {Li}}, \bibinfo {author} {\bibfnamefont {Y.}~\bibnamefont
  {Xia}}, \ and\ \bibinfo {author} {\bibfnamefont {J.}~\bibnamefont {Yin}},\
  }\href {\doibase 10.1088/1361-6455/50/1/015001} {\bibfield  {journal}
  {\bibinfo  {journal} {Journal of Physics B: Atomic, Molecular and Optical
  Physics}\ }\textbf {\bibinfo {volume} {50}},\ \bibinfo {pages} {015001}
  (\bibinfo {year} {2016}{\natexlab{b}})}\BibitemShut {NoStop}%
\bibitem [{\citenamefont {Corder}\ \emph {et~al.}(2015)\citenamefont {Corder},
  \citenamefont {Arnold},\ and\ \citenamefont {Metcalf}}]{Corder2015}%
  \BibitemOpen
  \bibfield  {author} {\bibinfo {author} {\bibfnamefont {C.}~\bibnamefont
  {Corder}}, \bibinfo {author} {\bibfnamefont {B.}~\bibnamefont {Arnold}}, \
  and\ \bibinfo {author} {\bibfnamefont {H.}~\bibnamefont {Metcalf}},\ }\href
  {\doibase 10.1103/PhysRevLett.114.043002} {\bibfield  {journal} {\bibinfo
  {journal} {Physical Review Letters}\ }\textbf {\bibinfo {volume} {114}},\
  \bibinfo {pages} {043002} (\bibinfo {year} {2015})}\BibitemShut {NoStop}%
\bibitem [{\citenamefont {Kozyryev}\ \emph {et~al.}(2018)\citenamefont
  {Kozyryev}, \citenamefont {Baum}, \citenamefont {Aldridge}, \citenamefont
  {Yu}, \citenamefont {Eyler},\ and\ \citenamefont {Doyle}}]{Kozyryev2018}%
  \BibitemOpen
  \bibfield  {author} {\bibinfo {author} {\bibfnamefont {I.}~\bibnamefont
  {Kozyryev}}, \bibinfo {author} {\bibfnamefont {L.}~\bibnamefont {Baum}},
  \bibinfo {author} {\bibfnamefont {L.}~\bibnamefont {Aldridge}}, \bibinfo
  {author} {\bibfnamefont {P.}~\bibnamefont {Yu}}, \bibinfo {author}
  {\bibfnamefont {E.~E.}\ \bibnamefont {Eyler}}, \ and\ \bibinfo {author}
  {\bibfnamefont {J.~M.}\ \bibnamefont {Doyle}},\ }\href {\doibase
  10.1103/PhysRevLett.120.063205} {\bibfield  {journal} {\bibinfo  {journal}
  {Physical Review Letters}\ }\textbf {\bibinfo {volume} {120}},\ \bibinfo
  {pages} {063205} (\bibinfo {year} {2018})}\BibitemShut {NoStop}%
\bibitem [{\citenamefont {Galica}\ \emph {et~al.}(2018)\citenamefont {Galica},
  \citenamefont {Aldridge}, \citenamefont {McCarron}, \citenamefont {Eyler},\
  and\ \citenamefont {Gould}}]{Galica2018}%
  \BibitemOpen
  \bibfield  {author} {\bibinfo {author} {\bibfnamefont {S.~E.}\ \bibnamefont
  {Galica}}, \bibinfo {author} {\bibfnamefont {L.}~\bibnamefont {Aldridge}},
  \bibinfo {author} {\bibfnamefont {D.~J.}\ \bibnamefont {McCarron}}, \bibinfo
  {author} {\bibfnamefont {E.~E.}\ \bibnamefont {Eyler}}, \ and\ \bibinfo
  {author} {\bibfnamefont {P.~L.}\ \bibnamefont {Gould}},\ }\href {\doibase
  10.1103/PhysRevA.98.023408} {\bibfield  {journal} {\bibinfo  {journal}
  {Physical Review A}\ }\textbf {\bibinfo {volume} {98}},\ \bibinfo {pages}
  {023408} (\bibinfo {year} {2018})}\BibitemShut {NoStop}%
\bibitem [{\citenamefont {Greenberg}\ \emph {et~al.}(2021)\citenamefont
  {Greenberg}, \citenamefont {Krohn}, \citenamefont {Bossert}, \citenamefont
  {Shyur}, \citenamefont {Macaluso}, \citenamefont {Fitch},\ and\ \citenamefont
  {Lewandowski}}]{Greenberg2021}%
  \BibitemOpen
  \bibfield  {author} {\bibinfo {author} {\bibfnamefont {J.}~\bibnamefont
  {Greenberg}}, \bibinfo {author} {\bibfnamefont {O.~A.}\ \bibnamefont
  {Krohn}}, \bibinfo {author} {\bibfnamefont {J.~A.}\ \bibnamefont {Bossert}},
  \bibinfo {author} {\bibfnamefont {Y.}~\bibnamefont {Shyur}}, \bibinfo
  {author} {\bibfnamefont {D.}~\bibnamefont {Macaluso}}, \bibinfo {author}
  {\bibfnamefont {N.~J.}\ \bibnamefont {Fitch}}, \ and\ \bibinfo {author}
  {\bibfnamefont {H.~J.}\ \bibnamefont {Lewandowski}},\ }\href {\doibase
  10.1063/5.0057859} {\bibfield  {journal} {\bibinfo  {journal} {Review of
  Scientific Instruments}\ }\textbf {\bibinfo {volume} {92}},\ \bibinfo {pages}
  {103202} (\bibinfo {year} {2021})}\BibitemShut {NoStop}%
\bibitem [{\citenamefont {Aggarwal}\ \emph {et~al.}(2021)\citenamefont
  {Aggarwal}, \citenamefont {Bethlem}, \citenamefont {Boeschoten},
  \citenamefont {Borschevsky}, \citenamefont {Esajas}, \citenamefont {Hao},
  \citenamefont {Hoekstra}, \citenamefont {Jungmann}, \citenamefont {Marshall},
  \citenamefont {Meijknecht}, \citenamefont {Mooij}, \citenamefont
  {Timmermans}, \citenamefont {Touwen}, \citenamefont {Ubachs}, \citenamefont
  {Willmann}, \citenamefont {Yin},\ and\ \citenamefont
  {Zapara}}]{Aggarwal2021}%
  \BibitemOpen
  \bibfield  {author} {\bibinfo {author} {\bibfnamefont {P.}~\bibnamefont
  {Aggarwal}}, \bibinfo {author} {\bibfnamefont {H.~L.}\ \bibnamefont
  {Bethlem}}, \bibinfo {author} {\bibfnamefont {A.}~\bibnamefont {Boeschoten}},
  \bibinfo {author} {\bibfnamefont {A.}~\bibnamefont {Borschevsky}}, \bibinfo
  {author} {\bibfnamefont {K.}~\bibnamefont {Esajas}}, \bibinfo {author}
  {\bibfnamefont {Y.}~\bibnamefont {Hao}}, \bibinfo {author} {\bibfnamefont
  {S.}~\bibnamefont {Hoekstra}}, \bibinfo {author} {\bibfnamefont
  {K.}~\bibnamefont {Jungmann}}, \bibinfo {author} {\bibfnamefont {V.~R.}\
  \bibnamefont {Marshall}}, \bibinfo {author} {\bibfnamefont {T.~B.}\
  \bibnamefont {Meijknecht}}, \bibinfo {author} {\bibfnamefont {M.~C.}\
  \bibnamefont {Mooij}}, \bibinfo {author} {\bibfnamefont {R.~G.~E.}\
  \bibnamefont {Timmermans}}, \bibinfo {author} {\bibfnamefont
  {A.}~\bibnamefont {Touwen}}, \bibinfo {author} {\bibfnamefont
  {W.}~\bibnamefont {Ubachs}}, \bibinfo {author} {\bibfnamefont
  {L.}~\bibnamefont {Willmann}}, \bibinfo {author} {\bibfnamefont
  {Y.}~\bibnamefont {Yin}}, \ and\ \bibinfo {author} {\bibfnamefont
  {A.}~\bibnamefont {Zapara}},\ }\href {\doibase 10.1063/5.0035568} {\bibfield
  {journal} {\bibinfo  {journal} {Review of Scientific Instruments}\ }\textbf
  {\bibinfo {volume} {92}},\ \bibinfo {pages} {033202} (\bibinfo {year}
  {2021})}\BibitemShut {NoStop}%
\bibitem [{\citenamefont {Shyur}\ \emph
  {et~al.}(2018{\natexlab{a}})\citenamefont {Shyur}, \citenamefont {Fitch},
  \citenamefont {Bossert}, \citenamefont {Brown},\ and\ \citenamefont
  {Lewandowski}}]{Shyur2018}%
  \BibitemOpen
  \bibfield  {author} {\bibinfo {author} {\bibfnamefont {Y.}~\bibnamefont
  {Shyur}}, \bibinfo {author} {\bibfnamefont {N.~J.}\ \bibnamefont {Fitch}},
  \bibinfo {author} {\bibfnamefont {J.~A.}\ \bibnamefont {Bossert}}, \bibinfo
  {author} {\bibfnamefont {T.}~\bibnamefont {Brown}}, \ and\ \bibinfo {author}
  {\bibfnamefont {H.~J.}\ \bibnamefont {Lewandowski}},\ }\href {\doibase
  10.1063/1.5040267} {\bibfield  {journal} {\bibinfo  {journal} {Review of
  Scientific Instruments}\ }\textbf {\bibinfo {volume} {89}},\ \bibinfo {pages}
  {084705} (\bibinfo {year} {2018}{\natexlab{a}})}\BibitemShut {NoStop}%
\bibitem [{\citenamefont {Shyur}\ \emph
  {et~al.}(2018{\natexlab{b}})\citenamefont {Shyur}, \citenamefont {Bossert},\
  and\ \citenamefont {Lewandowski}}]{Shyur2018a}%
  \BibitemOpen
  \bibfield  {author} {\bibinfo {author} {\bibfnamefont {Y.}~\bibnamefont
  {Shyur}}, \bibinfo {author} {\bibfnamefont {J.~A.}\ \bibnamefont {Bossert}},
  \ and\ \bibinfo {author} {\bibfnamefont {H.~J.}\ \bibnamefont
  {Lewandowski}},\ }\href {\doibase 10.1088/1361-6455/aad1b0} {\bibfield
  {journal} {\bibinfo  {journal} {Journal of Physics B: Atomic, Molecular and
  Optical Physics}\ }\textbf {\bibinfo {volume} {51}},\ \bibinfo {pages}
  {165101} (\bibinfo {year} {2018}{\natexlab{b}})}\BibitemShut {NoStop}%
\bibitem [{\citenamefont {Osterwalder}\ \emph {et~al.}(2010)\citenamefont
  {Osterwalder}, \citenamefont {Meek}, \citenamefont {Hammer}, \citenamefont
  {Haak},\ and\ \citenamefont {Meijer}}]{Osterwalder2010}%
  \BibitemOpen
  \bibfield  {author} {\bibinfo {author} {\bibfnamefont {A.}~\bibnamefont
  {Osterwalder}}, \bibinfo {author} {\bibfnamefont {S.~A.}\ \bibnamefont
  {Meek}}, \bibinfo {author} {\bibfnamefont {G.}~\bibnamefont {Hammer}},
  \bibinfo {author} {\bibfnamefont {H.}~\bibnamefont {Haak}}, \ and\ \bibinfo
  {author} {\bibfnamefont {G.}~\bibnamefont {Meijer}},\ }\href {\doibase
  10.1103/PhysRevA.81.051401} {\bibfield  {journal} {\bibinfo  {journal}
  {Physical Review A}\ }\textbf {\bibinfo {volume} {81}},\ \bibinfo {pages}
  {051401} (\bibinfo {year} {2010})}\BibitemShut {NoStop}%
\bibitem [{\citenamefont {Meek}\ \emph {et~al.}(2011)\citenamefont {Meek},
  \citenamefont {Parsons}, \citenamefont {Heyne}, \citenamefont
  {Platschkowski}, \citenamefont {Haak}, \citenamefont {Meijer},\ and\
  \citenamefont {Osterwalder}}]{Meek2011}%
  \BibitemOpen
  \bibfield  {author} {\bibinfo {author} {\bibfnamefont {S.~A.}\ \bibnamefont
  {Meek}}, \bibinfo {author} {\bibfnamefont {M.~F.}\ \bibnamefont {Parsons}},
  \bibinfo {author} {\bibfnamefont {G.}~\bibnamefont {Heyne}}, \bibinfo
  {author} {\bibfnamefont {V.}~\bibnamefont {Platschkowski}}, \bibinfo {author}
  {\bibfnamefont {H.}~\bibnamefont {Haak}}, \bibinfo {author} {\bibfnamefont
  {G.}~\bibnamefont {Meijer}}, \ and\ \bibinfo {author} {\bibfnamefont
  {A.}~\bibnamefont {Osterwalder}},\ }\href {\doibase 10.1063/1.3640413}
  {\bibfield  {journal} {\bibinfo  {journal} {Review of Scientific
  Instruments}\ }\textbf {\bibinfo {volume} {82}},\ \bibinfo {pages} {093108}
  (\bibinfo {year} {2011})}\BibitemShut {NoStop}%
\bibitem [{\citenamefont {van~den Berg}\ \emph {et~al.}(2014)\citenamefont
  {van~den Berg}, \citenamefont {Mathavan}, \citenamefont {Meinema},
  \citenamefont {Nauta}, \citenamefont {Nijbroek}, \citenamefont {Jungmann},
  \citenamefont {Bethlem},\ and\ \citenamefont {Hoekstra}}]{Berg2014}%
  \BibitemOpen
  \bibfield  {author} {\bibinfo {author} {\bibfnamefont {J.~E.}\ \bibnamefont
  {van~den Berg}}, \bibinfo {author} {\bibfnamefont {S.~C.}\ \bibnamefont
  {Mathavan}}, \bibinfo {author} {\bibfnamefont {C.}~\bibnamefont {Meinema}},
  \bibinfo {author} {\bibfnamefont {J.}~\bibnamefont {Nauta}}, \bibinfo
  {author} {\bibfnamefont {T.~H.}\ \bibnamefont {Nijbroek}}, \bibinfo {author}
  {\bibfnamefont {K.}~\bibnamefont {Jungmann}}, \bibinfo {author}
  {\bibfnamefont {H.~L.}\ \bibnamefont {Bethlem}}, \ and\ \bibinfo {author}
  {\bibfnamefont {S.}~\bibnamefont {Hoekstra}},\ }\href {\doibase
  10.1016/j.jms.2014.02.004} {\bibfield  {journal} {\bibinfo  {journal}
  {Journal of Molecular Spectroscopy}\ }\bibinfo {series} {Spectroscopic
  {Tests} of {Fundamental} {Physics}},\ \textbf {\bibinfo {volume} {300}},\
  \bibinfo {pages} {22} (\bibinfo {year} {2014})}\BibitemShut {NoStop}%
\bibitem [{\citenamefont {Shaw}\ \emph {et~al.}()\citenamefont {Shaw},
  \citenamefont {Semco}, \citenamefont {Wortley},\ and\ \citenamefont
  {McCarron}}]{Shaw2023}%
  \BibitemOpen
  \bibfield  {author} {\bibinfo {author} {\bibfnamefont {J.~C.}\ \bibnamefont
  {Shaw}}, \bibinfo {author} {\bibfnamefont {M.~A.}\ \bibnamefont {Semco}},
  \bibinfo {author} {\bibfnamefont {W.~B.}\ \bibnamefont {Wortley}}, \ and\
  \bibinfo {author} {\bibfnamefont {D.~J.}\ \bibnamefont {McCarron}},\
  }\href@noop {} {\bibinfo  {journal} {(in preparation)}\ }\BibitemShut
  {NoStop}%
\bibitem [{\citenamefont {Hunter}\ \emph {et~al.}(2012)\citenamefont {Hunter},
  \citenamefont {Peck}, \citenamefont {Greenspon}, \citenamefont {Alam},\ and\
  \citenamefont {DeMille}}]{Hunter2012}%
  \BibitemOpen
\bibfield  {journal} {  }\bibfield  {author} {\bibinfo {author} {\bibfnamefont
  {L.~R.}\ \bibnamefont {Hunter}}, \bibinfo {author} {\bibfnamefont {S.~K.}\
  \bibnamefont {Peck}}, \bibinfo {author} {\bibfnamefont {A.~S.}\ \bibnamefont
  {Greenspon}}, \bibinfo {author} {\bibfnamefont {S.~S.}\ \bibnamefont {Alam}},
  \ and\ \bibinfo {author} {\bibfnamefont {D.}~\bibnamefont {DeMille}},\ }\href
  {\doibase 10.1103/PhysRevA.85.012511} {\bibfield  {journal} {\bibinfo
  {journal} {Physical Review A}\ }\textbf {\bibinfo {volume} {85}},\ \bibinfo
  {pages} {012511} (\bibinfo {year} {2012})}\BibitemShut {NoStop}%
\bibitem [{\citenamefont {Hendricks}\ \emph {et~al.}(2014)\citenamefont
  {Hendricks}, \citenamefont {Holland}, \citenamefont {Truppe}, \citenamefont
  {Sauer},\ and\ \citenamefont {Tarbutt}}]{Hendricks2014}%
  \BibitemOpen
  \bibfield  {author} {\bibinfo {author} {\bibfnamefont {R.~J.}\ \bibnamefont
  {Hendricks}}, \bibinfo {author} {\bibfnamefont {D.~A.}\ \bibnamefont
  {Holland}}, \bibinfo {author} {\bibfnamefont {S.}~\bibnamefont {Truppe}},
  \bibinfo {author} {\bibfnamefont {B.~E.}\ \bibnamefont {Sauer}}, \ and\
  \bibinfo {author} {\bibfnamefont {M.~R.}\ \bibnamefont {Tarbutt}},\ }\href
  {\doibase 10.3389/fphy.2014.00051} {\bibfield  {journal} {\bibinfo  {journal}
  {Frontiers in Physics}\ }\textbf {\bibinfo {volume} {2}},\ \bibinfo {pages}
  {51} (\bibinfo {year} {2014})}\BibitemShut {NoStop}%
\bibitem [{\citenamefont {Hofsäss}\ \emph {et~al.}(2021)\citenamefont
  {Hofsäss}, \citenamefont {Doppelbauer}, \citenamefont {Wright},
  \citenamefont {Kray}, \citenamefont {Sartakov}, \citenamefont {Pérez-Ríos},
  \citenamefont {Meijer},\ and\ \citenamefont {Truppe}}]{Hofsass2021}%
  \BibitemOpen
  \bibfield  {author} {\bibinfo {author} {\bibfnamefont {S.}~\bibnamefont
  {Hofsäss}}, \bibinfo {author} {\bibfnamefont {M.}~\bibnamefont
  {Doppelbauer}}, \bibinfo {author} {\bibfnamefont {S.~C.}\ \bibnamefont
  {Wright}}, \bibinfo {author} {\bibfnamefont {S.}~\bibnamefont {Kray}},
  \bibinfo {author} {\bibfnamefont {B.~G.}\ \bibnamefont {Sartakov}}, \bibinfo
  {author} {\bibfnamefont {J.}~\bibnamefont {Pérez-Ríos}}, \bibinfo {author}
  {\bibfnamefont {G.}~\bibnamefont {Meijer}}, \ and\ \bibinfo {author}
  {\bibfnamefont {S.}~\bibnamefont {Truppe}},\ }\href {\doibase
  10.1088/1367-2630/ac06e5} {\bibfield  {journal} {\bibinfo  {journal} {New
  Journal of Physics}\ }\textbf {\bibinfo {volume} {23}},\ \bibinfo {pages}
  {075001} (\bibinfo {year} {2021})}\BibitemShut {NoStop}%
\bibitem [{\citenamefont {Berkeland}\ and\ \citenamefont
  {Boshier}(2002)}]{Berkeland2002}%
  \BibitemOpen
  \bibfield  {author} {\bibinfo {author} {\bibfnamefont {D.~J.}\ \bibnamefont
  {Berkeland}}\ and\ \bibinfo {author} {\bibfnamefont {M.~G.}\ \bibnamefont
  {Boshier}},\ }\href {\doibase 10.1103/PhysRevA.65.033413} {\bibfield
  {journal} {\bibinfo  {journal} {Physical Review A}\ }\textbf {\bibinfo
  {volume} {65}},\ \bibinfo {pages} {033413} (\bibinfo {year}
  {2002})}\BibitemShut {NoStop}%
\bibitem [{\citenamefont {Gray}\ \emph {et~al.}(1978)\citenamefont {Gray},
  \citenamefont {Whitley},\ and\ \citenamefont {Stroud}}]{Gray1978}%
  \BibitemOpen
  \bibfield  {author} {\bibinfo {author} {\bibfnamefont {H.~R.}\ \bibnamefont
  {Gray}}, \bibinfo {author} {\bibfnamefont {R.~M.}\ \bibnamefont {Whitley}}, \
  and\ \bibinfo {author} {\bibfnamefont {C.~R.}\ \bibnamefont {Stroud}},\
  }\href {\doibase 10.1364/OL.3.000218} {\bibfield  {journal} {\bibinfo
  {journal} {Optics Letters}\ }\textbf {\bibinfo {volume} {3}},\ \bibinfo
  {pages} {218} (\bibinfo {year} {1978})}\BibitemShut {NoStop}%
\bibitem [{\citenamefont {Shakhmuratov}\ \emph {et~al.}(2004)\citenamefont
  {Shakhmuratov}, \citenamefont {Odeurs}, \citenamefont {Coussement},\ and\
  \citenamefont {Szabo}}]{Shakhmuratov2004}%
  \BibitemOpen
  \bibfield  {author} {\bibinfo {author} {\bibfnamefont {R.~N.}\ \bibnamefont
  {Shakhmuratov}}, \bibinfo {author} {\bibfnamefont {J.}~\bibnamefont
  {Odeurs}}, \bibinfo {author} {\bibfnamefont {R.}~\bibnamefont {Coussement}},
  \ and\ \bibinfo {author} {\bibfnamefont {A.}~\bibnamefont {Szabo}},\
  }\href@noop {} {\bibfield  {journal} {\bibinfo  {journal} {Laser Physics}\
  }\textbf {\bibinfo {volume} {1}},\ \bibinfo {pages} {39} (\bibinfo {year}
  {2004})}\BibitemShut {NoStop}%
\bibitem [{\citenamefont {Barry}\ \emph {et~al.}(2012)\citenamefont {Barry},
  \citenamefont {Shuman}, \citenamefont {Norrgard},\ and\ \citenamefont
  {DeMille}}]{Barry2012}%
  \BibitemOpen
  \bibfield  {author} {\bibinfo {author} {\bibfnamefont {J.~F.}\ \bibnamefont
  {Barry}}, \bibinfo {author} {\bibfnamefont {E.~S.}\ \bibnamefont {Shuman}},
  \bibinfo {author} {\bibfnamefont {E.~B.}\ \bibnamefont {Norrgard}}, \ and\
  \bibinfo {author} {\bibfnamefont {D.}~\bibnamefont {DeMille}},\ }\href
  {\doibase 10.1103/PhysRevLett.108.103002} {\bibfield  {journal} {\bibinfo
  {journal} {Physical Review Letters}\ }\textbf {\bibinfo {volume} {108}},\
  \bibinfo {pages} {103002} (\bibinfo {year} {2012})}\BibitemShut {NoStop}%
\bibitem [{\citenamefont {Hemmerling}\ \emph {et~al.}(2016)\citenamefont
  {Hemmerling}, \citenamefont {Chae}, \citenamefont {Ravi}, \citenamefont
  {Anderegg}, \citenamefont {Drayna}, \citenamefont {Hutzler}, \citenamefont
  {Collopy}, \citenamefont {Ye}, \citenamefont {Ketterle},\ and\ \citenamefont
  {Doyle}}]{Hemmerling2016}%
  \BibitemOpen
  \bibfield  {author} {\bibinfo {author} {\bibfnamefont {B.}~\bibnamefont
  {Hemmerling}}, \bibinfo {author} {\bibfnamefont {E.}~\bibnamefont {Chae}},
  \bibinfo {author} {\bibfnamefont {A.}~\bibnamefont {Ravi}}, \bibinfo {author}
  {\bibfnamefont {L.}~\bibnamefont {Anderegg}}, \bibinfo {author}
  {\bibfnamefont {G.~K.}\ \bibnamefont {Drayna}}, \bibinfo {author}
  {\bibfnamefont {N.~R.}\ \bibnamefont {Hutzler}}, \bibinfo {author}
  {\bibfnamefont {A.~L.}\ \bibnamefont {Collopy}}, \bibinfo {author}
  {\bibfnamefont {J.}~\bibnamefont {Ye}}, \bibinfo {author} {\bibfnamefont
  {W.}~\bibnamefont {Ketterle}}, \ and\ \bibinfo {author} {\bibfnamefont
  {J.~M.}\ \bibnamefont {Doyle}},\ }\href {\doibase
  10.1088/0953-4075/49/17/174001} {\bibfield  {journal} {\bibinfo  {journal}
  {Journal of Physics B: Atomic, Molecular and Optical Physics}\ }\textbf
  {\bibinfo {volume} {49}},\ \bibinfo {pages} {174001} (\bibinfo {year}
  {2016})}\BibitemShut {NoStop}%
\bibitem [{\citenamefont {Yeo}\ \emph {et~al.}(2015)\citenamefont {Yeo},
  \citenamefont {Hummon}, \citenamefont {Collopy}, \citenamefont {Yan},
  \citenamefont {Hemmerling}, \citenamefont {Chae}, \citenamefont {Doyle},\
  and\ \citenamefont {Ye}}]{Yeo2015}%
  \BibitemOpen
  \bibfield  {author} {\bibinfo {author} {\bibfnamefont {M.}~\bibnamefont
  {Yeo}}, \bibinfo {author} {\bibfnamefont {M.~T.}\ \bibnamefont {Hummon}},
  \bibinfo {author} {\bibfnamefont {A.~L.}\ \bibnamefont {Collopy}}, \bibinfo
  {author} {\bibfnamefont {B.}~\bibnamefont {Yan}}, \bibinfo {author}
  {\bibfnamefont {B.}~\bibnamefont {Hemmerling}}, \bibinfo {author}
  {\bibfnamefont {E.}~\bibnamefont {Chae}}, \bibinfo {author} {\bibfnamefont
  {J.~M.}\ \bibnamefont {Doyle}}, \ and\ \bibinfo {author} {\bibfnamefont
  {J.}~\bibnamefont {Ye}},\ }\href {\doibase 10.1103/PhysRevLett.114.223003}
  {\bibfield  {journal} {\bibinfo  {journal} {Physical Review Letters}\
  }\textbf {\bibinfo {volume} {114}},\ \bibinfo {pages} {223003} (\bibinfo
  {year} {2015})}\BibitemShut {NoStop}%
\bibitem [{\citenamefont {Eckel}\ \emph {et~al.}(2022)\citenamefont {Eckel},
  \citenamefont {Barker}, \citenamefont {Norrgard},\ and\ \citenamefont
  {Scherschligt}}]{Eckel2022}%
  \BibitemOpen
  \bibfield  {author} {\bibinfo {author} {\bibfnamefont {S.}~\bibnamefont
  {Eckel}}, \bibinfo {author} {\bibfnamefont {D.~S.}\ \bibnamefont {Barker}},
  \bibinfo {author} {\bibfnamefont {E.~B.}\ \bibnamefont {Norrgard}}, \ and\
  \bibinfo {author} {\bibfnamefont {J.}~\bibnamefont {Scherschligt}},\ }\href
  {\doibase 10.1016/j.cpc.2021.108166} {\bibfield  {journal} {\bibinfo
  {journal} {Computer Physics Communications}\ }\textbf {\bibinfo {volume}
  {270}},\ \bibinfo {pages} {108166} (\bibinfo {year} {2022})}\BibitemShut
  {NoStop}%
\bibitem [{\citenamefont {Gordon}\ and\ \citenamefont
  {Ashkin}(1980)}]{Gordon1980}%
  \BibitemOpen
  \bibfield  {author} {\bibinfo {author} {\bibfnamefont {J.~P.}\ \bibnamefont
  {Gordon}}\ and\ \bibinfo {author} {\bibfnamefont {A.}~\bibnamefont
  {Ashkin}},\ }\href {\doibase 10.1103/PhysRevA.21.1606} {\bibfield  {journal}
  {\bibinfo  {journal} {Physical Review A}\ }\textbf {\bibinfo {volume} {21}},\
  \bibinfo {pages} {1606} (\bibinfo {year} {1980})}\BibitemShut {NoStop}%
\bibitem [{\citenamefont {Ungar}\ \emph {et~al.}(1989)\citenamefont {Ungar},
  \citenamefont {Weiss}, \citenamefont {Riis},\ and\ \citenamefont
  {Shu}}]{Ungar1989}%
  \BibitemOpen
  \bibfield  {author} {\bibinfo {author} {\bibfnamefont {P.~J.}\ \bibnamefont
  {Ungar}}, \bibinfo {author} {\bibfnamefont {D.~S.}\ \bibnamefont {Weiss}},
  \bibinfo {author} {\bibfnamefont {E.}~\bibnamefont {Riis}}, \ and\ \bibinfo
  {author} {\bibfnamefont {S.}~\bibnamefont {Shu}},\ }\href {\doibase
  10.1364/JOSAB.6.002058} {\bibfield  {journal} {\bibinfo  {journal} {Journal
  of the Optical Society of America B}\ }\textbf {\bibinfo {volume} {6}},\
  \bibinfo {pages} {2058} (\bibinfo {year} {1989})}\BibitemShut {NoStop}%
\bibitem [{\citenamefont {Shuman}\ \emph {et~al.}(2009)\citenamefont {Shuman},
  \citenamefont {Barry}, \citenamefont {Glenn},\ and\ \citenamefont
  {DeMille}}]{Shuman2009}%
  \BibitemOpen
  \bibfield  {author} {\bibinfo {author} {\bibfnamefont {E.~S.}\ \bibnamefont
  {Shuman}}, \bibinfo {author} {\bibfnamefont {J.~F.}\ \bibnamefont {Barry}},
  \bibinfo {author} {\bibfnamefont {D.~R.}\ \bibnamefont {Glenn}}, \ and\
  \bibinfo {author} {\bibfnamefont {D.}~\bibnamefont {DeMille}},\ }\href
  {\doibase 10.1103/PhysRevLett.103.223001} {\bibfield  {journal} {\bibinfo
  {journal} {Physical Review Letters}\ }\textbf {\bibinfo {volume} {103}},\
  \bibinfo {pages} {223001} (\bibinfo {year} {2009})}\BibitemShut {NoStop}%
\bibitem [{\citenamefont {Burau}\ \emph {et~al.}(2023)\citenamefont {Burau},
  \citenamefont {Aggarwal}, \citenamefont {Mehling},\ and\ \citenamefont
  {Ye}}]{Burau2023}%
  \BibitemOpen
  \bibfield  {author} {\bibinfo {author} {\bibfnamefont {J.~J.}\ \bibnamefont
  {Burau}}, \bibinfo {author} {\bibfnamefont {P.}~\bibnamefont {Aggarwal}},
  \bibinfo {author} {\bibfnamefont {K.}~\bibnamefont {Mehling}}, \ and\
  \bibinfo {author} {\bibfnamefont {J.}~\bibnamefont {Ye}},\ }\href {\doibase
  10.1103/PhysRevLett.130.193401} {\bibfield  {journal} {\bibinfo  {journal}
  {Physical Review Letters}\ }\textbf {\bibinfo {volume} {130}},\ \bibinfo
  {pages} {193401} (\bibinfo {year} {2023})}\BibitemShut {NoStop}%
\bibitem [{\citenamefont {Brickman}\ \emph {et~al.}(2007)\citenamefont
  {Brickman}, \citenamefont {Chang}, \citenamefont {Acton}, \citenamefont
  {Chew}, \citenamefont {Matsukevich}, \citenamefont {Haljan}, \citenamefont
  {Bagnato},\ and\ \citenamefont {Monroe}}]{Brickman2007}%
  \BibitemOpen
  \bibfield  {author} {\bibinfo {author} {\bibfnamefont {K.-A.}\ \bibnamefont
  {Brickman}}, \bibinfo {author} {\bibfnamefont {M.-S.}\ \bibnamefont {Chang}},
  \bibinfo {author} {\bibfnamefont {M.}~\bibnamefont {Acton}}, \bibinfo
  {author} {\bibfnamefont {A.}~\bibnamefont {Chew}}, \bibinfo {author}
  {\bibfnamefont {D.}~\bibnamefont {Matsukevich}}, \bibinfo {author}
  {\bibfnamefont {P.~C.}\ \bibnamefont {Haljan}}, \bibinfo {author}
  {\bibfnamefont {V.~S.}\ \bibnamefont {Bagnato}}, \ and\ \bibinfo {author}
  {\bibfnamefont {C.}~\bibnamefont {Monroe}},\ }\href {\doibase
  10.1103/PhysRevA.76.043411} {\bibfield  {journal} {\bibinfo  {journal}
  {Physical Review A}\ }\textbf {\bibinfo {volume} {76}},\ \bibinfo {pages}
  {043411} (\bibinfo {year} {2007})}\BibitemShut {NoStop}%
\bibitem [{\citenamefont {Press}\ \emph {et~al.}(1986)\citenamefont {Press},
  \citenamefont {Flannery}, \citenamefont {Teukolsky},\ and\ \citenamefont
  {Vetterling}}]{Press1986}%
  \BibitemOpen
  \bibfield  {author} {\bibinfo {author} {\bibfnamefont {W.}~\bibnamefont
  {Press}}, \bibinfo {author} {\bibfnamefont {B.}~\bibnamefont {Flannery}},
  \bibinfo {author} {\bibfnamefont {S.}~\bibnamefont {Teukolsky}}, \ and\
  \bibinfo {author} {\bibfnamefont {W.}~\bibnamefont {Vetterling}},\
  }\href@noop {} {\emph {\bibinfo {title} {Numerical Recipes: The Art of
  Scientific Computing}}}\ (\bibinfo  {publisher} {Cambridge University Press,
  Cambridge},\ \bibinfo {year} {1986})\BibitemShut {NoStop}%
\bibitem [{\citenamefont {Herzberg}(1950)}]{herzDspectra1950}%
  \BibitemOpen
  \bibfield  {author} {\bibinfo {author} {\bibfnamefont {G.}~\bibnamefont
  {Herzberg}},\ }\href@noop {} {\emph {\bibinfo {title} {Spectra of Diatomic
  Molecules}}},\ Molecular Spectra and Molecular Structure\ (\bibinfo
  {publisher} {Van Nostrand},\ \bibinfo {year} {1950})\ pp.\ \bibinfo {pages}
  {392--393}\BibitemShut {NoStop}%
\bibitem [{\citenamefont {Lu}\ \emph {et~al.}(2011)\citenamefont {Lu},
  \citenamefont {Rasmussen}, \citenamefont {Wright}, \citenamefont
  {Patterson},\ and\ \citenamefont {Doyle}}]{Lu2011}%
  \BibitemOpen
  \bibfield  {author} {\bibinfo {author} {\bibfnamefont {H.-I.}\ \bibnamefont
  {Lu}}, \bibinfo {author} {\bibfnamefont {J.}~\bibnamefont {Rasmussen}},
  \bibinfo {author} {\bibfnamefont {M.~J.}\ \bibnamefont {Wright}}, \bibinfo
  {author} {\bibfnamefont {D.}~\bibnamefont {Patterson}}, \ and\ \bibinfo
  {author} {\bibfnamefont {J.~M.}\ \bibnamefont {Doyle}},\ }\href {\doibase
  10.1039/c1cp21206k} {\bibfield  {journal} {\bibinfo  {journal} {Physical
  Chemistry Chemical Physics}\ }\textbf {\bibinfo {volume} {13}},\ \bibinfo
  {pages} {18986} (\bibinfo {year} {2011})}\BibitemShut {NoStop}%
\bibitem [{\citenamefont {Chae}\ \emph {et~al.}(2017)\citenamefont {Chae},
  \citenamefont {Anderegg}, \citenamefont {Augenbraun}, \citenamefont {Ravi},
  \citenamefont {Hemmerling}, \citenamefont {Hutzler}, \citenamefont {Collopy},
  \citenamefont {Ye}, \citenamefont {Ketterle},\ and\ \citenamefont
  {Doyle}}]{Chae2017}%
  \BibitemOpen
  \bibfield  {author} {\bibinfo {author} {\bibfnamefont {E.}~\bibnamefont
  {Chae}}, \bibinfo {author} {\bibfnamefont {L.}~\bibnamefont {Anderegg}},
  \bibinfo {author} {\bibfnamefont {B.~L.}\ \bibnamefont {Augenbraun}},
  \bibinfo {author} {\bibfnamefont {A.}~\bibnamefont {Ravi}}, \bibinfo {author}
  {\bibfnamefont {B.}~\bibnamefont {Hemmerling}}, \bibinfo {author}
  {\bibfnamefont {N.~R.}\ \bibnamefont {Hutzler}}, \bibinfo {author}
  {\bibfnamefont {A.~L.}\ \bibnamefont {Collopy}}, \bibinfo {author}
  {\bibfnamefont {J.}~\bibnamefont {Ye}}, \bibinfo {author} {\bibfnamefont
  {W.}~\bibnamefont {Ketterle}}, \ and\ \bibinfo {author} {\bibfnamefont
  {J.~M.}\ \bibnamefont {Doyle}},\ }\href {\doibase 10.1088/1367-2630/aa6470}
  {\bibfield  {journal} {\bibinfo  {journal} {New Journal of Physics}\ }\textbf
  {\bibinfo {volume} {19}},\ \bibinfo {pages} {033035} (\bibinfo {year}
  {2017})}\BibitemShut {NoStop}%
\bibitem [{\citenamefont {Xu}\ \emph {et~al.}(2019)\citenamefont {Xu},
  \citenamefont {Xia}, \citenamefont {Gu}, \citenamefont {Yin}, \citenamefont
  {Xu}, \citenamefont {Xia},\ and\ \citenamefont {Yin}}]{Xu2019b}%
  \BibitemOpen
  \bibfield  {author} {\bibinfo {author} {\bibfnamefont {S.}~\bibnamefont
  {Xu}}, \bibinfo {author} {\bibfnamefont {M.}~\bibnamefont {Xia}}, \bibinfo
  {author} {\bibfnamefont {R.}~\bibnamefont {Gu}}, \bibinfo {author}
  {\bibfnamefont {Y.}~\bibnamefont {Yin}}, \bibinfo {author} {\bibfnamefont
  {L.}~\bibnamefont {Xu}}, \bibinfo {author} {\bibfnamefont {Y.}~\bibnamefont
  {Xia}}, \ and\ \bibinfo {author} {\bibfnamefont {J.}~\bibnamefont {Yin}},\
  }\href {\doibase 10.1103/PhysRevA.99.033408} {\bibfield  {journal} {\bibinfo
  {journal} {Physical Review A}\ }\textbf {\bibinfo {volume} {99}},\ \bibinfo
  {pages} {033408} (\bibinfo {year} {2019})}\BibitemShut {NoStop}%
\bibitem [{\citenamefont {Xu}\ \emph {et~al.}(2021)\citenamefont {Xu},
  \citenamefont {Kaebert}, \citenamefont {Stepanova}, \citenamefont {Poll},
  \citenamefont {Siercke},\ and\ \citenamefont {Ospelkaus}}]{Xu2021}%
  \BibitemOpen
  \bibfield  {author} {\bibinfo {author} {\bibfnamefont {S.}~\bibnamefont
  {Xu}}, \bibinfo {author} {\bibfnamefont {P.}~\bibnamefont {Kaebert}},
  \bibinfo {author} {\bibfnamefont {M.}~\bibnamefont {Stepanova}}, \bibinfo
  {author} {\bibfnamefont {T.}~\bibnamefont {Poll}}, \bibinfo {author}
  {\bibfnamefont {M.}~\bibnamefont {Siercke}}, \ and\ \bibinfo {author}
  {\bibfnamefont {S.}~\bibnamefont {Ospelkaus}},\ }\href {\doibase
  10.1088/1367-2630/ac06e6} {\bibfield  {journal} {\bibinfo  {journal} {New
  Journal of Physics}\ }\textbf {\bibinfo {volume} {23}},\ \bibinfo {pages}
  {063062} (\bibinfo {year} {2021})}\BibitemShut {NoStop}%
\end{thebibliography}%

\end{document}